\documentclass[11pt]{article}
\usepackage[left=1in,top=1in,right=1in,bottom=1in,head=.1in,nofoot]{geometry}

\usepackage{setspace,url,bm,amsmath} 

\usepackage[T1]{fontenc}
 \usepackage[utf8]{inputenc}
\usepackage[titletoc]{appendix}
\usepackage{changepage}

\usepackage{tikz}
\usetikzlibrary{decorations.pathreplacing}
\usepackage{tikz} \usetikzlibrary{calc, arrows.meta, intersections, patterns, positioning, shapes.misc, fadings, through,decorations.pathreplacing}

\usepackage{caption}
\usepackage{subcaption}
\usepackage{adjustbox}

\usepackage{xcolor}
\definecolor{myLightGray}{RGB}{191,191,191}
\definecolor{myGray}{RGB}{160,160,160}
\definecolor{myDarkGray}{RGB}{144,144,144}
\definecolor{myDarkRed}{RGB}{167,114,115}
\definecolor{myLightRed}{RGB}{255,150,100}
\definecolor{myRed}{RGB}{255,58,70}
\definecolor{myGreen}{RGB}{0,255,71}

\usepackage{array}
\usepackage{titlesec} 
\titlelabel{\thetitle.\quad} 

\usepackage{bbm}
\usepackage{latexsym}
\usepackage{natbib}
 \bibpunct{(}{)}{;}{a}{,}{,}
\usepackage{graphicx}
\usepackage{amsmath,amssymb,amsthm}
\usepackage{bm}
\usepackage{rotating}
\usepackage{rotating}
\usepackage{multicol}
\usepackage{dsfont}
\usepackage{titlesec}
\usepackage{latexsym}
\usepackage{changepage}
\usepackage{multirow}
\usepackage{booktabs}
\usepackage{pbox}
\usepackage[affil-it]{authblk}
\usepackage[titletoc]{appendix}
\usepackage{float}
\usepackage{caption}
\usepackage[linktoc=page,colorlinks,linkcolor=blue,citecolor=blue,urlcolor=blue]{hyperref}
\usepackage{fancyhdr}
\usepackage{setspace}
\usepackage{enumitem}
\usepackage{ulem}
\usepackage[margin=20pt]{subcaption}

\newcommand{\GG}[1]{}

\usepackage{amsthm}
\usepackage{comment}
\theoremstyle{definition}
\newtheorem{assumption}{Assumption}
\newtheorem*{theorem*}{Theorem}

\newtheorem{proposition}{Proposition}

\newtheorem*{corollary*}{Corollary}

\theoremstyle{remark}
\newtheorem{remark}{Remark}

\usepackage{natbib} 
\bibpunct{(}{)}{;}{a}{}{,} 

\usepackage{etoolbox} 
\apptocmd{\sloppy}{\hbadness 10000\relax}{}{} 

\usepackage{color}
\usepackage{listings}

\newcommand{\vY}{\boldsymbol{Y}}
\newcommand{\vy}{\boldsymbol{y}}
\newcommand{\vX}{\boldsymbol{X}}
\newcommand{\vx}{\boldsymbol{x}}
\newcommand{\vZ}{\boldsymbol{Z}}
\newcommand{\vz}{\boldsymbol{z}}
\newcommand{\vs}{\boldsymbol{s}}
\newcommand{\vS}{\boldsymbol{S}}
\newcommand{\E}{\mathbb{E}}

\def\ind{\begin{picture}(9,8)
         \put(0,0){\line(1,0){9}}
         \put(3,0){\line(0,1){8}}
         \put(6,0){\line(0,1){8}}
         \end{picture}
        }

\begin{document}

\title{\bf 
Forecasting Causal Effects of 
Future Interventions:\\ Confounding and Transportability Issues
} 
 \author{Laura Forastiere$^*$} 
  \author{Fan Li $^*$} 
 \author{Michela Baccini$^{**}$} 
 \affil{$^*$ Department of Biostatistics, Yale University, $^{**}$Department of Statistics, Informatics, Applications, University of Florence}
\date{}

\maketitle
\begin{abstract}
Recent developments in causal inference allow us to transport a causal effect of a time-fixed treatment from a randomized trial to a target population across space but within the same time frame. In contrast to transportability across space, transporting causal effects across time or forecasting causal effects of future interventions is more challenging due to time-varying confounders and time-varying effect modifiers. In this article, we seek to formally clarify the causal estimands for forecasting causal effects over time and the structural assumptions required to identify these estimands. Specifically, we develop a set of novel nonparametric identification formulas---g-computation formulas---for these causal estimands, and lay out the conditions required to accurately forecast causal effects from a past observed sample to a future population in a future time window. Our overarching objective is to leverage the modern causal inference theory to provide a theoretical framework for investigating whether the effects seen in a past sample would carry over to a new future population. Throughout the article, a working example addressing the effect of public policies or social events on COVID-related deaths is considered to contextualize the developments of analytical results.
\end{abstract}


\section{Introduction}
\subsection{Background}
Exposure to biochemical and physical agents in the environment, as well as bacteria or viruses circulating within human communities, can produce a wide range of adverse health consequences.  
When, on the basis of epidemiological and/or biological considerations, the harmful effect of such agents is recognized or suspected, designing and implementing policies aimed at reducing the population exposure and, in turn, its negative consequences on community health should be considered as a priority \citep{Rothman2008}.
Once a public health policy (e.g., lockdown, school closure, travel ban, vaccination campaign, traffic ban, recycling policy) has been implemented it is crucial to conduct a policy evaluation 
to assess  whether the expected results have actually been achieved, and to indicate unintended effects or unforeseen results \citep{evidencebased,mckenzie2022planning}.
In supporting evidence-based policy,
it is equally important to evaluate
the impact of circumstances, events or socio-economic interventions that are likely to increase the exposure to harmful agents (e.g. installation of power plants, incinerators or industries, mass events and social gatherings during an epidemic).

Impact assessment, conducted using causal inference methods on data from randomized experiments or observational studies, 
informs on the effectiveness of public health policies or the detrimental effect of events and interventions \citep{imbens2015causal, whatif}. 
Impact evaluation is typically carried out  by comparing two counterfactual conditions, one where the intervention or event of interest is present (treatment condition) and one where it is not (control condition). 
Common causal inference methods 
estimate causal effects by implicitly or explicitly imputing what would have happened if units under the control condition were actually assigned to treatment and vice versa \citep{imbens2015causal}.

Governments and policy-makers often rely on evidence from such evaluations to guide the design and implementation of future policies \citep{evidencebased}. When a policy is shown to be effective, it motivates governments and policy-makers to pursue its implementation and scale-up. Conversely, if evidence indicates that an event or a socio-economic intervention is harmful, efforts should be made to prevent its recurrence. However, traditional policy evaluation and health assessments of social events and socio-economic interventions are typically conducted on past interventions or observed events within the sampled population and at the time data were collected. Thus, they cannot directly inform future interventions or preventions. In fact, 
they do not address a key question: ``\textit{Will the intervention be effective when applied to the same or another target population in a future implementation?}'' 
This question is central to evaluating any intervention. While assessing whether an intervention worked as intended is important, informing decisions about its future use is equally critical. Yet, answers to such forward-looking questions often rely on informal, qualitative assessments that overlook the possibility that the estimated impact may differ in future implementations.

Oftentimes, effects vary for different types of people and in different contexts. A common concern when generalizing the effect estimated on one sample from a given population to a different population is that there may be characteristics or circumstances that differ between the two populations, modifying the treatment effect \citep{imai2008misunderstandings}. 
When using the estimated effect of an intervention to forecast the effect of implementing the same intervention in a future time, a similar concern applies. In fact, even assuming that the population were the same (same individuals living in a specific geographical area), the effects estimated from an observed sample in the past may not reflect the effects that would be seen if the intervention were implemented in the future.

Nevertheless, when there is available evidence on the casual health effect of an intervention that has already been implemented or of an event that has been observed, policy-makers and scientists advising them usually make their decisions based on the informal qualitative assumption that the impact observed in the past will be similar in the future. By doing so they rely on the (often implicit) assumption that everything will stay the same, including population behaviors and other concurrent interventions or events.
They may or may not sense that their predictions are susceptible to error due to a context change.

A more quantitative approach to forecasting the effect of a current event or a current implementation of an intervention exists and typically relies on model-based simulations, such as system dynamic models or agent-based models. 
Such model-based simulations have also been used to produce projections of hypothetical scenarios by adjusting specific model parameters. For instance, hypothetical scenarios may represent public health policies to reduce social contacts and mitigate the spread of an emerging infectious disease, explore different vaccination coverage scenarios, or examine interventions to reduce greenhouse gas emissions. 
By simulating the dynamic phenomenon under investigation, with some parameters estimated from the data (or assumed based on substantive knowledge) and other parameters modified to represent hypothetical scenarios, one can predict future dynamics under hypothetical interventions, potential future circumstances, human behaviors, or environmental emissions 
\citep[e.g.,][]{forecasting_book, frontiers2021, Covid_predictions_Jewell, VicedoCabrera2019HandsonTO}.
Such model-based projections have been used  extensively in environmental epidemiology \citep[e.g.,][]{kendrovski, Baccini64, GASPARRINI2017, adaptation2} and infectious diseases epidemiology \citep[e.g.,][]{ferguson2020impact, Flaxman2020, davies2020effects,prem2020effect,di2020expected, anderson2020will,tuite2020mathematical, naturemedicine2021}. 
Although not always, projections often deal to some extent with the problem of population characteristics changing over time by assuming hypothetical scenarios of the evolution of the population structure, or by estimating a population dynamic model based on the current data and using it to predict the population characteristics in the future time window under which the intervention is assumed to be implemented \citep{VicedoCabrera2019HandsonTO, murray2017comparison, sanchez}. 
Agent-based models, as opposed to system dynamic models, are particularly well-suited for this task given their focus on micro-level simulations that can more easily incorporate heterogeneity \citep{sonnenberg1993markov,marshall2015formalizing}.
However, this approach relies on the stability of the models, specified for time-varying covariates and outcome, across different populations and time, 
and does not account for the possibility that time-varying characteristics are also affected by the hypothesized interventions.
Furthermore, the estimation of models underpinning such simulation-based projection is not usually conducted under a causal framework; the explicit causal estimands under projection and the structural assumptions required for their nonparametric identification have not been explicitly characterized in the previous literature. 

%

\subsection{Objectives}

Although little prior work has formally described identification conditions for forecasting causal effects into future time periods, a related body of literature addresses the problem of generalizing or transporting causal effects from a randomized trial to a target population. This literature---typically concerned with point treatments and baseline effect modifiers—--has inspired numerous contributions on identifying conditions and developing statistical methods to address the challenge of trial participants differing from those in the target population \citep{cole_stuart_2010,stuart,pearl_2105,tipton,tipton14,buchanan_hudgens_2018,li2021note,li2022generalizing,dahabreh2020extending}. In this context, generalizability refers to extending inferences from a study sample to a target population when the sample is a subset of the population, whereas transportability refers to settings where the study sample is not contained within the target population \citep{westreich2017transportability,dahabreh2019extending}. In both cases, the target population often has a different distribution of baseline effect modifiers compared with individuals in the study sample. Yet the central goal for existing generalizability and transportability methods remains to assess what would have happened if a point treatment (as opposed to control) were assigned to the target population in the same time frame as the one where the experiment was conducted in the trial sample.

In contrast to the existing problem of extending inferences from a trial to a target populations, possibly across space \citep{cole_stuart_2010,tipton,hartman2015sample,muirch,rudolph2017robust}, transporting causal effects across time is more challenging due to time-varying confounders and time-varying effect modifiers. In addition, most works on generalizability to a target population are usually concerned with time-fixed treatments. Conversely, when forecasting causal effects across time, time-varying treatments (c.f. Section 19 of \citet{whatif}) should also be considered. More importantly, 
compared to generalizing to a target population where we observe the baseline effect modifiers, generalizing to a target future time window generally requires assumptions on the evolution of observable time-varying characteristics, which is not observed in the future, as well as the evolution of non-observable time-varying factors that may affect causal effects. Therefore, forecasting causal effects into the future requires a new theoretical framework to account for time-varying processes and dynamic shifts in effect modifiers that may occur beyond the time frame from which the observed data are collected.

Our primary objective in this article is to provide a potential outcomes framework to forecasting the casual effect of an intervention, that is, to investigating the extent to which the effects seen in a past sample would carry over to a new future population. Specifically, we build on the emerging literature on `generalizing' and `transporting' causal effects from randomized trials to a target population to lay out the conditions required to accurately forecast causal effects from a past observed sample to a future population, provide nonparametric identification results, and discuss the conceptual issues associated with such predictions. We first address the evaluation and the prediction of the impact of a specific intervention or event that is assumed to affect the exposure to a harmful agent. 
The term \textit{evaluation} refers to the assessment of the impact of the implemented intervention or observed event in the observed sample and in the observed time frame, while the term \textit{prediction} refers to the forecast of the impact of the same intervention or event occurring in a future time window.
Throughout, we will use the terms `prediction', `forecast', and `projection' interchangeably.
For instance, one could be interested in estimating the effect of COVID-19 restrictions implemented in the spring of 2020, and informing policy-makers on the potential impact of implementing these same restrictions in a different stage of the epidemic, when social distancing 
behaviors might have changed.
We then extend the framework to the assessment of a hypothetical intervention that modifies the exposure level, either by setting it to a specific value or by shifting the exposure distribution below a certain threshold.
Using the estimated exposure-response function, we can evaluate the effect of shifting the distribution of the exposure, and investigate what would happen if such a shift were achieved in a future situation with a possibly different exposure-response function.


The remainder of the paper is organized as follows. After introducing the notation in Section \ref{sec:notation}, we discuss in Section \ref{sec:ex} the motivating example of COVID-19. 
In Section \ref{sec:simple_scenario} we develop a framework for assessing and predicting the impact of a point intervention: we first define the causal estimands and review the ignorability assumption required to evaluate the effect of the intervention in the observed sample; we then introduce the temporal transportability assumptions required to be able to carry over the estimated effect to a future time.
In Section \ref{sec:intervention}, we extend the framework to settings with time-varying treatments, with and without duration, under the presence or absence of time-varying confounders affected by previous treatments. 
We discuss the implications of the identification conditions and the possible issues that would invalidate them in Section \ref{sec:issues}. In Section \ref{sec:hyp} of the Appendix, we extend this framework to hypothetical interventions on the exposure distribution. Section \ref{sec:dis} concludes the paper with some discussion.

\section{Notation and Set Up}
\label{sec:notation}

We consider a sample $\mathcal{I}$ of $I$ units, indexed by $i=1,\ldots,I$, observed over a temporal window $\mathcal{T}$ of $T$ time points (e.g., days, years), $t\in \mathcal{T}=[1, \ldots,T]$.  To contextualize the notation without loss of generality, we can consider the units in the sample as geographical areas (e.g., cities, states, or countries). 
Let $U^{obs}=\{it: i\in \mathcal{I}, t\in \mathcal{T}\}$ be the set of observations on the $I$ units in the temporal window $\mathcal{T}$.
We also assume that the set of units $\mathcal{I}$ corresponds to our target population where we would like to assess the effect of an intervention.


Let us denote with $S_{it} \in \mathcal{S}$ a continuous (univariate or multivariate) exposure, measured on unit $i$ at time $t$,
that is supposed to potentially affect some outcome of interest.
We consider an intervention that is designed to set the value or shift the distribution of such an exposure. We define $Z_{it}\in\{0,1\}$ as the intervention indicator, being 1 if a specific intervention of interest is implemented on the unit $i$ at time $t$ and 0 otherwise.
Alternatively, $Z_{it}$ may be the 
indicator of an event that could affect the distribution of the exposure $S_{it}$ and, in turn, could have an effect on the outcome of interest. Thus, throughout, we refer to $Z_{it}$ as the `treatment' indicator and we use the terms intervention, event or treatment, interchangeably. 
We let $\vZ_{t}$ be the vector of the treatment indicators at time $t$ for the target population, $\overline{\vZ}_i=\{Z_{i1},\dots,Z_{iT}\}$ the vector of treatment indicators for the area $i$ during the observed time window, and $\vZ$ the treatment matrix in the whole observed sample. 
Moreover, we denote by $\overline{\vZ}_{it}^{M,L}=\{Z_{i,(t-M-
L)},Z_{i,(t-M-
L+1)},\dots,Z_{i,(t-M)}\}$
the treatment history for unit $i$ in the time window between $t-M-L$ and $t-M$, with $L,M\geq0$. 

We let $Y_{it}^{obs}$ be the observed outcome for unit $i$ at time $t$, 
which is presumed to be a function of previous exposure history 
and, in turn, by previous treatments
.
In the case of units being geographical areas, the treatment and exposure variables are defined at the area-level, and the outcome variable $Y_{it}^{obs}$ is usually an aggregate function of individual-level outcomes (e.g. daily number of deaths, number of hospital admissions). 
We let $\overline{\vY}_{it}^{M,L}=\{Y^{obs}_{i,(t-M-
L)},Y^{obs}_{i,(t-M-L+1)},\dots,Y^{obs}_{i,(t-M)}\}$ be the outcome history of unit $i$ in the time window $[t-M-L, t-M]$. 
To maintain a causal order between variables,
we assume that, for each time point $t$, the outcome 
is actually measured sometime after the exposures
are collected or the interventions 
are implemented, but before time point $t+1$. 

To complete the notation specification, we let $\vX_{it}$ be a vector of covariates measured on unit $i$ during time $t$. 
The characteristics collected in $\vX_{it}$ could be confounders of the relationship between the intervention and the outcome and/or effect modifiers; we do not distinguish between them for notational simplicity. 
The specific role as confounders or modifiers will be explicit as they are included in the different conditional sets of the identifying assumptions. We assume that the covariates corresponding to time $t$ are collected just prior to the exposures and the implementation of the interventions at time $t$.
We let $\vX_{i0}$ be the vector of time-invariant baseline covariates of unit $i$.
Finally, we denote by $\overline{\vX}_{it}^{M,L}=\{\vX_{i,(t-M-
L)},\vX_{i,(t-M-
L+1)},\dots,\vX_{i,(t-M)}\}$ the collection of time-varying covariates of unit $i$ in the time window $[t-M-L, t-M]$.

Finally, throughout, we use a stochastic (or model-based) perspective to causal inference, where all variables including exposures, treatments, outcomes, and covariates, are considered as random variables \citep[e.g.,][]{whatif, Li2022BayesianCI,Zigler2020BipartiteIA}. However, here our causal estimands will be based on finite sample averages of mean potential outcomes conditioned on the time-invariant baseline covariates $\vX_{i0}$, as our goal is to forecast on the same locations $\mathcal{I}$ the effect of an intervention or event observed in the past in a future time window where the distribution of time-varying covariates may be different. Furthermore, we assume that the conditional distribution of potential outcomes is homogeneous across units.

\section{A Motivating Example: The Effect of Public Policies or Social Events on Covid-related Deaths}
\label{sec:ex}
The global COVID-19 pandemic has underscored the critical role of modeling infectious disease transmission dynamics, not only to characterize the epidemiological features of the disease and assess the current state of an outbreak, but also to forecast its future trajectory. Simulation-based methods have been used to make projections about the impact of possible interventions \citep[e.g.,][]{davies2020effects, prem2020effect,anderson2020will}. 
The outputs of these models have helped healthcare systems make operational plans and public health officials make decisions about mitigation measures.
Predicting the health effects of mitigation actions can further help policy makers prioritize investments. For instance, in 2020, model-based forecasts about the spread of COVID-19 led to national lockdowns and border closures \citep{adam2020}.

Given the widespread use of model-based projections during the COVID-19 pandemic, we use this setting as a motivating example.
Suppose that we want to investigate the effect of an intervention or event that could affect the spread of SARS-CoV-2 infections and, in turn, subsequent COVID19-related mortality. 
Suppose that our observed sample is the United States population and that the observed temporal window spans from January to July 2020, corresponding to the first wave of the SARS-CoV-2 epidemic. We could be interested in the effect of a public policy aimed at reducing the number of potential contagious interactions (e.g. the introduction of social distancing measures, gathering restrictions, school closures and distance learning, stay-at-home orders or lockdowns), or in the effect of planned or spontaneous social events that could have increased the number of potential contagious interactions (e.g. rallies, protests, in-person elections). 

In this setting, the presence or absence of the intervention or event of interest in a given area $i$ (e.g., region, county or state) at time $t$ defines the intervention indicator $Z_{it}$, while the number of potentially contagious interactions between infectious and susceptible people in $i$ at time $t$, is the exposure variable $S_{it}$, which should be affected intentionally by the intervention or unintentionally by social events -- $S_{it}$ is proportional to the effective reproduction number $R_t$ and the number of circulating infected individuals in the area \citep{adam2020guide, Inglesby2020PublicHM,Rubin2020AssociationOS, BrooksPollock2021MappingSD}. We assume that the outcome of interest $Y_{it}$ is the daily number of COVID-19 related deaths 
in the area $i$ at time $t$. 
Potential confounders and/or effect modifiers to be included in $\vX_{it}$ are: variables related to population composition and population health (e.g. percentage of elderly, percentage of people with chronic diseases), variables related to local risk factors (e.g. air pollution level, temperature, humidity, population density, use of public transportation) and to characteristics of the healthcare system (e.g. number of hospital beds, number of intensive care beds, indicators of accessibility to care), the epidemic status before time $t$ (e.g. number of notified infections), as well as the tendency of the population to engage in risky behaviors or follow public health recommendations (e.g., mask wearing). 

During the first wave of the SARS-CoV-2 epidemic, several state-level and national-level restrictions were put in place to slow the spread of infections. Meanwhile, social events that gathered large number of people in limited spaces occurred during the same months, whether as a result of discontent with the imposed restrictions, as a response to socio-political issues, or as part of the national electoral calendar.
Researchers from different fields mobilized rapidly to estimate the impact that the implemented restrictions and the co-occurring events have had on the epidemic dynamics, in order to provide insights into contagion dynamics and inform policy-makers about the effectiveness of the implemented measures. Here, the causal question of interest concerns the effect of a time-varying exposure or treatment that has been observed in a sample of the population. 
A methodological framework addressing these issues has already been developed in both fields of infectious disease epidemiology and causal inference. To answer these COVID-19-related questions, researchers have employed a range of estimation approaches, including mechanistic epidemiological models \citep[e.g.,][]{Chinazzi2020TheEO,Zhang2022EvaluatingTI} and causal inference methods such difference-in-differences \citep[e.g.,][]{Palguta2021DoEA, VELIAS2022114538}.

After the summer, with the resumption of the epidemic, it became crucial to use evidence from the spring to answer a different question: ``\textit{Will the policies applied during the spring 2020 have the same effect months later, during the second epidemic wave? Or will their impact change?}'' A prediction of the causal impact of applying the same restrictions in the future, possibly during a second wave of COVID-19, 
was of paramount importance for informing actions to control the second outbreak. 
Similarly, still with reference to the epidemic dynamics, it was of interest to predict the impact that social events that had already happened in the past, would have if repeated in the future. 
To answer this kind of research question, one might be tempted to assume that the effects estimated for the spring of 2020 would hold constant when the same interventions or events recur in the future, as long as they apply to the same population.
The problem with this assumption is that the context in which new restrictions or elections are implemented may differ meaningfully from the context in which they were first applied.
Moreover, the altered context could reflect not only the natural course of epidemic dynamics but also the cumulative impact of earlier interventions.
In Section \ref{sec:pred}, we clarify the assumptions needed to identify predicted causal effects and highlight potential challenges and possible solutions.

Furthermore, policy-makers may want to plan new restrictions designed to produce a pre-specified reduction in the number of contagious interactions between infected and susceptible. In this case, we are dealing with the evaluation of a hypothetical intervention that is assumed to modify the distribution of the exposure $S$. To inform decision-makers, several epidemiologists have used epidemiological models to predict the reduction in COVID-19-related morbidity and mortality that could result from hypothetical interventions leading to a certain reduction in the average number of contagious contacts to a certain value  \citep{di2020expected, prem2020effect, davies2020effects, ferguson2020impact, anderson2020will, naturemedicine2021}.
However, the use of epidemiological models to answer these kinds of questions relies on strong parametric assumptions. Moreover, the actual aim of these analyses was to predict effect of hypothetical interventions in the future, not within the observed temporal window. Therefore, the aforementioned issues related to predictions of causal effects remain relevant in this context.




\section{Assessing and Predicting the Impact of an Actual Point Treatment}
\label{sec:simple_scenario}
We begin by considering the simple scenario where a point-treatment is observed for some units in $\mathcal{I}$ at time $S$, with $1<S<T$. For instance, the point treatment could be a protest that occurred in some geographical areas in the same day. 
Here, for the sake of simplicity, we assume that the treatment has an immediate effect on the outcome, and we are interested in estimating such a short-term effect. In our motivating example, this means that the occurrence of a protest could affect COVID-related deaths immediately after the protest. 

\subsection{Definition of potential outcomes}
We define $Y_{iS}(\vz)$ as the potential outcome of unit $i$ at time $S$ under a treatment matrix $\vZ$ collecting the treatments for the whole sample equal to $\vz$. In this paper, we assume that there is no interference between areas, that is, the potential outcome of unit $i$ at time $S$ can only depend at most on the treatments at location $i$, i.e., $\overline{\vZ}_i$ \citep{rubin1980discussion}.
The no-interference assumption is usually valid when population units do not interact with each other or when their interaction does not lead to interference mechanisms.  In the case of infectious diseases, although the spread of the disease does occur through physical interactions invalidating the no-interference assumption across individuals, oftentimes we can still assume the absence of interference across geographical areas \citep{Hudgens2008TowardCI}.
This is the case when the geographical areas in the sample $\mathcal{I}$ are confined and distant from each other or when there is little movement between areas. 
During the COVID-19 emergency in spring 2020, given national travel restrictions and limited movement, it is likely that any intervention implemented in one state did not have an effect on the level of contagion in other states \citep[e.g.,][]{Feltham2020MassGF}. For this reason, in our motivating example we use states as units of analysis, indexed by $i$. 
Under the above no-interference assumption, each potential outcomes can be indexed by the area-specific treatment vector, i.e., $Y_{iS}(\vz_i)$. 
However, here we consider a point treatment that occurred for some units at time $S$. Therefore, the treatment sequence over time for unit $i$ is of the form $\vz=[0, \dots, 0, z,0, \dots, 0]$. In this case, we can simply index the potential outcome at time $S$ only by the point treatment at time $S$, i.e., $Y_{iS}(z)$.

\subsection{Actual impact on the observed sample}
Suppose that we want to evaluate the effect of an intervention or event that occurred for some units at time $S$ on the outcome measured immediately after. 
The impact of an intervention or event can be defined by several causal estimands. Here, we focus on the average treatment effect on the treated (ATT), but other causal estimands could also be of interest, e.g., the average treatment effect (ATE) or their ratio counterparts. 
Formally, the ATT is the average comparison between the two potential outcomes under $Z_{iS}=0$ and $Z_{iS}=1$ across treated units:
\begin{equation}
\label{eq:ATT_4}
\text{ATT}=
\frac{1}{N_1^{obs}}\sum_{i\in{U_1^{obs}}}
\Big(\E\Big[Y_{iS}(1)| \vX_{iS}, \vX_{i0}\Big]-\E\Big[Y_{iS}(0)| \vX_{iS}, \vX_{i0}\Big]\Big)
\end{equation}
where $U_1^{obs}=\{i \in \mathcal{I}: Z_{iS}=1\}$ is the set of units that received the treatment at time $S$, and $N_1^{obs}=|U_1^{obs}|$ is the number of treated units. 
Under our stochastic perspective, where potential outcomes are seen as random variables, the expectations in  \eqref{eq:ATT_4} are
taken over the 
conditional 
distribution of potential outcomes, conditional on each unit's baseline covariates $\vX_{i0}$. As already mentioned, we assume that the conditional expectation $\E\Big[Y_{iS}(z)| \vX_{iS}, \vX_{i0}\Big]$ is homogeneous across units in $\mathcal{I}$ with the same value of covariates $\vX_{iS}$ and  $\vX_{i0}$.
The estiimand in \eqref{eq:ATT_4} can be seen as a finite sample estimand under a stochastic perspectives and again is due to our focus on the sample of units $\mathcal{I}$. Oftentimes, it is written as an expectation over the empirical distribution of covariates \citep{Zigler2020BipartiteIA}.
With reference to the case in which the outcome is a count of sanitary events (e.g. number of deaths) and the treatment is a protest, Equation \eqref{eq:ATT} quantifies the average change in the number of sanitary events in the areas where protests took place at time $S$ that occurred right after the protests and that are attributable solely to the protests. 

\subsection{Identification under ignorability}
\label{sec:ign}
The fundamental problem to identify the average treatment effect on the treated (Equation \eqref{eq:ATT_4}) is that the potential outcomes $Y_{iS}(0)$ are not observed for the units in $U_1^{obs}$. In fact, in the observed sample $U^{obs}$ we only observe the potential outcome corresponding to the treatment that was actually received. 
In particular, for the estimation of the ATT, we need to impute $Y_{iS}(0)$ for the treated units in $U_1^{obs}$.
This imputation can be done implicitly or explicitly in different ways, depending on the identifying assumption that we are willing to make. 
Different causal inference methods rely on different identifying assumptions (e.g., the parallel trend assumption for the difference-in-difference estimator) \citep{whatif}. Here, we briefly review results under the most commonly used exchangeability assumption. 

In the presence of control units who did not received the treatment (i.e., the intervention was not implemented or the event did not occur), we can impute the potential outcomes $Y_{iS}(0)$, missing for the treated units, using the distribution of the observed outcomes on the control units. This imputation relies on the assumption that treated and control units are exchangeable, that is, the distribution of the potential outcome are independent of the actual treatment status. This exchangeability assumption, also called treatment ignorability or unconfoundedness, holds whenever the treatment is randomized \citep{whatif, imbens2015causal}. On the other hand, in non-randomized observational studies, ignorability may be more plausible conditional on covariates. 
Formally, we write the following assumption.

\begin{assumption}[Treatment Ignorability]
\label{ass:Ign_4}
\textit{The treatment status $Z_{iS}$ at time $S$ is independent of potential outcomes given baseline and most recent covariates, i.e.,
\begin{align}
\hspace{1cm}
Y_{iS}(z) \ind Z_{iS} |  \vX_{iS}, \vX_{i0},~~~z\in\{0,1\}
\end{align}}
\end{assumption}
\noindent Assuming ignorability within strata of covariates implies assuming that there are no unmeasured confounders between exposure and outcome. 
Note that in this section, for simplicity,  in Assumption \ref{ass:Ign_4} we assume that 
the relationship between the treatment and outcome at time $S$ is only confounded at most by baseline covariates, $\vX_{i0}$, and covariates at the same time point $S$, although measured before the treatment and outcome, and not by past trends.
In our motivating examples on protests, potential confounders 
of the relationship between the occurrence of a protest and COVID-related deaths may include variables related to population size and composition, including political beliefs, epidemic status, as well as public health interventions and recommendations together with the population compliance information. 
Note that these variables can also be effect modifiers, especially the epidemic status and population behaviors.

Under Assumption \eqref{ass:Ign_4}, the average treatment effect on the treated is identified from the observed data as follows:
\begin{equation}
\label{eq:att_id_4}
ATT=\frac{1}{N_1^{obs}}\sum_{i\in{U_1^{obs}}}\Big(\E\Big[Y_{iS}^{obs}| Z_{iS}=1,\vX_{iS}, \vX_{i0}\Big]-
\E\Big[Y_{iS}^{obs}| Z_{iS}=0,\vX_{iS}, \vX_{i0}\Big]\Big)
\end{equation}
Under Equation \eqref{eq:att_id_4}, 
different regression methods can be used for the estimation of the ATT \citep[e.g.,][]{whatif}.
If we could rule out the presence of unobserved time-varying factors affecting the distribution of potential outcomes under control, the conditional expectation 
$\E\Big[Y_{iS}(0)| \vX_{iS}, \vX_{i0}\Big]$
could also be identified by borrowing information on the observed outcome across the whole observed time window $\mathcal{T}$, where the treatment was not observed.

\subsection{Predicting the impact on future observations}
\label{sec:pred_4}

\subsubsection{Causal estimands}
Now, let us assume that we have already estimated the effect of a point intervention or a point event observed in $\mathcal{I}$ at time $S$. We now wish to predict the impact of the same point intervention or point event if it was replicated on the same population $\mathcal{I}$ but at a future time point $T+F$. We recognize that the term `prediction' could lead to some confusion. To clarify, we are not interested in forecasting the long-term effect of a treatment or event that we have already observed. Instead, we are interested in forecasting the effect of an intervention or an event that we wish to administer or that could occur at a future time. In order to achieve this objective, we assume that the intervention or event has been already observed on a past sample and its effect has been estimated (Section \ref{sec:ign}). 
It is also worth noting that here we are not dealing with the problem of generalizing a causal effect to a different population of units, a problem that has been studied elsewhere \citep{cole_stuart_2010,tipton,hartman2015sample,muirch}. Instead, here we address the issue of transporting the causal effect to future observations of the same population. Although possible, an extension of our results to generalize an estimated causal effect to future observations but in a different population is possible but beyond the scope of this paper.

At the end of the first wave of COVID-19, policy-makers were interested in designing effective strategies that they would have implemented in the case of a second wave. Moreover, at the time where we did see a surge in COVID-19 in the Fall 2020, policy-makers would have benefited from insights into the possible effect of implementing the same restrictions as in the Spring 2020. Similarly, they might want to prevent social events that occurred in that period and were estimated to have an effect on the spread of COVID-19, or allow the ones that were proven not to have an effect. As an example, the estimated effect of primary elections that took place during the first wave has informed on whether to hold in-person voting during the US presidential elections in Fall 2020 \citep{Feltham2020}. 
The prototype of these questions is to forecast the effect of some treatment---the impact of which has already been assessed on an observed sample---if it were to be assigned in the same geographical area but in a future time point. 

We therefore define the average causal effect of the treatment in the future time point $T+F$ as:
\begin{equation}
\label{eq: ATTF_4}
\text{ATT}_\text{F}=\frac{1}{|\mathcal{I}_F|}\sum_{i\in\mathcal{I}_F}
\Big(\E\Big[Y_{iT+F}(1)| \vX_{i0}\Big]-\E\Big[Y_{iT+F}(0)| \vX_{i0}\Big]\Big)
\end{equation}
where $\mathcal{I}_F\subseteq \mathcal{I}$ is a subset of the set of the sample locations where the intervention or event is assumed to occur at time $T+F$. For instance, one can set $\mathcal{I}_F=U_1^{obs}$ such that ATT$_\text{F}$ is defined for the same set of locations that received the treatment at time $S$. Notice that here, as opposed to Equation \eqref{eq:ATT_4}, we do not condition on time-varying covariates at time $T+F$ as they are not measured. The expectation in \eqref{eq: ATTF_4} is then taken also with respect to these covariates.

\begin{figure}[t]
\centering

\begin{tikzpicture}[%
    every node/.style={
        font=\scriptsize,
        text height=1ex,
        text depth=.25ex,
    },
]

\draw[->] (0,0) -- (7.5,0);

\foreach \x in {0,1,...,7}{
    \draw (\x cm,3pt) -- (\x cm,0pt);
}

    \draw (0cm,3pt) -- (0 cm,-3pt);
    \draw (4cm,3pt) -- (4 cm,-3pt);
    \draw[dashed] (4cm,2cm) -- (4 cm,-2cm);
    \draw (6cm,3pt) -- (6 cm,-3pt);

\draw[-latex] (2 cm,20pt) -- (2 cm,0pt);
\node[text width=0.5cm] at (1.9,0.9) {$Z_{iS}$};
\draw (3 cm,20pt) -- (3 cm,10pt);
\node[text width=0.5cm] at (2.9,0.9) {$Y_{iS}(z)$};
\draw[-latex] (6 cm,20pt) -- (6 cm,0pt);
\node[text width=0.5cm] at (5.7,0.9) {$Z_{iT+F}$};
\draw (7 cm,20pt) -- (7 cm,10pt);
\node[text width=0.5cm] at (6.9,0.9) {$Y_{iT+F}(z)$};

\node[anchor=north] at (0,0) {$0$};
\node[anchor=north] at (2,0) {$S$};
\node[anchor=north] at (4,0) {$T$};
\node[anchor=north] at (6,0) {$T\!+\!F$};
\node[anchor=north] at (7.5,0) {time};

\fill[myGray] (0,0.25) rectangle (2,0.4);
\fill[myRed] (2,0.25) rectangle (3,0.4);
\fill[myGray] (3,0.25) rectangle (4,0.4);
\filldraw[draw=myLightGray,fill=white] (4,0.25) rectangle (7,0.4);

\filldraw[draw=myRed, dashed,fill=white, line width=1pt] (6,0.25) rectangle (7,0.4);

\draw[decorate,decoration={brace,amplitude=5pt}] (4,-0.6) -- (0,-0.6)
    node[anchor=south,midway,below=4pt] {Observed window}
    node[anchor=south,midway,below=15pt] {$\mathcal{T}$};
;
\end{tikzpicture}
 \caption{Timeline and notation for point-treatments.
}
\label{fig:future}
\end{figure}

\subsubsection{Identification under temporal transportability}
\label{sec:transp_4}
The first problem that arises in forecasting a causal effect is that both potential outcomes $Y_{iT+F}(1)$ and $Y_{iT+F}(0)$ in \eqref{eq: ATTF_4} are missing for all locations in $\mathcal{I}_F$. In fact, 
we have not yet applied the treatment on any units in the future window. $Y_{iT+F}(0)$ is also missing because we actually do not observe the outcome in the future time point, not even under the control condition. 
The simplest solution that is commonly  taken is to assume that the effect of a treatment, estimated on an observed sample, will be constant if replicated on the same population in the future. 
This is what we tend to assume in our daily life when we expose ourselves to the same event that benefited us in the past, predicting that it will have the same effect on another occasion.
This is also what it is implicitly done when the effect of policies or interventions is estimated to with the purpose of informing policy-makers on the best strategy.  
The main problem of making this assumption of constant effect over time is that a unit's response to a treatment 
usually depends on the context, which could change over time. 
Intuitively, if the treatment effect only depends upon time-invariant characteristics, the average treatment effect estimated on a population of units would be constant if the treatment were assigned on the same population in the future. 
In reality, the causal effect of a treatment, estimated in a past time period, could reflect the time-varying characteristics of that period and is subject to change in a future time when those characteristics evolve over time. It is, in fact, the presence of time-varying effect modifiers that makes the prediction of causal effects challenging.

Given that data is only observed in $\mathcal{T}$, one would need to leverage potential outcomes identified and estimated in the past to predict the causal effect at a future time point. We introduce here the assumptions and identification strategy that justify common approaches.

To formalize the transportability of causal effects, we define a mixture population where the time of observation is treated as a random variable. 

The general version of the temporal transportability assumption in a temporal domain $\mathcal{D}$ is stated as follows.

\begin{assumption}[Temporal Transportability of Potential Outcomes] 
\label{ass:transp_4}
\textit{Let $A_i$ be a random selection variable taking values in a temporal domain $\mathcal{D}$. 
We define the composite potential outcome $Y_i(z)$ and the composite covariate vector $X_i$ as:
\begin{equation*}
        Y_i(z) = \sum_{t \in \mathcal{D}} Y_{it}(z) \cdot \mathbb{I}(A_i = t)
\qquad
    X_i = \sum_{t \in \mathcal{D}} X_{it} \cdot \mathbb{I}(A_i = t)
\end{equation*}
For all units $i \in \mathcal{I}_F$, the selection variable $A_i$ is independent of the potential outcomes given the observed time-varying covariates and baseline history:}
\begin{equation}
    Y_i(z) \ind A_i \mid X_i, X_{i0} 
\end{equation}
\end{assumption}

In the simplest case, we consider transportability between the past observed time point $S$ and the future target time point $T+F$, such that $\mathcal{D} = \{S, T+F\}$. 
\noindent Under Assumption \ref{ass:transp_4}, the distribution of potential outcomes is the same between the observed time point $S$ and the future time point $T+F$ for units with the same covariates $\vX_{it}$ and $\vX_{i0}$, i.e., $P(Y_{iS}(z) \mid X_{iS}, X_{i0}) = P(Y_{i,T+F}(z) \mid X_{i,T+F}, X_{i0})$.
These covariates play the role of effect modifiers. Assumption \ref{ass:transp_4} then implies that all effect modifiers are included in $\vX_{it}$ and $\vX_{i0}$.
As a consequence,
 both missing potential outcomes in \eqref{eq: ATTF} can be imputed, implicitly or explicitly, conditional on effect modifiers from a set of control and treated units observed in the past. Formally, under Assumption \ref{ass:transp_4}, conditional expectations of  potential outcomes for units in the subset $\mathcal{I}_F$ are identified as in the following proposition.

\begin{proposition}[Identification via Temporal Transportability]
\label{prop:transp_4}
\textit{
Under Assumption \ref{ass:transp_4}, and further assuming the following conditions for all $i \in \mathcal{I}_F$:}
\begin{enumerate}
    \item {Consistency:} $Y^{obs}_{iS} = Y_{iS}(z)$ if  $Z_{iS} = z$;
    \item {Positivity:} $0 < P(Z_{iS} = 1 \mid \mathbf{X}_{iS}, \mathbf{X}_{i0}) < 1$;
\end{enumerate}
\textit{the expected potential outcome at time $T+F$ is identified as:}
\begin{equation}
\begin{aligned}
\label{eq:predATT_4}
    \mathbb{E}\big[Y_{i,T+F}(z) \mid \mathbf{X}_{i0}, i \in \mathcal{I}_F\big] = 
    \sum_{\mathbf{x}} \mathbb{E}\big[Y^{obs}_{iS} \mid Z_{iS}=z, \mathbf{X}_{i,S}=\mathbf{x}, \mathbf{X}_{i0}\big] f_{\mathbf{X}_{i,T+F}}(\mathbf{x} \mid \mathbf{X}_{i0}, i \in \mathcal{I}_F) 
\end{aligned}
\end{equation}
\textit{where the marginalizing distribution $f_{\vX_{iT+F}}(\vx|\vX_{i0}, i\in \mathcal{I}_F)$ is the conditional distribution for time-varying covariates at the future time point $T+F$\footnote{For simplicity, we present the identification results for discrete covariates using summations and density functions. These results generalize to continuous or mixed covariate spaces by replacing the summations with integrals with respect to the appropriate probability measures.}.}
\end{proposition}
\noindent Proposition \ref{prop:transp_4} confirms that the outcome mechanism estimated at time $S$ can be validly applied to the covariate distribution expected at time $T+F$.


If we instead let $\mathcal{D} = \mathcal{T} \cup \{T+F\}$, Assumption \ref{ass:transp_4} states that the distribution of potential outcomes is stable across the entire observed history and the future point. Under this broader independence assumption, the conditional expectation of potential outcomes is invariant for any $t \in \mathcal{D}$. 
This allows us to identify the potential outcome under control at time $T+F$ by pooling information across the entire observed window $\mathcal{T}$, i.e.,:
\begin{equation}
\begin{aligned}
\label{eq:predATT0_4}
    \mathbb{E}\big[Y_{i,T+F}(0) \mid \mathbf{X}_{i0}, i\in \mathcal{I}_F\big] = 
    \sum_{\mathbf{x}} \mathbb{E}\big[Y^{obs}_{it} \mid Z_{it}=0, \mathbf{X}_{it}=\mathbf{x}, \mathbf{X}_{i0}, t \in \mathcal{T}\big] f_{\mathbf{X}_{i,T+F}}(\mathbf{x} \mid \mathbf{X}_{i0}, i \in \mathcal{I}_F)
\end{aligned}
\end{equation}

Nevertheless, the identification of future causal effects based on the temporal transportability assumption (Assumption \ref{ass:transp_4}) relies on the effect modifiers being known in both the observed time window and the future period.
However, the effect modifiers in the future time point, i.e., $\vX_{iT+F}$, are not observed.
In the literature related to prediction, when the effect is not assumed constant but time-varying effect modifiers are assumed to be present, two approaches are used. The first approach considers hypothetical  scenarios that could determine a different predicted effect of a treatment in the future, i.e., effect modifiers ${\vX}_{iT+F}$  are set or drawn from a distribution \citep{guidelines, VicedoCabrera2019HandsonTO}. 
However, setting the effect modifiers to hypothetical values leads to a predicted causal effect under hypothetical scenarios that are only qualitatively informed by past data and that are not predicted from past outcome, covariate and treatment history. In a way, this approach does not forecast a causal effect in the future, but rather estimates the causal effect of both implementing the intervention of interest and setting the context in which the intervention will be rolled out. 
On the contrary, a second approach forecasts the evolution of effect modifiers using a dynamic model . Then, in turn, we predict the effect of implementing a treatment in the future relying on the estimated treatment effect heterogeneity \citep{sonnenberg1993markov,marshall2015formalizing}. Importantly, this approach hinges on the assumption that effect modifiers follow the same model as in  the past observed sample. We formalize here the conditions under which this approach is valid, and we provide the identification result supporting this approach.

To identify future causal effects, we must account for the evolution of time-varying effect modifiers. We define the historical context for each unit $i$ at time $t$ using the variable $\mathcal{H}_{it} = \{ \overline{\vX}_{it}^{1, L_{xx}}, \overline{Z}_{it}^{1, L_z}, \vX_{i0} \}$, where $\overline{\vX}_{it}^{1, L_{xx}}$ and $\overline{Z}_{it}^{1, L_z}$ represent the covariate and treatment history up to lags $L_{xx}$ and $L_z$.

\begin{assumption}[Temporal Transportability of Time-varying Modifiers]
\label{ass:conf_transp_4}
\textit{
Let $A_i$ be a random selection variable taking values in a temporal domain $\mathcal{D}$. For any $t \in \mathcal{D}$, we define the composite covariate vector $X_i$ and composite history as:
\begin{equation*}
    X_i = \sum_{t \in \mathcal{D}} X_{it} \cdot \mathbb{I}(A_i = t);
    \qquad
    \mathcal{H}_i = \sum_{t \in \mathcal{D}} \mathcal{H}_{it} \cdot \mathbb{I}(A_i = t).
\end{equation*}
Then, for all units $i \in \mathcal{I}_F$, the selection variable $A_i$ is independent of the current time-varying covariates given the preceding covariate and treatment  history:}
\begin{equation}
    X_i \ind A_i \mid \mathcal{H}_i
\end{equation}
\end{assumption}
\noindent Assumption \ref{ass:conf_transp_4} asserts that the conditional transition distribution of the covariates is invariant across the domain $\mathcal{D}$. 
Note that $L_z$ and $L_{xx}$ define the presence of a lagged effect of past treatments and covariates on future covariates, respectively, and need to be specified as part of Assumption \ref{ass:conf_transp_4}.  Thus, Assumption \ref{ass:conf_transp_4} 
rules out the presence of events that would change the way effect modifiers evolve over time, and 
 implies that their evolution learned from the observed past is valid for the unobserved future.
Note that in the setting with a point-treatment only observed for some units at time $S$, the treatment history $\overline{Z}_{it}^{1, L_z}$ will simply be $\overline{Z}_{it}^{1, L_z}=[0,\dots, 0, Z_{iS}, 0, \dots, 0]$ if $S\in [t-1-L_z,t-1]$ or $\overline{Z}_{it}^{1, L_z}=\mathbf{0}$ otherwise.

Under Assumption \ref{ass:conf_transp_4}, the unknown distribution of effect modifiers at time $T+F$, conditional on the subset $\mathcal{I}_F$ and baseline covariates $\vX_{i0}$, which is needed in the identifying quantity in Equation \eqref{eq:predATT_4}, 
is given by the following recursive identification.

\begin{proposition}[Identification of Future Covariate Distributions]
\label{prop:recursive_covariates}
\textit{Suppose Assumption \ref{ass:conf_transp_4} holds for the domain $\mathcal{D}=\mathcal{T}\cup\{T+1, \ldots, T+F\}$. Further assume the recursive positivity condition: for all $i \in \mathcal{I}_F$, the support of the predicted history is contained within the observed historical support, $\text{supp}(f_{\mathcal{H}_{it}} \mid  t \in \{T+1, \ldots, T+F\}) \subseteq \text{supp}(f_{\mathcal{H}_{it} \mid t \in \mathcal{T}})$. Then, the distribution of the covariates at the future time point $T+F$ is identified by:}
\begin{equation}
\label{eq:predX_4}
\begin{aligned}
f_{\mathbf{X}_{i,T+F}}(\mathbf{x} \mid \mathbf{X}_{i0}, i \in \mathcal{I}_F) = \sum_{\mathbf{v}} & f_{\mathbf{X}_{it}}(\mathbf{x} \mid \overline{\mathbf{X}}_{i,T+F}^{1, L_{xx}} = \overline{\mathbf{v}}_{i,T+F}^{1, L_{xx}}, \overline{Z}_{i,T+F}^{1, L_z}, \mathbf{X}_{i0}, i \in \mathcal{I}_F, t \in \mathcal{T}) \\
& \times \prod_{\ell=1}^{T+F-1} f_{\mathbf{X}_{it}}(\mathbf{v}_{\ell} \mid \overline{\mathbf{X}}_{i\ell}^{1, L_{xx}} = \overline{\mathbf{v}}_{i\ell}^{1, L_{xx}}, \overline{Z}_{i\ell}^{1, L_z}, \mathbf{X}_{i0}, i \in \mathcal{I}_F, t \in \mathcal{T})
\end{aligned}
\end{equation}
\textit{where $\mathbf{v}=[\mathbf{v}_1, \dots, \mathbf{v}_{T+F-1}]$ is a realization of the time-varying covariates from time $1$ to $T+F-1$, while $f_{\vX_{it}}(\cdot| \dots, \vX_{i0}, i\in \mathcal{I}_F, t\in \mathcal{T})$ is the conditional distribution of covariates measured in the subset of locations $\mathcal{I}_F$ during the observed time window $\mathcal{T}$.}
\end{proposition}
\noindent The identification formula in Proposition \ref{prop:recursive_covariates} unrolls the recursive evolution of the system. The product term represents the joint probability of a covariate path, with each factor identified using the stable evolution process observed in the observed period $\mathcal{T}$.

In the special case where the evolution of time-varying effect modifiers is a stationary process that depends only on baseline covariates, exhibits no temporal trend, and is not affected by previous treatments, Assumption \ref{ass:conf_transp_4} simplifies significantly. Under these conditions, the distribution at the future time point $T+F$ is equivalent to the distribution observed at any past time point $t \in \mathcal{T}$: 
\begin{equation}
\label{eq:predX_4b}
f_{\vX_{i,T+F}}(\vx \mid \vX_{i0}, i \in \mathcal{I}_F) = f_{\vX_{it}}(\vx \mid \vX_{i0}, i \in \mathcal{I}_F)
\end{equation}
for any $t \in \in[1, \dots, T, \dots, T+F]$. If, in addition to this stationarity, the target population $\mathcal{I}_F$ coincides with the observed treated population $U_1^{obs}$ (i.e., the sample where the intervention was observed at time $S$ with $Z_{iS}=1$), then the forecasted effect reduces to the observed effect: $\text{ATT}_{\text{F}} = \text{ATT}$.

Conversely, when the time-varying effect modifiers follow a temporal trend or exhibit complex dependencies on past history, this equivalence no longer holds. In such settings, the identification result in Proposition \ref{prop:recursive_covariates} (Equation \ref{eq:predX_4}), derived from Assumption \ref{ass:conf_transp_4}, provides the necessary bridge. It allows us to replace the unknown distribution of future effect modifiers $f_{\vX_{i,T+F}}(\vx \mid \vX_{i0}, i \in \mathcal{I}_F)$ in the identifying quantity for the potential outcomes, as established in Equation \eqref{eq:predATT_4} (or Equation \eqref{eq:predATT0_4}). 
By substituting the identified covariate distribution into the causal estimand, the future treatment effect is rendered identifiable as a functional of the observed historical data, even in the presence of shifting environmental states and carryover effects.

The formal identification established in Propositions \ref{prop:transp_4} and \ref{prop:recursive_covariates} provides the theoretical justification for a practical Monte Carlo imputation procedure to estimate future causal effects. This approach proceeds in two sequential stages: first, projecting the state of the system, and second, applying the transportable outcome mechanism.
In the first stage, we use the historical data from the observed window $t\in\mathcal{T}$ and $i\in \mathcal{I}_F$ to estimate a generative model for the evolution of the time-varying effect modifiers, $f_{\vX_{it}}(\cdot| \dots, \vX_{i0}, i\in \mathcal{I}_F, t\in \mathcal{T})$. This model allows us to simulate covariate trajectories for units in the future target population.
Depending on the application, researchers can employ diverse modeling strategies, ranging from standard time-series models with geographical fixed or random effects to complex mechanistic models, such as compartmental models in infectious disease epidemiology \citep[e.g.,][]{tuite2020mathematical, ferguson2020impact, Flaxman2020}.
%
Let
$\widehat{{\vX}}_{iT+F}$
be the vector of covariates for unit $i$ at the future time $T+F$, predicted using an estimated transition model. 
For a future time point $T+F$, 
the potential outcomes under both the treatment and control condition for a unit $i$ in $\mathcal{I}_F$ with baseline covariates $\vX_{i0}$ can be imputed from the observed outcomes at time $S$ of units in the observed sample subject to the treatment or the control condition, respectively, with the same baseline characteristics $\vX_{i0}$ and with time-varying covariates equal to $\widehat{{\vX}}_{iT+F}$ at time $S$ right before that condition was observed. 
This imputation can be implemented, more or less explicit, through various estimation frameworks, including non-parametric matching, semi-parametric weighting, or more common parametric outcome modeling \citep[e.g.,][]{Flaxman2020, davies2020effects, ferguson2020impact, naturemedicine2021, VicedoCabrera2019HandsonTO}.

\section{Assessing and Predicting the Impact of Actual Time-Varying Treatments}
\label{sec:intervention}
In this section, we extend the discussion and results of Section \ref{sec:simple_scenario} to settings with time-varying treatments: one-day events occurring at different times in different locations (e.g., US primary elections in the Spring 2020), and multi-day interventions or events with a continuous or intermittent duration (e.g., protests, stay-at-home orders, school closures, mask guidelines). Here, we also consider a latency between treatments and outcomes, lagged effects of time-varying covariates and past outcomes, both seen as potential effect modifiers, and potentially time-dependent covariates affected by past treatments.

\subsection{Definition of Potential Outcomes}

In addition to no-interference across locations, we assume that the outcome at time $t$ can at most depend on interventions over a period from $t-B-K$ to $t-B$, where $0\leq B<T$ and $ 0 \leq K<t-B$. This assumption allows carry-over effects over a period of length $K$ and rules out that an intervention implemented after time $t-B$ may have a short-term effect on the outcome at time $t$. In epidemiology $B$ is sometimes called latency time. It is worth noting that the absence of carry-over effects corresponds to the case with $K=0$, while if $B=0$, immediate effects are possible. Let us define the lagged intervention vector of interest as $\overline{\vZ}_{it}^{B,K}=[Z_{i(t-B)}, Z_{i(t-B-1)}, \ldots, Z_{i(t-B-K)}]$.
 We also introduce a function $h(\cdot)$ which summarizes $\overline{\vZ}_{it}^{B,K}$ and maps it into a binary treatment variable $D_{it}=h(\overline{\vZ}_{it}^{B,K}) \in \{0,1\}$. For example, $h(\cdot)$ can be defined such that $D_{it}$ is $1$ if area $i$ is treated at least once over the period $[t-B-K, t-B]$ and $0$ otherwise, that is, $D_{it}=\mathds{I}(\sum_{k=0}^{K}{Z_{i(t-B-k)}}\geq 1)$. It is worth noting that when there are no carry-over effects, i.e., $K=0$,  $h(\cdot)$ is simply the identity function, that is, $D_{it}=Z_{i(t-B)}$.
 Given the mapping function $h(\cdot)$, we assume that the potential outcome of unit $i$ at time $t$ only depends on the value of the binary variable $D_{it}$, summarizing the treatment vector over a period from $t-B-K$ to $t-B$ corresponding to location $i$.
 
 Formally, the previous considerations are expressed by the following Stable Unit Treatment Value Assumption (SUTVA).
\begin{assumption}[SUTVA]
\label{ass:SUTVA_E}
Let $h(\cdot)$ be a function $h: \mathcal{R}^{K+1} \rightarrow \{0,1\}$. $\forall \vz$, $\vz'\in \{0,1\}^{I\times T}$
such that $h(\overline{\vz}_{it}^{B,K})=h(\overline{\vz}_{it}^{B,K'})$, the following equality holds:
\[ \qquad Y_{it}(\vz) = Y_{it}(\vz').\]
\end{assumption}
\noindent Under this assumption we can index the potential outcome by the binary variable $D_{it}=h(\overline{\vZ}_{it}^{B,K})$, i.e., $Y_{it}(D_{it})$.
Throughout this section, we will refer to $Y_{it}(d)$,  as the potential outcome that we would observe at unit $it$ if the lagged treatment $D_{it}$ were set equal to $d$. Figure \ref{fig:timeline} represents the timeline and notation under SUTVA (Assumption \ref{ass:SUTVA_E}), including the carry-over period and the latency period for a time point $t$.

\begin{figure}[t]
\centering
\begin{tikzpicture}[%
    every node/.style={
        font=\scriptsize,
        text height=1ex,
        text depth=.25ex,
    },
]
\draw[->] (0,0) -- (9.5,0);

\foreach \x in {0,1,...,9}{
    \draw (\x cm,3pt) -- (\x cm,0pt);
}
    \draw[-latex] (8 cm,20pt) -- (8 cm,0pt);

\node[anchor=north] at (1,0) {$t\!-\!B\!-\!K$};
\node[anchor=north] at (5,0) {$t\!-\!B$};
\node[anchor=north] at (8,0) {$t$};
\node[anchor=north] at (9.5,0) {time};

\fill[pattern color=myRed!50, pattern=north east lines, line width = 1pt, very thick] (0,0.25) rectangle (1,0.4);
\fill[myDarkGray] (2,0.25) rectangle (3,0.4);
\fill[myDarkGray] (4,0.25) rectangle (5,0.4);
\fill[myRed] (1,0.25) rectangle (2,0.4);
\fill[myRed] (3,0.25) rectangle (4,0.4);
\fill[pattern color=myRed!50, pattern=north east lines, line width = 1pt, very thick] (5,0.25) rectangle (6,0.4);
\fill[pattern color=myGray!50, pattern=north east lines, line width = 1pt, very thick] (6,0.25) rectangle (8,0.4);

\draw[dashed,thick,-latex] (3.15,0.9) -- (4,0.9);
\node[draw,text width=0.5cm] at (4.5,0.9) {$h(\cdot)$};
\draw[dashed,thick,-latex] (5,0.9) -- (5.4,0.9);
\node[text width=0.5cm] at (5.7,0.9) {$D_{it}$};
\draw[dashed,thick,-latex] (6,0.9) -- (7.6,0.9);
\node[text width=0.5cm] at (7.9,0.9) {$Y_{it}(D_{it})$};

\draw[decorate,decoration={brace,amplitude=5pt}] (1,0.45) -- (5,0.45)
    node[anchor=south,midway,above=6pt] {$\overline{Z}_{it}^{B,K}$};
\draw[decorate,decoration={brace,amplitude=5pt}] (5,-0.6) -- (1,-0.6)
    node[anchor=south,midway,below=4pt] {Carry-over period}
    node[anchor=south,midway,below=15pt] {$K$};
\draw[decorate,decoration={brace,amplitude=5pt}] (8,-0.6) -- (5,-0.6)
    node[anchor=north,midway,below=4pt] {Latency period}
    node[anchor=south,midway,below=15pt] {$B$};;
\end{tikzpicture}
\caption{Timeline and notation for time-varying treatments.}
\label{fig:timeline}
\end{figure}

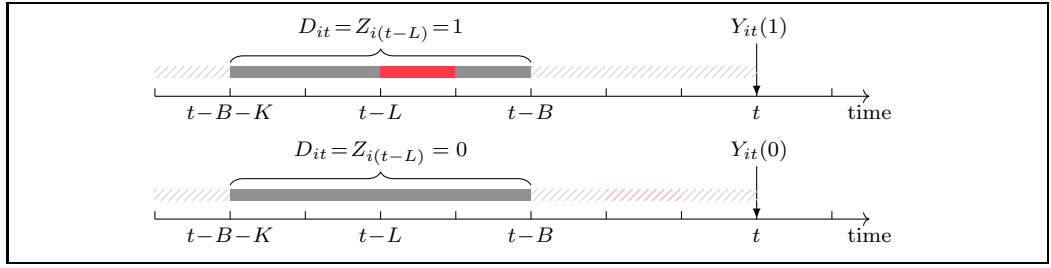
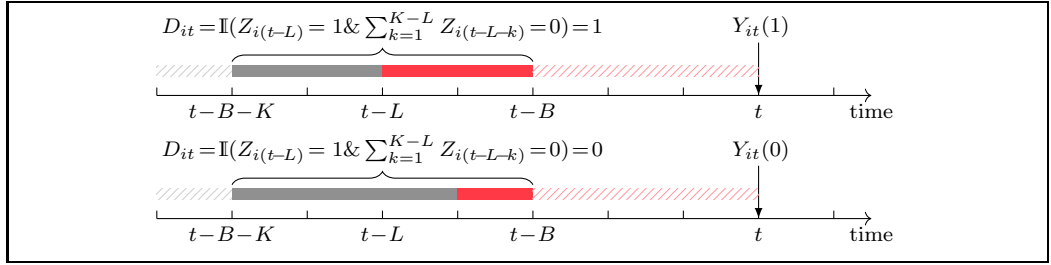
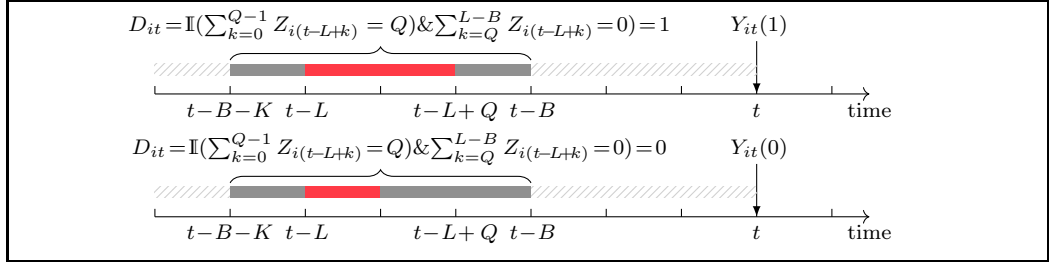
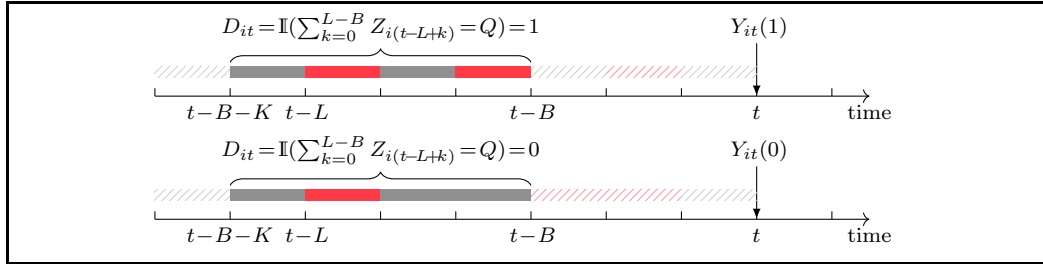
\begin{figure}[htbp]
\begin{subfigure}[a]{1.03\textwidth} 
\centering
  \begin{adjustbox}{minipage=0.8\linewidth, cfbox=black}
    \centering
\begin{tikzpicture}[%
    every node/.style={
        font=\scriptsize,
        text height=1ex,
        text depth=.25ex,
    },
]
\draw[->] (0,0) -- (9.5,0);

\foreach \x in {0,1,...,9}{
    \draw (\x cm,3pt) -- (\x cm,0pt);
}
    \draw[-latex] (8 cm,20pt) -- (8 cm,0pt);

\node[anchor=north] at (1,0) {$t\!-\!B\!-\!K$};
\node[anchor=north] at (3,0) {$t\!-\!L$};
\node[anchor=north] at (5,0) {$t\!-\!B$};
\node[anchor=north] at (8,0) {$t$};
\node[anchor=north] at (9.5,0) {time};

\fill[pattern color=myGray!50, pattern=north east lines, line width = 1pt, very thick] (0,0.25) rectangle (1,0.4);
\fill[myDarkGray] (1,0.25) rectangle (3,0.4);
\fill[myRed] (3,0.25) rectangle (4,0.4);
\fill[myDarkGray] (4,0.25) rectangle (5,0.4);
\fill[pattern color=myGray!50, pattern=north east lines, line width = 1pt, very thick] (5,0.25) rectangle (8,0.4);

\node[text width=0.5cm] at (7.9,0.9) {$Y_{it}(1)$};


\draw[decorate,decoration={brace,amplitude=5pt}] (1,0.45) -- (5,0.45)
    node[anchor=south,midway,above=6pt] {$D_{it}\!=\!Z_{i(t-L)}\!=\!1$};
\end{tikzpicture}

\begin{tikzpicture}[%
    every node/.style={
        font=\scriptsize,
        text height=1ex,
        text depth=.25ex,
    },
]
\draw[->] (0,0) -- (9.5,0);

\foreach \x in {0,1,...,9}{
    \draw (\x cm,3pt) -- (\x cm,0pt);
}
    \draw[-latex] (8 cm,20pt) -- (8 cm,0pt);

\node[anchor=north] at (1,0) {$t\!-\!B\!-\!K$};
\node[anchor=north] at (3,0) {$t\!-\!L$};
\node[anchor=north] at (5,0) {$t\!-\!B$};
\node[anchor=north] at (8,0) {$t$};
\node[anchor=north] at (9.5,0) {time};

\fill[pattern color=myGray!50, pattern=north east lines, line width = 1pt, very thick] (0,0.25) rectangle (1,0.4);
\fill[myDarkGray] (1,0.25) rectangle (3,0.4);
\fill[myDarkGray] (3,0.25) rectangle (5,0.4);
\fill[pattern color=myGray!50, pattern=north east lines, line width = 1pt, very thick] (5,0.25) rectangle (6,0.4);
\fill[pattern color=myRed!50, pattern=north east lines, line width = 1pt, very thick] (6,0.25) rectangle (7,0.4);
\fill[pattern color=myGray!50, pattern=north east lines, line width = 1pt, very thick] (7,0.25) rectangle (8,0.4);

\node[text width=0.5cm] at (7.9,0.9) {$Y_{it}(0)$};


\draw[decorate,decoration={brace,amplitude=5pt}] (1,0.45) -- (5,0.45)
    node[anchor=south,midway,above=6pt] {$D_{it}\!=\!Z_{i(t-L)}=0$};
\end{tikzpicture}
\end{adjustbox}
\caption{One-day intervention: effect of the one-day intervention}
\label{subfig:oneday}
\end{subfigure}

\begin{subfigure}[b]{1.03\textwidth}  
\centering
  \begin{adjustbox}{minipage=0.8\linewidth, cfbox=black}
\centering
\begin{tikzpicture}[%
    every node/.style={
        font=\scriptsize,
        text height=1ex,
        text depth=.25ex,
    },
]
\draw[->] (0,0) -- (9.5,0);

\foreach \x in {0,1,...,9}{
    \draw (\x cm,3pt) -- (\x cm,0pt);
}
    \draw[-latex] (8 cm,20pt) -- (8 cm,0pt);

\node[anchor=north] at (1,0) {$t\!-\!B\!-\!K$};
\node[anchor=north] at (3,0) {$t\!-\!L$};
\node[anchor=north] at (5,0) {$t\!-\!B$};
\node[anchor=north] at (8,0) {$t$};
\node[anchor=north] at (9.5,0) {time};

\fill[pattern color=myGray!50, pattern=north east lines, line width = 1pt, very thick] (0,0.25) rectangle (1,0.4);
\fill[myDarkGray] (1,0.25) rectangle (3,0.4);
\fill[myRed] (3,0.25) rectangle (5,0.4);
\fill[pattern color=myRed!50, pattern=north east lines, line width = 1pt, very thick] (5,0.25) rectangle (8,0.4);

\node[text width=0.5cm] at (7.9,0.9) {$Y_{it}(1)$};

\draw[decorate,decoration={brace,amplitude=5pt}] (1,0.45) -- (5,0.45)
    node[anchor=south,midway,above=6pt] {$D_{it}\!=\!
    \mathds{I}(Z_{i(t\!-\!L)}\!=1\& \sum_{k=1}^{K-L} Z_{i(t\!-\!L\!-\!k)}\!=\!0)\!=\!
    1$};
\end{tikzpicture}

\begin{tikzpicture}[%
    every node/.style={
        font=\scriptsize,
        text height=1ex,
        text depth=.25ex,
    },
]
\draw[->] (0,0) -- (9.5,0);

\foreach \x in {0,1,...,9}{
    \draw (\x cm,3pt) -- (\x cm,0pt);
}
    \draw[-latex] (8 cm,20pt) -- (8 cm,0pt);

\node[anchor=north] at (1,0) {$t\!-\!B\!-\!K$};
\node[anchor=north] at (3,0) {$t\!-\!L$};
\node[anchor=north] at (5,0) {$t\!-\!B$};
\node[anchor=north] at (8,0) {$t$};
\node[anchor=north] at (9.5,0) {time};

\fill[pattern color=myGray!50, pattern=north east lines, line width = 1pt, very thick] (0,0.25) rectangle (1,0.4);
\fill[myDarkGray] (1,0.25) rectangle (4,0.4);
\fill[myRed] (4,0.25) rectangle (5,0.4);
\fill[pattern color=myRed!50, pattern=north east lines, line width = 1pt, very thick] (5,0.25) rectangle (8,0.4);

\node[text width=0.5cm] at (7.9,0.9) {$Y_{it}(0)$};


\draw[decorate,decoration={brace,amplitude=5pt}] (1,0.45) -- (5,0.45)
    node[anchor=south,midway,above=6pt] {$D_{it}\!=\!
    \mathds{I}(Z_{i(t\!-\!L)}\!=1\& \sum_{k=1}^{K-L} Z_{i(t\!-\!L\!-\!k)}\!=\!0)\!=\!0$};
\end{tikzpicture}
\end{adjustbox}
\caption{Multi-day intervention: effect of initiation time}
\label{subfig:initiation}
\end{subfigure}

\begin{subfigure}[c]{1.03\textwidth}  \centering
  \begin{adjustbox}{minipage=0.8\linewidth, cfbox=black}
\centering
\begin{tikzpicture}[%
    every node/.style={
        font=\scriptsize,
        text height=1ex,
        text depth=.25ex,
    },
]
\draw[->] (0,0) -- (9.5,0);

\foreach \x in {0,1,...,9}{
    \draw (\x cm,3pt) -- (\x cm,0pt);
}
    \draw[-latex] (8 cm,20pt) -- (8 cm,0pt);

\node[anchor=north] at (1,0) {$t\!-\!B\!-\!K$};
\node[anchor=north] at (2,0) {$\,t\!-\!L$};
\node[anchor=north] at (4,0) {$t\!-\!L\!+Q$};
\node[anchor=north] at (5,0) {$\,t\!-\!B$};
\node[anchor=north] at (8,0) {$t$};
\node[anchor=north] at (9.5,0) {time};

\fill[pattern color=myGray!50, pattern=north east lines, line width = 1pt, very thick] (0,0.25) rectangle (1,0.4);
\fill[myDarkGray] (1,0.25) rectangle (2,0.4);
\fill[myRed] (2,0.25) rectangle (4,0.4);
\fill[myDarkGray] (4,0.25) rectangle (5,0.4);
\fill[pattern color=myGray!50, pattern=north east lines, line width = 1pt, very thick] (5,0.25) rectangle (8,0.4);

\node[text width=0.5cm] at (7.9,0.9) {$Y_{it}(1)$};


\draw[decorate,decoration={brace,amplitude=5pt}] (1,0.45) -- (5,0.45)
    node[anchor=south,above=6pt] {$\hspace{-3.5cm}D_{it}\!=\!\mathds{I}(\sum_{k=0}^{Q-1} Z_{i(t\!-\!L\!+\!k)}=Q) \& \! \sum_{k=Q}^{L-B} Z_{i(t\!-\!L\!+\!k)}\!=\!0)\!=\!1$};
\end{tikzpicture}

\begin{tikzpicture}[%
    every node/.style={
        font=\scriptsize,
        text height=1ex,
        text depth=.25ex,
    },
]
\draw[->] (0,0) -- (9.5,0);

\foreach \x in {0,1,...,9}{
    \draw (\x cm,3pt) -- (\x cm,0pt);
}
    \draw[-latex] (8 cm,20pt) -- (8 cm,0pt);

\node[anchor=north] at (1,0) {$t\!-\!B\!-\!K$};
\node[anchor=north] at (2,0) {$\,t\!-\!L$};
\node[anchor=north] at (4,0) {$t\!-\!L\!+Q$};
\node[anchor=north] at (5,0) {$\,t\!-\!B$};
\node[anchor=north] at (8,0) {$t$};
\node[anchor=north] at (9.5,0) {time};

\fill[pattern color=myGray!50, pattern=north east lines, line width = 1pt, very thick] (0,0.25) rectangle (1,0.4);
\fill[myDarkGray] (1,0.25) rectangle (2,0.4);
\fill[myRed] (2,0.25) rectangle (3,0.4);
\fill[myDarkGray] (3,0.25) rectangle (5,0.4);
\fill[pattern color=myGray!50, pattern=north east lines, line width = 1pt, very thick] (5,0.25) rectangle (8,0.4);

\node[text width=0.5cm] at (7.9,0.9) {$Y_{it}(0)$};


\draw[decorate,decoration={brace,amplitude=5pt}] (1,0.45) -- (5,0.45)
    node[anchor=south,above=6pt] {$\hspace{-3,5cm}D_{it}\!=\!\mathds{I}(\sum_{k=0}^{Q-1} Z_{i(t\!-\!L\!+\!k)}\!=\!Q) \& \! \sum_{k=Q}^{L-B} Z_{i(t\!-\!L\!+\!k)}\!=\!0)\!=\!0$};
\end{tikzpicture}
\end{adjustbox}
\caption{Multi-day continuous intervention: effect of Q (e.g., =2) consecutive intervention days}
\label{subfig:duration}
\end{subfigure}

\begin{subfigure}[d]{1.03\textwidth}
  \centering
  \begin{adjustbox}{minipage=0.8\linewidth, cfbox=black}
\centering
\begin{tikzpicture}[%
    every node/.style={
        font=\scriptsize,
        text height=1ex,
        text depth=.25ex,
    },
]
\draw[->] (0,0) -- (9.5,0);

\foreach \x in {0,1,...,9}{
    \draw (\x cm,3pt) -- (\x cm,0pt);
}
    \draw[-latex] (8 cm,20pt) -- (8 cm,0pt);

\node[anchor=north] at (1,0) {$t\!-\!B\!-\!K$};
\node[anchor=north] at (2,0) {$\,t\!-\!L$};
\node[anchor=north] at (5,0) {$\,t\!-\!B$};
\node[anchor=north] at (8,0) {$t$};
\node[anchor=north] at (9.5,0) {time};

\fill[pattern color=myGray!50, pattern=north east lines, line width = 1pt, very thick] (0,0.25) rectangle (1,0.4);
\fill[myDarkGray] (1,0.25) rectangle (2,0.4);
\fill[myRed] (2,0.25) rectangle (3,0.4);
\fill[myDarkGray] (3,0.25) rectangle (4,0.4);
\fill[myRed] (4,0.25) rectangle (5,0.4);
\fill[pattern color=myGray!50, pattern=north east lines, line width = 1pt, very thick] (5,0.25) rectangle (6,0.4);
\fill[pattern color=myRed!50, pattern=north east lines, line width = 1pt, very thick] (6,0.25) rectangle (7,0.4);
\fill[pattern color=myGray!50, pattern=north east lines, line width = 1pt, very thick] (7,0.25) rectangle (8,0.4);

\node[text width=0.5cm] at (7.9,0.9) {$Y_{it}(1)$};


\draw[decorate,decoration={brace,amplitude=5pt}] (1,0.45) -- (5,0.45)
    node[anchor=south,midway,above=6pt] {$D_{it}\!=\!\mathds{I}(\sum_{k=0}^{L-B} Z_{i(t\!-\!L\!+\!k)}\!=\!Q)\!=\!1$};
\end{tikzpicture}

\begin{tikzpicture}[%
    every node/.style={
        font=\scriptsize,
        text height=1ex,
        text depth=.25ex,
    },
]
\draw[->] (0,0) -- (9.5,0);

\foreach \x in {0,1,...,9}{
    \draw (\x cm,3pt) -- (\x cm,0pt);
}
    \draw[-latex] (8 cm,20pt) -- (8 cm,0pt);

\node[anchor=north] at (1,0) {$t\!-\!B\!-\!K$};
\node[anchor=north] at (2,0) {$\,t\!-\!L$};
\node[anchor=north] at (5,0) {$\,t\!-\!B$};
\node[anchor=north] at (8,0) {$t$};
\node[anchor=north] at (9.5,0) {time};

\fill[pattern color=myGray!50, pattern=north east lines, line width = 1pt, very thick] (0,0.25) rectangle (1,0.4);
\fill[myDarkGray] (1,0.25) rectangle (2,0.4);
\fill[myRed] (2,0.25) rectangle (3,0.4);
\fill[myDarkGray] (3,0.25) rectangle (5,0.4);
\fill[pattern color=myRed!50, pattern=north east lines, line width = 1pt, very thick] (5,0.25) rectangle (7,0.4);
\fill[pattern color=myGray!50, pattern=north east lines, line width = 1pt, very thick] (7,0.25) rectangle (8,0.4);

\node[text width=0.5cm] at (7.9,0.9) {$Y_{it}(0)$};


\draw[decorate,decoration={brace,amplitude=5pt}] (1,0.45) -- (5,0.45)
    node[anchor=south,midway,above=6pt] {$D_{it}\!=\!\mathds{I}(\sum_{k=0}^{L-B} Z_{i(t\!-\!L\!+\!k)}\!=\!Q)\!=\!0$};
\end{tikzpicture}
\end{adjustbox}
\caption{Multi-day intermittent intervention: effect of $Q$ (e.g.,=2) possibly nonconsecutive intervention days}
\label{subfig:intermittent}
\end{subfigure}
\caption{Potential treatment vectors $\overline{Z}_{it}^{B,K}$ and corresponding treatment indicator $D_{it}$ for a time point $t$  for different intervention/event settings and effects of interest.}
\label{fig:settings}
\end{figure}

Let us suppose that we were interested in evaluating the effect of a one-day event that could have affected the spread of SARS-CoV-2 infection, such as US primary in-person elections, on the COVID-related hospital admissions after $L$ days, and in particular on the $L$th day, with $B \leq L\leq B+K$ 
($L$ could reasonably range from 10 to 30 days). Under our set-up, in order to answer this question we would set the lagged treatment variable to
$D_{it}=Z_{i(t-L)}$, for each $t\in [L+1, \dots, T]$, that is, for each time point where the event is observable (see Figure \ref{subfig:oneday}). 
Conversely, in the case 
of interventions or events that lasted more than one day, e.g. lockdowns or stay-at-home orders, the definition of the lagged treatment variable would depend on the causal effect of interest. 
If we were interested in the average effect of starting a lockdown on the increase of COVID-19 related deaths after $L$ days,  with $B \leq L\leq B+K$, we would set the lagged treatment indicator
equal to 1 if the intervention started at day $t-L$ and was not in place before that day during the carry-over period, that is,   $D_{it}=\mathds{I}(Z_{i(t-L)}=1 \,\& \, \sum_{k=1}^{K-L} Z_{i(t-L-k)}=0)$ (see Figure \ref{subfig:initiation}). 
\footnote{
We could, instead, estimate the average effect of a one-day occurrence of the multi-day intervention or event, averaged across the initiating time and duration, by setting $D_{it}=[\vZ_{it}^{B-1,K},D^{\star}_{it}=\mathds{I}(Z_{i(t-B)}=1)]$. In this case, our treatment of interest is $D^{\star}_{it}$, while we would control for and average across the lagged intervention vector $\vZ_{it}^{B-1,K}$. For simplicity, given the non-binary nature of $D_{it}$, we do not consider this case here, although identification issues discussed in this paper would still apply.
}
If we, instead, wanted to evaluate the effect of keeping a lockdown in place for $Q$ days, during the period $[t-L,t-L+Q-1]$, on the outcome after $L$ days, with $B+Q \leq L\leq B+K$,  we should set $D_{it}=\mathds{I}(\sum_{k=0}^{Q-1} Z_{i(t-L+k)}=Q) \,\& \, \sum_{k=Q}^{L-B} Z_{i(t-L+k)}=0)$ (see Figure \ref{subfig:duration}).
Finally, as another example, we could assess the effect of $Q$ days of protests, which might or might not be continuous,  on the spread of COVID-19 after $L$ days of the first day of protest, with $B+Q \leq L\leq B+K$, by setting the lagged treatment variable equal to 1 if in the temporal window $[t-L,t-B]$ there were Q days of protests in location $i$, i.e.,  $D_{it}=\mathds{I}(\sum_{k=0}^{L-B} Z_{i(t-L+k)}=Q)$ (see Figure \ref{subfig:intermittent}).


\subsection{Actual impact on the observed sample}
\label{sec:actual}

\subsubsection{Causal Estimands}
Under Assumption \ref{ass:SUTVA_E}, let us assume the we want to evaluate the causal effect of a specific intervention that has been implemented in some areas or of a specific event that occurred in some locations over the observed temporal window.  
Formally, let 
$U_1^{obs}=\{it\in U^{obs}: D_{it}=1\}$ be the set of the $N_1^{obs}=|U_1^{obs}|$ treated units, that is, the set of location-time units $it$ whose treatment indicator $D_{it}$ is equal to 1. Notice that, depending on the carry-over period and the definition of the treatment indicator for the causal effect of interest, the treatment indicator may not be observed if defined based on time points outside of the observed time window. For example the effect of starting a lockdown on COVID-related deaths after 20 days cannot be evaluated using death counts measured at time points $t<20$ in the observed time window. Since the treated set $U_1^{obs}$ is defined as the set of units whose treatment indicator is 1, location-time units whose treatment indicator is not observed would not be included in this set.
Here, the ATT is the average comparison between the two potential outcomes under $D_{it}=0$ and $D_{it}=1$ across treated units:
\begin{equation}
\label{eq:ATT}
ATT=\frac{1}{N_1^{obs}}\sum_{it\in{U_1^{obs}}}
\Big(\E\Big[Y_{it}(1)| \vX_{i0}\Big]-\E\Big[Y_{it}(0)| \vX_{i0}\Big]\Big)
\end{equation}

In the COVID-19 example, the ATT could be the number of COVID19-related deaths prevented by initiating lockdowns, stay-at-home orders or school closures, or the number of COVID19-related deaths attributable to having Q days of protests, or in-person elections rallies, 
in areas where these interventions or events actually happened.

\subsubsection{Identification
under Sequential Ignorability}

In settings with time-varying treatment and time-varying covariates, 
the ignorability assumption must hold sequentially: for each time $t$ the corresponding potential outcomes $Y_{it}(d)$ are independent from the current treatment $D_{it}$ conditional on the treatment, outcome and covariate history. This assumption is called in the literature sequential ignorability \citep[e.g.,][]{Robins2000MarginalSM}. 
Let 
$\overline{\mathbf{R}}_{it}$ be the vector of pre-treatment time-varying confounders,
including covariates in the pre-treatment window $[t-B-L_x, t-B-K]$ with $L_x \geq K$, and possibly lagged outcomes in the pre-treatment window $[t-B-L_y, t-B-K-1]$ with $L_y>K$, i.e., $\overline{\mathbf{R}}_{it}=[\overline{\vX}_{it}^{B+K,L_x-K}, \overline{\vY}_{it}^{B+K+1,L_y-K-1}]$.
Let $\overline{\mathbf{V}}_{i\ell}$ be the set of time-varying covariates and lagged outcomes in the cross-over treatment window until a time point $\ell$,
i.e., $\overline{\mathbf{V}}_{i\ell}=[\overline{\vX}_{it}^{0,\ell-1-H}, \overline{\vY}_{it}^{1,\ell-1-H}]$, with $H=t-B-K$, $\overline{\vX}_{i\ell}^{0,\ell-1-H}$ being the covariate history between $t-B-K+1$ and $\ell$, and $\overline{\vY}_{i\ell}^{1, \ell-1-H}$ being the outcome history between $t-B-K$ and $\ell-1$. We can now formalize the sequential ignorability assumption as follows:

\begin{assumption}[Sequential Ignorability - no unmeasured confounders]
\label{ass:Ign}
\begin{align}
\hspace{0cm}
Y_{it}(d) \ind Z_{i\ell} | \overline{\vZ}_{i\ell}^{1, \ell-1-H}=\overline{\vz}^{\ell-1}, 
\overline{\mathbf{V}}_{i\ell}, \overline{\mathbf{R}}_{it}, \vX_{i0} \quad
\forall it\in U\subseteq U^{obs}, \forall \, d, \forall \, \ell\in [t-B-K, t-B], \forall \overline{\vz}:D_{it}=d
\end{align}
\end{assumption}
\noindent where 
$\overline{\vZ}_{i\ell}^{1, \ell-1-H}$ is the treatment history from $t-B-K$ until $\ell-1$, 
$\overline{\vz}^k=\{z_H,...,z_k\}$ 
is the  sub-vectors of any realization $\overline{\vz}$ such that $D_{it}=d$, 
and $U$ is a subset of units contained in $U^{obs}$ where Assumption \ref{ass:Ign} holds.
The sequential ignorability assumption states that for each unit in $U\subseteq U^{obs}$ the distribution of the potential outcomes at time $t$ is independent of the treatment $Z_{i\ell}$ received at each time points $\ell$ in the cross-over time window $[t-B-K, t-B]$ defining the lagged treatment $D_{it}$ (i.e, $Z_{i\ell}$ is as good as randomized), conditional on the treatment history until the first treatment $Z_{i(t-B-K)}$ in the lagged treatment window, the covariate and the outcome history before and during the cross-over period, possibly until previous time points $t-B-L_x$ and $t-B-L_y$. Indeed, the evolution of covariates as well as the outcome history could confound the causal effect of a treatment long before the treatment can start having an effect. 
For example,
if were interested in estimating the effect of $Q=30$ days of restrictions to control the spread of COVID-19 on mortality after $L=15$ days (Figure \ref{subfig:duration}), in addition to area-specific characteristics that could confound the effect, we should also control for the epidemic dynamics before and during the time window of interest. In fact,
the decision of policy-makers to start restrictions may depend on the trend of COVID-19 cases and deaths in the previous month, i.e., $\overline{\mathbf{R}}_{it}$, but the decision to keep them for longer than few days and up to 30 days (instead of lifting them) can in principle be affected each day $\ell$ by the epidemic trend during this time, i.e., $\overline{\mathbf{V}}_{i\ell}$. Because 
the epidemic status of each day $\ell$ during this 30-day period of interest depends on the restrictions implemented up to that day, we also need to control for the treatment vector $\overline{\vZ}_{i\ell}^{1, \ell-1-H}$ up to day $\ell$.  
Nevertheless,
under the sequential ignorability assumption (Assumption \ref{ass:Ign}), the identification of the ATT (Equation \ref{eq:ATT}), that is, the types of confounders we should control for and how,  depends on the definition of the lagged treatment vector $D_{it}$ and on the presence of time-dependent confounders.

First, let us assume that we wanted to evaluate the effect of a one-day intervention or event that could have happened in different days depending on the location (e.g., elections) on deaths after $L$ days, with $B \leq L\leq B+K$ (Figure \ref{subfig:oneday}). The treatment indicator would be defined as $D_{it}=Z_{i(t-L)}$, for $t\in [L+1, \dots, T]$.
In this simple setting, there are no time-dependent confounders during the treatment window because its duration is one day. As a consequence, 
we only need to adjust for 
baseline covariates and the evolution of time-varying covariates sometime before the treatment during a time window $[t-B-L_x, t-B]$, i.e., $\overline{\mathbf{R}}_{it}=\overline{\vX}_{it}^{B,L_x}$. In this case, under Assumption \ref{ass:Ign}, the average potential outcome under the control condition for treated units 
is identified by:
\begin{equation}
\begin{aligned}
\label{eq:adj}
    \E\Big[Y_{it}(0)| D_{it}=1, \vX_{i0}
    \Big]=
     \sum_{\overline{\mathbf{c}}}\E\Big[Y^{obs}_{it}|D_{it}=0, \overline{\mathbf{C}}_{it}=\overline{\mathbf{c}}, \vX_{i0}\Big]    f_{\overline{\mathbf{C}}_{it}}(\overline{\mathbf{c}}| D_{it}=1, \vX_{i0})  
    \end{aligned}
\end{equation}
with $\overline{\mathbf{C}}_{it}=\overline{\mathbf{R}}_{it}$. 
Thus, the average treatment effect on the treated is identified from the observed data as follows:
\begin{equation}
\label{eq:att_id}
ATT=\frac{1}{N_1^{obs}}\sum_{it\in{U_1^{obs}}}
\Big(\E\Big[Y_{it}^{obs}| D_{it}=1,\vX_{i0}\Big]-
 \sum_{\overline{\mathbf{c}}}\E\Big[Y^{obs}_{it}|D_{it}=0, \overline{\mathbf{C}}_{it}=\overline{\mathbf{c}}, \vX_{i0}\Big]    f_{\overline{\mathbf{C}}_{it}}(\overline{\mathbf{c}}| D_{it}=1, \vX_{i0})
\Big)
\end{equation}
Different approaches exist to adjust for trends in panel or longitudinal data, e.g. matching or stratification \citep[e.g.,][]{perrakis2014controlling, Kim2021MatchingMF}.
\footnote{
The same consideration applies to the effect of having one more day of an intervention or event, regardless of its initiation time and duration, on the outcome after $L$ days, with $D_{it}=[\vZ_{it}^{L+1,K},\mathds{I}(Z_{i(t-L)}=1)]$. In this case, we would also adjust for lagged treatments $\vZ_{it}^{L+1,K}$ that can be regarded as confounders, i.e. $\overline{\mathbf{R}}_{it}=[\overline{\vX}_{it}^{B,L_x}, \vZ_{it}^{L+1,K}]$.
}
When previous outcomes also affect the propensity of having the one-day event, then we would also need to adjust for the outcome evolution over a sufficient time window $[t-L-L_y, t-L]$, that is, $\overline{\mathbf{R}}_{it}=[\overline{\vX}_{it}^{L,L_x}, \overline{\vY}_{it}^{L+1,L_y}]$.
For instance, if researchers wanted to evaluate the effect of in-person elections that occurred in the US during the first wave of COVID-19, they would need to control for the epidemic dynamics during the month prior to the elections, which could have influenced the decision of holding the election in-person as well as their turnout.    

On the other hand, 
multi-day interventions or events, with a cross-over effect over a period of time of $K>1$ days, complicate the identification of the ATT when time-dependent confounders are present. 
This is the case when 
we observe an intervention or event that lasted multiple days (e.g., lockdowns, stay-at-orders, school closures, protests), with a varying number of days across areas, and we want to evaluate the effect of its initiation time (Figure \ref{subfig:initiation}) or the effect of its duration (Figures \ref{subfig:duration} and \ref{subfig:intermittent}).  
Time-dependent confounders are defined as variables that (i) are caused by (or share a common cause with) previous treatments and (ii) are confounders for the effect of a subsequent treatment on the outcome. These variables could be time-varying covariates that, in addition to affecting subsequent treatments and the outcome, are also affected by previous treatments. Time-dependent confounders also include previous outcomes affecting subsequent treatments, given that lagged outcomes are always also caused by previous treatments. 
The identification of the ATT depends on whether time-dependent confounders are present or not. When they are not present -- i.e., 
when previous outcomes do not confound the effect of a subsequent treatment on a subsequent outcome, by not affecting future treatments, and when time-varying covariates in $\vX_{it}$ 
are not affected by previous treatments --
under Assumption \ref{ass:Ign}, the ATT is identified by the observed data using the simple identification results 
in Equations \eqref{eq:adj} and \eqref{eq:att_id},
simply conditioning for all type of confounders, including baseline covariates and outcomes as well as the evolution of time-varying covariates and outcomes pre-treatment 
and during the cross-over treatment window.
Specifically, the conditioning set is given by $\vX_{i0}$ and
$\overline{\mathbf{C}}_{it}=[\overline{\mathbf{R}}_{it}, \overline{\mathbf{V}}_{it}]$, where $\overline{\mathbf{R}}_{it}=[\overline{\vX}_{it}^{B+K,L_x-K}, \overline{\vY}_{it}^{B+K+1,L_y-K}]$ is the set of pre-treatment confounders and $\overline{\mathbf{V}}_{it}=[\overline{\vX}_{it}^{B,K-1},\overline{\vY}_{it}^{B,K}]$ is the set of time-varying confounders during the treatment window $t-B-K, t-B$.
When the duration of the intervention is decided a priori based on the population characteristics and, say, the epidemic dynamics up until the initiation time, the evolution of the epidemic during the duration of the intervention should not be considered as a time-varying confounder and the adjustment set must only include previous covariates that led to the decision of initiating the intervention with a pre-specified duration.

On the contrary, when time-dependent confounders exist,
including previous outcomes affecting subsequent treatments, 
methods that match or stratify on $\overline{\vX}_{it}^{B,K}$ and $\overline{\mathbf{Y}}_{it}^{B,K}$ are biased.
In this situation, simply adjusting for the evolution of covariates and outcomes after time $t-B-K$ would lead to selection bias \citep{robins1999association}. Instead, we need to condition on the values drawn from the distribution of confounders and outcomes that would have been observed under the counterfactual scenario. 
This is the case 
when researchers wish to evaluate the effect of the implementation of COVID-19 restrictions or the occurrence of social events that lasted several days, with its duration (continuous or intermittent) being influenced by the daily/weekly status of the epidemic that was, in turn, affected by the restrictions being already in place.
For example, 
stay-at-home orders were imposed in the Spring 2020 at different times across states and lasted from few weeks to several months. The decision of imposing new restrictions depended upon the population characteristics, the government political views, and the epidemic status at that point in time (also affected by not imposing restrictions earlier). After few weeks or months, the decision of lifting stay-at-home orders depended on similar factors, including the epidemic dynamics clearly affected by maintaining the restrictions until that point. If we wanted to evaluate two different strategies implemented by different states, with a different initiation time and a different duration, we must consider the fact that epidemic dynamics influenced both the initiation and duration of the stay-at-home orders, but were also affected by them. 
Therefore, intuitively each state with strategy 1 must be compared to states with strategy 0, that would have the same baseline covariates but also a similar evolution of the epidemic under strategy 0.
Similarly, 
in the Fall 2020, several US states (e.g, California) and other countries around the world (e.g, India, Spain, Italy)
shifted their COVID-19 control plan from a uniform lockdown to more locally tailored restrictions implemented by local governments. 
For instance, Italy implemented a color code classification system, classifying the COVID-19 criticality of each region and indicating the strictness of restrictions accordingly. Color codes were weekly assigned by the government and Minister of Health and depended on the degree of risk and local statistics reported in each region. Let us suppose that were interested in evaluating the impact of implementing the strict measures 
in the region classified as `red zones' (high risk) compared to those associated to the orange classification (medium risk) \citep{italycolors, Pelagatti2021AssessingTE}. 
If we wanted to evaluate the impact of the first week of implementation (on COVID19- related deaths after 15-30 days) we must control for the epidemic status and the occupancy rate of hospital beds before the regions shifted from `orange' to `red'. Instead, if we were also interested in the impact of keeping for an extra week the strict measures associated to the `red zones', 
we must consider the fact that the epidemic status and risk level that led to decision of keeping those regions colored in `red' were affected by the strict measured implemented in the first week. 
These settings make the identification and estimation of such causal effects more complicated due to the presence of time-dependent confounders affected by previous events.
Methods referred to as
g-methods have been proposed to deal with issue \citep{naimi2017introduction}.

Formally, 
under the sequential ignorability assumption \ref{ass:Ign}, the average potential outcome for treated units under the control condition $D_{it}=0$, conditional on baseline covariates,
can be identified by the following g-formula:

\begin{equation}
\begin{aligned}
\label{eq:gformula}
    \E\Big[Y_{it}(0)| 
    D_{it}=1, \vX_{i0}\Big]
    =\sum_{\overline{\vz}:D_{it}=0}
    &\sum_{\overline{\mathbf{r}}} 
    \sum_{\overline{\vx}} \sum_{\overline{\vy}} \E\Big[Y^{obs}_{it}|\overline{\vZ}_{it}^{B,K}=\overline{\vz}, \overline{\mathbf{R}}_{it}=\overline{\mathbf{r}}, \vX_{i0},    \overline{\vX}_{it}^{B, K-1}=\overline{\vx}, \overline{\vY}_{it}^{B, K}=\overline{\vy}\Big]\\
    &f_{\overline{\vX}_{it}^{B, K-1}, \overline{\vY}_{it}^{B, K}}(  \overline{\vX}_{it}^{B, K-1}=\overline{\vx}, \overline{\vY}_{it}^{B, K}=\overline{\vy}|\overline{\vZ}_{it}^{B,K}=\overline{\vz}, \overline{\mathbf{R}}_{it}=\overline{\mathbf{r}}, \vX_{i0})\times\\
     &
    f_{\overline{\mathbf{R}}_{it}}(\overline{\mathbf{r}}| D_{it}=1, \vX_{i0})  
    \end{aligned}
\end{equation}
In Equation \ref{eq:gformula} we identify the mean potential outcome under the control condition, 
for locations with baseline covariates $\vX_{i0}$ treated at some time $t$,
by taking the mean outcome of the control units averaged over 
the distribution of covariate and outcome history $\overline{\mathbf{R}}_{it}$ before the carry-over period $t-B-K$ conditional on the treated, as well as over the distribution of the evolution of outcomes and covariates during the carry-over period under the control condition. In addition, given that $D_{it}=0$ corresponds to multiple treatment vectors $\overline{Z}_{it}^{B,K}$ through the mapping function $h(\cdot)$, we take the average of the treatment history over all the values that map into $D_{it}=0$.
For example, when evaluating the effect of keeping an Italian region classified as `red zone' for two weeks and imposing the stricter curbs as opposed to the ones associated to the orange zones, in addition to adjusting for 
the epidemic status and the occupancy rate of hospital beds before the shift from `orange' to `red' (i.e, $\overline{\mathbf{R}}_{it}$), we must compare the regions that were classified as `red zones' for at least two weeks to those classified as `orange zones' during the same period, controlling for the propensity of having a similar evolution of the epidemic and level of risk (i.e, $\overline{\mathbf{Y}}_{it}^{B,K}$ and $\overline{\mathbf{X}}_{it}^{B,K}$) if the region were assigned an  `orange code' instead of a `red code'.
Under the sequential ignorability assumption, given the identification result in \ref{eq:gformula} or a rework of it, researchers have developed methods for the estimation of average treatment effects, appropriately adjusting for time-varying confounders that are affected by past treatments. In the literature, these methods are broadly referred to as \textit{g-methods} and include inverse-probability-of-treatment weighting, parametric g-formula, and g-estimation \citep{whatif}.

\subsection{Predicting the impact on future observations}
\label{sec:pred}

\subsubsection{Causal Estimands}
Now, let us assume that we have already assessed the impact of an intervention or event that we had observed in a subsample $U^{obs}_1$ of locations in $\mathcal{I}$ at some time points during the observed temporal window $\mathcal{T}$. We now wish to predict the impact of the same  intervention or event if were able to replicate it in the future on the same population $\mathcal{I}$.

Let $\mathcal{T}_Z^F=[T+F,\ldots,T+F+T_Z^F]$,  with $F \geq 1, T_Z^F\geq 0$, the future time window where we wish to apply the intervention or event of interest (future treatment window), and let 
$\mathcal{T}^F=[T+F+L, \ldots, T+F+T^F]$, with $B \leq L\leq B+K$ and $L \leq T^F\leq T_Z^F+K$,
be the future time window
where we wish to forecast the causal effect (future outcome window) (see Figure \ref{fig:future}).
Note that the duration of the future treatment window, i.e., $T_Z^F$, depends on whether the intervention of interest is with or without duration and on the type of causal effect we wish to predict. For instance, if we wish to predict the effect of a multi-day intervention lasting $Q=3$ days, then we would set $T_Z^F=3$.  
\footnote{Note that if 
$0\leq F<K$ and $L<K-F$, 
for any $t\leq T+B+K \in \mathcal{T}^F$
the cross-over treatment period $[t-B-K, t-B]$ overlaps with the observed time window $\mathcal{T}$ and we would estimate the effect of switching the observed treatments and imposing new treatments such that $D_{it}=h(\overline{\vZ}_{it}^{B,K})=0$ or $1$.
} 

\begin{figure}[t]
\centering

\begin{tikzpicture}[%
    every node/.style={
        font=\scriptsize,
        text height=1ex,
        text depth=.25ex,
    },
]

\draw[->] (0,0) -- (15.5,0);

\foreach \x in {0,1,...,15}{
    \draw (\x cm,3pt) -- (\x cm,0pt);
}

    \draw (0cm,3pt) -- (0 cm,-3pt);
    \draw (4cm,3pt) -- (4 cm,-3pt);
    \draw[dashed] (4cm,2cm) -- (4 cm,-2cm);
    \draw (6cm,3pt) -- (6 cm,-3pt);
    \draw (9cm,3pt) -- (9 cm,-3pt);
    \draw (12cm,3pt) -- (12 cm,-3pt);
    \draw (14cm,3pt) -- (14 cm,-3pt);

\node[anchor=north] at (0,0) {$0$};
\node[anchor=north] at (4,0) {$T$};
\node[anchor=north] at (6,0) {$T\!+\!F$};
\node[anchor=north] at (9,0) {$T\!+\!F\!+\!T_Z^F$};
\node[anchor=north] at (12,0) {$T\!+\!F\!+\!L$};
\node[anchor=north] at (14,0) {$T\!+\!F\!+\!T^F$};
\node[anchor=north] at (15.5,0) {time};

\fill[myRed] (0,0.25) rectangle (3,0.4);
\fill[myGray] (3,0.25) rectangle (4,0.4);
\filldraw[draw=myLightGray,fill=white] (4,0.25) rectangle (15,0.4);

\filldraw[draw=myRed, dashed,fill=white, line width=1pt] (6,0.25) rectangle (9,0.4);

\draw[decorate,decoration={brace,amplitude=5pt}] (4,-0.6) -- (0,-0.6)
    node[anchor=south,midway,below=4pt] {Observed window}
    node[anchor=south,midway,below=15pt] {$\mathcal{T}$};
;
\draw[decorate,decoration={brace,amplitude=5pt}] (9,-0.6) -- (6,-0.6)
    node[anchor=south,midway,below=4pt] {Future treatment window}
    node[anchor=south,midway,below=15pt] {$\mathcal{T}_Z^F$};
;
\draw[decorate,decoration={brace,amplitude=5pt}] (14,-0.6) -- (12,-0.6)
    node[anchor=south,midway,below=4pt] {Future outcome window}
    node[anchor=south,midway,below=15pt] {$\mathcal{T}^F$};
;
\end{tikzpicture}
\caption{Timeline and notation for predicting the impact of an intervention if applied to a future treatment window on a future outcome window. The time window with a dashed red border represents the hypothetical 3-day intervention that we wish to apply in the future.}
\label{fig:future}
\end{figure}
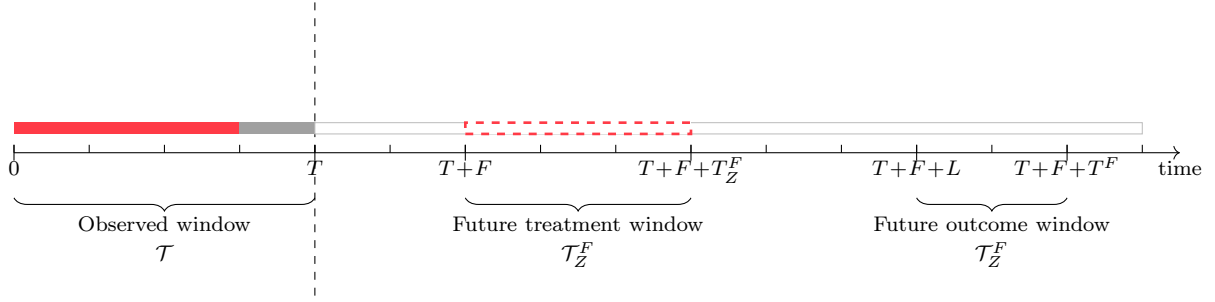

We define the average causal effect of the treatment in the future as the following quantity:
\begin{equation}
\label{eq: ATTF}
ATT_F=
\frac{1}{|\mathcal{I}_F|\times |\mathcal{T}^F|}\sum_{i\in\mathcal{I}_F}\sum_{t\in \mathcal{T}^F}
\Big(\E\Big[Y_{it}(1)| \vX_{i0}\Big]-\E\Big[Y_{it}(0)| \vX_{i0}\Big]\Big)
\end{equation}
where $\mathcal{I}_F\subseteq \mathcal{I}$ is a subset of the set of the sample locations where the interventions or events are assumed to occur in the future window $\mathcal{T}_Z^F$, and $U_1^F=\{it: i\in \mathcal{I}_F, t\in \mathcal{T}^F\}$.
The $ATT_F$ compares the two counterfactual scenarios where the lagged treatment vector  $\overline{\vZ}_{it}^{B,K}$ is such that $D_{it}=0$ or $D_{it}=1$ on all units $it\in U_1^F$.
The sub-sample $U_1^F$ where we wish to predict the causal effect $ATT^F$ depends on the question of interest.
We could focus on one future time point $\mathcal{T}^F=T+F+L$, with $B \leq L \geq K+B$ and predict the effect of initiating an intervention or having a single-day event at time $T+F$, in all areas in $\mathcal{I}$ or a subset of them, i.e., $U^F_1=\{it: i\in \mathcal{I}^{\star}\subset \mathcal{I}, t=T+F+L\}$. For example, 
after the US primary elections, federal and state government were anxious to gain insight into the possible effect of the general elections to be held on November 3rd, 2020. Similarly, during the surge of COVID-19 cases in Europe in October 2020, European governments had to decide on whether to impose restrictions to control the spread of the coronavirus and based their decision on the predicted effect of such regulations.
Similar questions can arise about the effect 
of having multi-day events (e.g., lockdowns, protests) lasting Q days within the time window $\mathcal{T}_Z^F=[T+F,\ldots,T+F+T_Z^F]$,  with $F \geq 1, T_Z^F\geq Q$ on a future day or days 
$\mathcal{T}^F=[T+F+L, \ldots, T+F+T^F]$, with $B+Q \leq L\leq B+Q+K$ and $T^F\leq T_Z^F+K$.
Predicting the effect of such events would allow policy-makers to decide on the length of lockdowns or on preventing long-lasting social events. 
Another interesting question could be about the predicted effect of implementing an intervention, not on a specific day, but during those days in a time window where characteristics reached a level similar to the one that determined the implementation of the intervention in the past \citep{naturemedicine2021}. Formally, 
$U^F_1=\{it \in \mathcal{I}\times\mathcal{T}^F: \overline{\mathbf{R}}_{it}\sim \mathcal{R}(U_1^{obs})\}$, 
where $\mathcal{R}(U_1^{obs})$ is the empirical distribution of past covariates and outcomes in the past treated sample $U_1^{obs}$.
This case refers to the question that policy-makers were asking at the end of the first wave about what they should have done in the case of a second wave with similar epidemic dynamic to the one in the Spring.
Finally, we could be interested in predicting the effect of administering a regulation under a pre-specified scenario of interest
$\mathcal{R}^{\star}$ for the covariate and outcome history before implementing the regulation, i.e., $U^F_1=\{it \in \mathcal{I}\times\mathcal{T}^F: \overline{\mathbf{R}}_{it}\in \mathcal{R}^{\star}\}$. For example, we could be interested in predicting the causal effect of imposing strict curbs when the weekly incidence and mortality exceed a threshold.

\subsubsection{Identification under Temporal Transportability}
Similarly to the discussion in Section \ref{sec:transp_4},
the first problem that arises in forecasting a causal effect is that both potential outcomes $Y_{it}(1)$ and $Y_{it}(0)$ in \eqref{eq: ATTF} are missing for all units in $U^F$. 
Here, we extend the temporal transportability assumptions required to forecast causal effects of time-varying treatments in the future.

Let us assume for now that all the time-varying effect modifiers were included in the covariate vector $\vX_{i\ell}$ and possibly in the outcome variable $Y_{i\ell}$, with their histories before the treatment and during the treatment period being potential modifiers.
Under the hypothetical assumption that all the effect modifiers were `measured' in both the observed time window and the unobserved future period, we formalize the temporal transportability assumption for potential outcomes under time-varying treatments. 
We first define the time-specific path-dependent history. For any unit $i$ and time $t$, let $d$ denote a treatment path and define the history $\mathcal{H}_{it}(d)$ as
$
    \mathcal{H}_{it}(d) = \{ \overline{\mathbf{Z}}_{it}^{B, K}: \mathbf{D}_{it} = d, \overline{\mathbf{X}}_{it}^{B, L_x}, \overline{\mathbf{Y}}_{it}^{B, L_y}, \mathbf{X}_{i0} \}
$, 
where $\overline{\mathbf{Z}}_{it}^{B, K}$ is the treatment history summary consistent with path $d$, $\overline{\mathbf{X}}_{it}^{B, L_x}$ and $\overline{\mathbf{Y}}_{it}^{B, L_y}$ are the lagged covariate and outcome histories with respective maximum lags $L_x$ and $L_y$, and $\mathbf{X}_{i0}$ represents baseline characteristics.

\begin{assumption}[Temporal Transportability of Potential Outcomes] 
\label{ass:transp}
   \textit{
Let $A_i$ be a random selection variable for unit $i$ taking values in a temporal domain $\mathcal{D}$. Define $\mathcal{H}_i(d)$ as the path-dependent history evaluated at the selected time $A_i$ and $Y_i(d)$ as the corresponding potential outcome under treatment path $d$:}
\begin{equation*}
        Y_i(d) = \sum_{t \in \mathcal{D}} Y_{it}(d) \cdot \mathbb{I}(A_i = t);
\qquad
    \mathcal{H}_i(d) = \sum_{t \in \mathcal{D}} \mathcal{H}_{it}(d)\cdot \mathbb{I}(A_i = t).
\end{equation*}
\textit{For all units $i \in \mathcal{I}_F$, the selection variable $A_i$ is independent of the potential outcomes given the path-dependent history:}
    \begin{equation}
        Y_i(d) \ind A_i \mid \mathcal{H}_i(d)
    \end{equation}
\end{assumption}
\noindent In words, Assumption \ref{ass:transp} states that, for units with the same baseline covariates, $\vX_{i0}$, that would have the same evolution of outcomes $\overline{\mathbf{Y}}_{it}^{B, L_y}$ and covariates $\overline{\vX}_{it}^{B,L_x}$ under a treatment history $\overline{Z}_{it}^{B, K}$ consistent with a treatment $d$, the distribution of potential outcomes under $d$ is the same across the temporal domain $\mathcal{D}$, regardless of whether unit $it$ were observed 
or it belongs to a future time window.
This means that Assumption \ref{ass:transp} states that all the potential effect modifiers are included in the covariate and outcome history and rules out the presence of unobserved time-varying effect modifiers that could be different between the observed sample and the future period. 
Notice that we also allow the evolution of covariates and outcomes during the cross-over treatment period, i.e.,  $\overline{\vX}_{it}^{B,K-1}$ and $\overline{\vY}_{it}^{B,K}$, to be effect modifiers.
As a consequence, both missing potential outcomes in \eqref{eq: ATTF} for future time points in $\mathcal{D}$ can be imputed, implicitly or explicitly, conditional on effect modifiers from a set of control and treated units observed in the past in the time window $\mathcal{D}$. 
 
In general, neither the evolution of outcomes $\overline{\mathbf{Y}}_{it}^{B, L_y}$ nor that of time-varying covariates $\overline{\vX}_{it}^{B,L_x}$ are observed (or fully observed) for units in the future set  $U^F_1$.
The temporal transportability of time-varying modifiers, including covariates and outcomes, with time-varying treatments can be formalized as follows, extending Assumption \ref{ass:conf_transp_4}.

\begin{assumption}[Temporal Transportability of Time-Varying Modifiers] 
\label{ass:conf_transp}
\textit{Let $\mathcal{H}_{it}^X=\{\overline{\mathbf{Z}}_{it}^{1, L_z}, \overline{\vX}_{it}^{1, L_{xx}},\overline{\vY}_{it}^{1, L_{xy}}, \vX_{i0}\}$ and 
$\mathcal{H}^Y_{it}=\{\overline{Z}_{it}^{B, K}, \overline{\vX}_{it}^{B, L_x},\overline{\mathbf{Y}}_{it}^{B, L_y}, \vX_{i0} \}$. Let $A_i$ be a random selection variable for unit $i$ taking values in a temporal domain $\mathcal{D}$. Define} 
\begin{equation*}
        X_i = \sum_{t \in \mathcal{D}} X_{it} \cdot \mathbb{I}(A_i = t);
\quad
    \mathcal{H}^X_i = \sum_{t \in \mathcal{D}} \mathcal{H}^X_{it}\cdot \mathbb{I}(A_i = t);
    \quad
    \mathcal{H}^Y_i = \sum_{t \in \mathcal{D}} \mathcal{H}^Y_{it}\cdot \mathbb{I}(A_i = t);
    \quad
    \mathbf{Y}_i = \sum_{t \in \mathcal{D}} \mathbf{Y}_{it} \cdot \mathbb{I}(A_i = t).
\end{equation*}
\textit{For every $i\in \mathcal{I}_F$, we have}
\begin{enumerate}[label={\textbf{Part~\Alph*}}, ref={\theassumption.\Alph*}, leftmargin=4\parindent]
\item \label{ass-A}%
 (Confounders): 
$\qquad\qquad\qquad\mathbf{X}_i\ind A_i \mid \mathcal{H}_i^X$
\item \label{ass-B}
(Lagged Outcomes):
$\qquad\qquad Y_i\ind A_i\mid\mathcal{H}_i^Y$
\end{enumerate}
\end{assumption}

\noindent Assumption \ref{ass:conf_transp} captures the temporal stability of time-varying modifiers across a temporal domain $\mathcal{D}$. In particular, 
if $\mathcal{D}$ spans across past and future time points
 Assumption \ref{ass-A} states that the distribution of time-varying covariates at each time point $t$ is the same between the past observed sample 
and the future time period, 
conditional on the past covariate, outcome, and treatment history.
Thus, Assumption \ref{ass-A}
rules out the presence of events that would change the way covariates evolve over time, and allows the prediction of time-varying covariates during the window $[t-B-L_x, t-B]$ that could affect the potential outcome $Y_{it}(d)$. 
Assumption \ref{ass-B} on the lagged outcomes is similar to Assumption \ref{ass:transp}, but it is defined as an independence of observed outcomes rather than potential outcomes. It states that the distribution of outcomes at each time point $t$ is the same between the observed time period and the future time period,
conditional on the past `observed' covariate, outcome and treatment history. 

Because
the treatment history in the conditioning sets of Assumptions \ref{ass-A} and \ref{ass-B}, i.e., $\overline{\vZ}_{i\ell}^{1,L_z}$ or $\overline{\vZ}_{i\ell}^{B,K}$, is in general not actually observed, we consider a pre-specified treatment vector $\mathbf{Z}_{i1:t-B-K-1}= \{Z_{i1}, \dots, Z_{i,t-B-K-1}\}=\tilde{\mathbf{z}}$ for the period before a future time point $t\in \mathcal{T}^F$. 
We would typically set set $Z_{ik}$ to: i) 0 for $k\in [T+1, \ldots, T+F-1]$, ii) the observed value for $k\in \mathcal{T}$.
Under Assumptions \ref{ass:transp} and \ref{ass:conf_transp}, we can prove the following proposition for the expected value of a potential outcome under future treatment level $d$ at a future time point defining $ATT_F$.

\begin{proposition}[Identification via Temporal Transportability]
\label{prop:transp}
\textit{
Assume the following conditions hold for all $i \in \mathcal{I}_F$ under a fixed pre-treatment trajectory $\mathbf{Z}_{i1:t-B-K-1}= \{Z_{i1}, \dots, Z_{i,t-B-K-1}\}=\tilde{\mathbf{z}} $ and a future treatment level $d$:}
\begin{enumerate}
\item Assumption \ref{ass:SUTVA_E} holds;
    \item {Consistency:} $Y^{obs}_{it} = Y_{it}(d)$ if  $\overline{\vZ}_{it}^{B, K} : D_{it} = d, \text{ for } t \in \mathcal{T}$;
   \item Assumption \ref{ass:transp} holds for the domain $\mathcal{D} = \mathcal{T} \cup \mathcal{T}^F$;
        \item Assumption \ref{ass:conf_transp} holds for  the domain $\mathcal{D} =  \mathcal{T} \cup \{T+1, \dots, T+F+T^F\}$;
        \item Recursive Positivity: 
    $
    \text{supp}(f_{\overline{\mathbf{R}}_{it}, \overline{\mathbf{G}}_{it} }(\cdot\mid \tilde{\mathbf{z}}, t \in \mathcal{T}^F, \mathbf{X}_{i0})) \subseteq \text{supp}(f_{\overline{\mathbf{R}}_{it}, \overline{\mathbf{G}}_{it}  }(\cdot\mid \tilde{\mathbf{z}}, t \in \mathcal{T}, \mathbf{X}_{i0}))$,
    where 
    $\mathbf{G}_{it} = \{X_{ik}, Y_{ik}, \tilde{z}_{ik}\}_{k=t-B-K-L_{max}}^{t-B-K-1}$, with $L_{max} = \max(L_x, L_y, L_{xx}, L_{xy})$, is an  intermediate vector of covariates, outcomes, and treatments.
\end{enumerate}
\textit{Then, the expected potential outcome at time $t\in \mathcal{T}^F$ is identified as:}
\begin{equation}
\begin{aligned}
\label{eq:ident_ATTF}
    \E\Big[Y_{it}(d) \mid t \in \mathcal{T}^F, \vX_{i0}\Big]=
 \sum_{\overline{\mathbf{r}}} \sum_{\overline{\mathbf{g}}} &
  \E\Big[Y^{obs}_{it} \mid t \in \mathcal{T}, D_{it}=d, \overline{\mathbf{G}}_{it}=\overline{\mathbf{g}}, \vX_{i0}\Big] \\&
 \times f_{\overline{\mathbf{R}}_{it},\overline{\mathbf{G}}_{it}}(\overline{\mathbf{r}},\overline{\mathbf{g}} \mid \tilde{\mathbf{z}}, t \in \mathcal{T}^F, \vX_{i0}) 
    \end{aligned}
\end{equation}
where
\begin{equation}
\label{eq:predR}
\begin{aligned}
f(\overline{\mathbf{r}}, \overline{\mathbf{g}} \mid \mathbf{\tilde{z}}, t \in \mathcal{T}^F, \mathbf{X}_{i0}) = \sum_{\mathcal{H}_{iT}, \mathcal{H}_{gap} \setminus \{\overline{\mathbf{r}}, \overline{\mathbf{g}}\}} \Bigg[ &\prod_{k=T+1}^{t-B-K-1} f(x_{ik} \mid \mathcal{H}_{ik}^X(\mathbf{\tilde{z}}), \tilde{z}{ik}, t \in \mathcal{T}) f(y_{ik} \mid \mathcal{H}_{ik}^Y(\mathbf{\tilde{z}}), t \in \mathcal{T})\Bigg]
\end{aligned}
\end{equation}
\textit{where $\mathcal{H}_{iT} = \{\mathbf{x}_{ik}, \mathbf{y}_{ik}, \tilde{\mathbf{z}}_{ik}\}_{k=1}^{T}$ and $\mathcal{H}_{gap} = \{\mathbf{x}_{ik}, \mathbf{y}_{ik}, \tilde{\mathbf{z}}_{ik}\}_{k=T+1}^{t-B-K-1}$ are the historical and intermediate gap trajectories under fixed pre-treatment values $\tilde{z}_{ik}$, $\mathcal{H}_{ik}^X(\tilde{\mathbf{z}})$ and $\mathcal{H}_{ik}^Y(\tilde{\mathbf{z}})$ represent the {sliding-window conditioning histories} constructed using the evolving realizations of modifiers, satisfying the kernel lags $L_{max}$, and the fixed values from the vector $\tilde{\mathbf{z}}$, and
 the transition kernels $f(\cdot \mid \dots, t \in \mathcal{T})$ are taken from the observed historical sample $\mathcal{T}$.}
       
\noindent Proof is reported in the Appendix.
\end{proposition}

\noindent Proposition \ref{prop:transp} states that the expected potential outcome $Y_{it}(d)$ at a future time point  $t\in \mathcal{T}^F$ is identified by the expected value of the observed outcomes observed in the past time window conditional on treatment level $d$, baseline covariates, and averaged over the distribution of effect modifiers before the carry-over treatment period up to $t-B-K-L_{max}$, also evaluated in the past, conditional on the pre-specfied treatment vector $\mathbf{\tilde{z}}$, and obtained recursively until baseline.

Proposition also \ref{prop:transp} implies that we only need to predict effect modifiers before the treatment starts at time $t-B-K$ .
The evolution of post-treatment effect modifiers $\overline{\vX}^{B,K-1}_{it}$ and
$\overline{\vY}^{B,K}_{it}$, that are affected by the counterfactual scenario,  does not need to be predicted because it will be averaged over, conditional on pre-treatment covariates.
The intuition is that the evolution of the time-varying modifiers affected by the counterfactual treatment scenario is just a mechanism through which the treatment has an effect on the final outcome.
If we assume that these time-varying modifiers can be predicted by their observed history (Assumption \ref{ass:conf_transp}), this means that this mechanism 
can be explained by the variables collected in $X$ (and $Y$), ruling out other unmeasured factors that could be responsible for a different trend of these modifiers as a response of the treatment. 

On the other hand, the identification result in Proposition \ref{prop:transp} still requires a marginalization over the distribution of time-varying covariates and outcomes before the carry-over window. This marginalization could be performed using a Monte-Carlo simulations as commonly done in G-computation approach or with multiple imputation techniques \citep{Bartlett}.
Assumptions \ref{ass-A} and \ref{ass-B} would allow us to `impute' the covariate and outcome history in the future time period before the carry-over treatment window under the counterfactual scenario $D_{it}=d$, using a model fitted to data observed in $U^{obs}$, as shown in Equation \eqref{eq:predR}.
In practice, this can be done recursively as more explicitly suggested in Equation \eqref{eq:predR}.
For instance, if US federal and state officials were interested in predicting the possible effect of holding the general elections in-person after observing the primary elections during the first wave of COVID-19 in the Spring 2020, they would have had to impute the epidemic dynamics until November 3rd.
It is worth mentioning that units in $U_1^F$ are `matched' with units in $U^{obs}$ across time but also across space. In fact, a future time point $t\in \mathcal{T}^F$ at a geographical area $i\in \mathcal{I}$ could be matched with an observed time point $t'\in \mathcal{T}$ at another geographical area. 
\begin{remark}[Additional observations]We have been denoting by $\mathcal{T}$ the observed time window used for the estimation of the effect of the intervention or event of interest that occurred on some units, e.g., the primary elections in the Spring 2020.
However, 
at the time of predicting the effect  on a future period we could have also observed the trend of deaths and cases -- as well as the treatment history --
beyond the observed time window $\mathcal{T}$ used for estimation, e.g. in the Summer 2020.  This set of observations could still be used to estimate the model for the evolution of time-varying modifiers and predict them to forecast the causal effect of the general election in the Fall 2020. 
\end{remark}

\begin{remark}[Alternative estimand: conditioning on observed history]
In Equation \ref{eq:predR}, time-varying effect modifiers in the pre-treatmnet window are recursively obtained by averaging their distribution in the intermediate gap as well as in the observed time period. This is because the expected potential outcomes of interested are conditional on baseline covariate, but not on the history.  

An alternative estimand of practical interest is the history-conditioned future average treatment  effect, which conditions the expectations on the full observed history $\mathcal{H}_{iT} = \{\mathbf{X}_{ik}, \mathbf{Y}_{ik}, \mathbf{Z}_{ik}\}_{k=1}^{T}$, in addition to baseline $\mathbf{X}_{i0}$, i.e., $\mathbb{E}[Y_{it}(d) \mid \mathcal{H}_{iT}, \mathbf{X}_{i0}, t \in \mathcal{T}^F]$.
This formulation is particularly relevant for real-time forecasting as it treats the study's end $T$ as a fixed initial state, thereby capturing the latent unit-specific risks accumulated during the historical period.

This conditioning simplifies the identification by treating the historical path as a fixed initial condition rather than a random variable to be marginalized. 
In particular, identification is simplified by initiating the recursive unrolling in Propositions \ref{prop:transp} strictly from $T+1$,
eliminating the need for the marginalization over the observed past, $f(\mathcal{H}_{iT} \mid \mathbf{X}_{i0})$.
\end{remark}

\begin{remark}[Forecasting in the immediate future]
Consider the setting where 
we observed all the time-varying effect modifiers until the day before the time we wish to assess the possible effect of an intervention. Assume that we are interested in the history-conditioned future average treatment  effect. 

In this case, the temporal transportability assumptions for time-varying modifiers (Assumptions \ref{ass-A} and \ref{ass-B}) are not required for a single-day intervention or event, but are still required for multi-day interventions. This is because we would estimate the effect in a future time window using data from the past sample
averaging over the evolution of post-treatment modifiers that would be seen for these past units. 
\end{remark}

\section{Potential issues invalidating temporal transportability}
\label{sec:issues}
 It is worth stressing the implications of the temporal transportability assumptions and the possible situations that would invalidate them. 
The temporal transportability of potential outcomes assumptions (Assumptions \ref{ass:transp_4} and \ref{ass:transp}) rule out the presence of unmeasured time-varying effect modifiers. 
Thus, when predicting the effect on mortality rate of a COVID regulation or event that could affect the spread of the virus, we must consider all time-varying factors that could potentially affect such effect.
Such time-varying effect modifiers could be the weather and other environmental factors, population preventive behaviors (e.g., social distancing and mask-wearing), the status of COVID-19 infections and hospitalizations, health care preparedness (e.g., medical supplies and access to personal protective equipment), as well as the infectiousness and deadliness of the virus which could change over time due to mutations. 
Some of these factors, such as weather, are not affected by COVID-19 restrictions and can be predicted using estimated trends. On the contrary, other factors, such as the epidemic status, are influenced by previous regulations or events and should be predicted under the observed interventions/events and hypothetical interventions/events. 
Nevertheless, other factors, such as preventive behaviors and virus mutations, are hard to measure and, in general, will not be included in X.

On the other hand, the temporal transportability assumptions of time-varying modifiers (Assumptions \ref{ass:conf_transp_4} and \ref{ass:conf_transp}) require the absence of unobserved events or time-varying factors that would have affected their trend. In the case of COVID-19 effect modifiers, similar unmeasured factors to the ones invalidating the temporal transportability assumption of potential outcomes (e.g., virus mutations and preventive behaviors) would also invalidate the prediction of other time-varying effect modifiers (e.g., epidemic status) from the observed data. In addition, we should also be careful about other interventions and events that could possibly affect the COVID-19 transmission. In the presence of such phenomenons between our observed time window and the future period where we want to make predictions, the temporal transportability of effect modifiers will not hold. 


In general, the closer we are to the future time window $\mathcal{T}^F$ the more plausible the temporal transportability assumptions will be and, in turn, the more accurate the prediction of time-varying modifiers and causal effects can be. 
In the presence of a large gap between the observed period and the future window of interest, i.e., $F>>1$, it is likely that some of these genetic, behavioral and environmental factors, or the occurrence of social events or restrictions other than the ones defining the treatment, have changed the transmission dynamics of COVID-19. 

In practice, 
the accuracy of predictions also depend on the accuracy of the models, including those for the effect modifiers and possibly the outcome model.
Such accuracy will depend on the availability of historical data as well as on the gap between the observed and future time windows. Our ability to predict the future context may also depend on current uncertain circumstances and on our understanding of the phenomenon.
For example: sometimes the amount of medical supplies and equipment that will be at hand is hard to predict; despite advances in meteorological modeling technology, weather forecasts often fail; predicting virus mutations is still a difficult task.

If we were to use the effect estimated for the primary elections in the US in the Spring 2020 to predict the effect of the general elections on November 3rd, we would need to predict and control for all the aforementioned effect modifiers, as well as in-person turnout. 
Social distancing and mask-wearing behaviors have likely changed over time, from the first to the second wave. Over the summer, we saw in Europe a decreasing adherence of the population to preventive guidelines.
Another difference between the Spring and the Fall 2020 is in the shortage of medical supplies and limited access to personal protective equipment, that contributed to a higher mortality rate during the first wave. 
If these factors were not taken into account, it is possible that the predicted effect of elections 
would be biased.
If this exercise were conducted in the Summer 2020, it would have been harder to predict the spread of COVID-19 in the Fall 2020, preventive behaviors, health care preparedness, and in-person turnout. Conversely, if such analysis were carried out in early October, we could have better predicted in-person turnout and heath care preparedness. At that time, we also had a good sense of the different adherence to preventive guidelines in each area between the Spring and the Fall, and we could have predicted mask wearing and social distancing behaviors during and outside the polls. However, it would have been harder to forecast the epidemic dynamics, because temperatures were going down, in-door gatherings were becoming more frequent, and we were still uncertain about the potential effect of school opening.

\section{Conclusions and discussion}
\label{sec:dis}
The goal of policy evaluation is commonly stated as ``to determine the impact of a policy on some  outcome of interest'' \citep{heckman2007econometric}. This statement can be viewed in two ways: 1) we would like to assess the effect of the policy on the outcome of interest in the population and at the time the policy was actually implemented; 2) we would like to investigate whether the implemented policy would also be effective in another population and in a future time. The first type of causal question is actually what we have in mind when we design an evaluation study. However, the answer to the second question  would be the one that should inform future policies.
During the SARS-CoV-2 emergency in summer 2020, such question appeared to be crucial for public health officials trying to 
decide whether restrictions similar to those implemented in the spring should be replicated and whether similar social events should still be allowed. 
For this reason we use the evaluation of COVID-19 interventions as a motivating example throughout the paper. 

We have first reviewed the identification of the causal effect of an intervention, observed or hypothetical, in the observed time window, distinguishing between point intervention, intervention without duration, with continuous duration and intermittent occurrence. Then, we have dealt with the problem of predicting the effect of an intervention in a future time window. Due to the presence of possible time-varying effect modifiers, we have extended existing results on generalizability across populations to a problem of generalizability from the past to the future. We have clarified the assumptions required to identify predicted causal effects, including those needed to predict time-varying effect modifiers, which cannot be observed in a future time, and those that would allow model-based or non-parametric predictions of causal effects under the predicted modified context. By decomposing the identification process into the temporal transportability of potential outcomes and the sequential projection of effect modifiers, we bridge the latent gap between observed data and future intervention windows. The unifying logic across this framework is that while the future state of a population may be unobserved, its causal structures remain identifiable as long as the underlying dynamic mechanisms of effect modification—the rules governing the evolution of covariates, outcomes, and exposures—remain time-invariant.

In the simplest point-treatment setting, identification is achieved by applying a stable historical outcome mapping to the recursively identified future density of effect modifiers. This allows for a straightforward projection of treatment effects into a future period, provided the evolution of modifiers from baseline is captured by historical trends. However, as we move to the time-varying treatment setting, the complexity increases due to the presence of path-dependency and carry-over effects. In this context, the identification results rely on a temporal buffer, $\mathbf{G}_{it}$, which carries the system’s memory from the end of the observed history through the intermediate gap.
The most general case addressed in the Appendix  involves hypothetical interventions that modify the exposure distribution. Identifying the Average Exposure Effect in the future requires a dual characterization of the future: the ``natural course'' of the exposure and the counterfactual ``interventional course''. A critical distinction of this third setting is the additional requirement for the transportability of the exposure distribution itself. Unlike treatment settings where the ``natural course" of the treatment is either observed or assumed to be absent, forecasting the effect of an intervention modifying the exposure
  necessitates the sequential projection of the exposure's natural evolution. By assuming the stationarity of the exposure mechanism, we can identify the likely path of the system without intervention.
In the case of multi-day interventions, we have shown that time-varying modifiers only need to be predicted for the time window between the observed one and the first day in the future where we assume that the intervention of interest will be implemented. That is, we do not need to explicitly predict effect modifiers during the carry-over window. Because the transportability of the outcome mechanism is assumed to hold conditional on the pre-treatment history,  by averaging over the historical distribution of post-treatment modifiers observed under equivalent realizations, we effectively marginalize over the mechanism through which the intervention exerts its effect.
This is also true for hypothetical interventions (see Appendix), except when they depend on previous outcomes or exposures. We further distinguish between baseline-conditioned and history-conditioned estimands, providing flexibility to anchor forecasts to a unit’s actual realized state.
In Table \ref{tab:table} we summarize identification results for causal effects in the observed or future sample by intervention type and presence of time-dependent confounders. 


\begin{table}[t]
  \renewcommand{\arraystretch}{2.5}
    \centering
    \caption{Summary of identification results by time window, intervention type, and whether time-dependent confounders are present. Note that  $\overline{\mathbf{R}}_{it}$: pre-treatment confounders and $\overline{\mathbf{V}}_{it}$: post-treatment confounders during the treatment window}
    \label{tab:table}
      \begin{adjustwidth}{-0.9cm}{0cm}
    \begin{tabular}{|c|c|c|c|c|@{\,\,\,\,\,\,}l@{\,\,\,\,\,\,}|}
    \toprule
         Time Window & Causal Effect& Intervention Type& \pbox{3cm}{Time-dependent\\ Confounders}& Assumptions &Identification \\[0.5cm]
         \hline
                  observed & Eq.\eqref{eq:ATT_4} &point treatment &NA &Ass.\ref{ass:Ign_4}& Eq.
\eqref{eq:att_id_4}\\
         observed & Eq.\eqref{eq:ATT} &one-day &NA &Ass.\ref{ass:Ign}& \pbox{3cm}{Eq. \eqref{eq:adj}, \eqref{eq:att_id}\\
         \hspace{0.5cm}$C_{it}=\overline{\mathbf{R}}_{it}$}
       \\
         observed &Eq.\eqref{eq:ATT} &multi-day  & no&Ass.\ref{ass:Ign}&   
         \pbox{3cm}{Eq. \eqref{eq:adj}, \eqref{eq:att_id}\\
         $C_{it}\!\!=\!\![\overline{\mathbf{R}}_{it}\!,\! \overline{\mathbf{V}}_{it}]$}\\
         observed &Eq.\eqref{eq:ATT} &multi-day & yes&Ass.\ref{ass:Ign}&
         Eq.\eqref{eq:att_id}, \eqref{eq:gformula}\\
        future & Eq.\eqref{eq: ATTF_4} &point treatment &NA &Ass.\ref{ass:transp_4}, \ref{ass:conf_transp_4}& \pbox{3.5cm}{Prop.\ref{prop:transp_4}: Eq.\eqref{eq:predATT_4}, \eqref{eq:predATT0_4}, \\Prop.\ref{prop:recursive_covariates}: Eq.\eqref{eq:predX_4}, \eqref{eq:predX_4b} }\\

         future & Eq.\eqref{eq: ATTF} & \pbox{3cm}{time-varying \\treatment}  &yes/no& Ass.\ref{ass:transp}, \ref{ass:conf_transp}& Prop.\ref{prop:transp}: Eq.\eqref{eq:ident_ATTF}, \eqref{eq:predR}\\
         future & Eq.\eqref{eq: AEEF} & \pbox{3cm}{time-varying \\treatment}  &yes/no& Ass.\ref{ass:transp_exp}, \ref{ass:transp_exp_dis}, \ref{ass:conf_transp_exp}& \pbox{3.5cm}{Prop.\ref{prop:transp_exp}: Eq.\eqref{eq:ident_future_ERF_collapsed}, \eqref{eq:p3_unrolling_fixed_background_final}\\
         Prop.\ref{prop:AEEF_identification}: Eq.\eqref{eq:natural_S_unrolling}}\\
    \bottomrule
    \end{tabular}
\end{adjustwidth}
\end{table}

Furthermore, we have discussed several issues invalidating the prediction of causal effects: the problem of unmeasured effect modifiers (e.g., preventive behaviors, viral mutations) that cannot be observed and predicted, 
possible phenomenons occurring between the observed and the future time windows, and the inaccuracy of the models used for both time-varying modifiers and potential outcomes. As mentioned, such issues depend on the data quality at hand, the availability of historical data, the gap between the observed and future time windows, current uncertain circumstances, and on our understanding of the phenomenon. Moreover, one must ensure recursive positivity, verifying that the projected trajectories of effect modifiers do not drift into regions of the state space never witnessed in the historical sample.

Although we have discussed the problem of causal prediction using the example of COVID-19 interventions, our causal framework, identification results, and discussion on the underlying assumptions, apply to all fields where there is interest in forecasting causal effects of interventions or event.
Model-based projections have been used  extensively in environmental epidemiology, and, in particular, in the field of climate change to predict long-term health effects of implemented or hypothetical interventions to reduce greenhouse emissions
\citep{Baccini64, kendrovski,GASPARRINI2017, guidelines, adaptation, VicedoCabrera2019HandsonTO,
adaptation2}.
For instance, we could be interested in estimating the effect of  traffic bans or the installation of scrubbers on power plants, 
designed to reduce the emission of fine particulate matter (PM$_{10}$ or PM$_{2.5}$), and informing policy-makers on the potential impact of implementing these interventions in a different time when 
mobility behaviors might have changed.
On the other hand, as a hypothetical interventions, we could think of a emissions reduction policy that would be able keep the annual average levels of PM$_{10}$ below a recommended threshold or a law limit (e.g., 20 $\mu$g$/$m$^3$) \citep{forastiere_envhp2020}.

Many limitations of the use of model-based projections to forecast the effect of actual or hypothetical interventions in the future are known among researchers \citep{wronguseful2020, kim_gu_yu_wang_wang_2021}.
For example, in the climate change literature, changes in the population structure as well as physiological adaptation to varying temperatures have been identified as two important issues that should be discussed and possibly addressed with sensitivity analyses when performing model-based projections \citep{VicedoCabrera2019HandsonTO, guidelines, adaptation, adaptation2}.
Here, we wish to provide formal evidence of such limitations and give further insights into the underlying assumptions that we make whenever such predictions are conducted. 
By doing so, we hope to encourage researchers to be more transparent and clear about their assumptions and to discuss their plausibility and possible circumstances that could invalidate them.
In spite of the limitations, model-based projections, constrained by what we have observed and what we assume, can still be a necessary input 
to public policy decisions,
if used appropriately and with an understanding of their weaknesses.


Future research should prioritize the development of sensitivity analysis frameworks to quantify the robustness of forecasted causal effects against violations of the core transportability assumptions. 
On the estimation front, while the current framework supports a generative Monte Carlo imputation procedure, there is significant opportunity to explore more efficient and robust estimators.

Among the issues of forecasting a causal effect in the future, we have not discussed two main problems that would also invalidate predictions from a past observed sample to a future time. The first is the fact that the implementation of the intervention in the future might be different from the way it was implemented in the past. This is because, usually the implementation of an intervention is context-specific and it not easy to replicate the same operational steps. For example, in Europe the way the same COVID-19 preventive guidelines were imposed in the Fall was very different from the stricter controls we have seen in the Spring.
Another issue is that the no-interference assumption that could have been valid during the Spring, where strict mobility restrictions were imposed, might not be plausible in the Fall 2020, where inter-state mobility has begun to rise again. These questions are beyond the scope of this article but will be addressed in future research.



\bibliographystyle{elsarticle-harv}
\bibliography{prediction.bib}

@article{cole_stuart_2010,
author = {Cole, Stephen and Stuart, Elizabeth},
year = {2010},
month = {07},
pages = {107-15},
title = {Generalizing Evidence From Randomized Clinical Trials to Target Populations. The ACTG 320 Trial},
volume = {172},
journal = {American journal of epidemiology},
doi = {10.1093/aje/kwq084}
}

@article{buchanan_hudgens_2018,
author = {Buchanan, Ashley and Hudgens, Michael and Cole, Stephen and Mollan, Katie and Sax, Paul and Daar, Eric and Adimora, Adaora and Eron, Joseph and Mugavero, Michael},
year = {2018},
month = {02},
pages = {},
title = {Generalizing Evidence from Randomized Trials using Inverse Probability of Sampling Weights},
volume = {181},
journal = {Journal of the Royal Statistical Society: Series A (Statistics in Society)},
doi = {10.1111/rssa.12357}
}

@article{imai2008misunderstandings,
  title={Misunderstandings between experimentalists and observationalists about causal inference},
  author={Imai, Kosuke and King, Gary and Stuart, Elizabeth A},
  journal={Journal of the royal statistical society: series A (statistics in society)},
  volume={171},
  number={2},
  pages={481--502},
  year={2008},
  publisher={Wiley Online Library}
}

@article{marshall2015formalizing,
  title={Formalizing the role of agent-based modeling in causal inference and epidemiology},
  author={Marshall, Brandon DL and Galea, Sandro},
  journal={American journal of epidemiology},
  volume={181},
  number={2},
  pages={92--99},
  year={2015},
  publisher={Oxford University Press}
}

@article{murray2017comparison,
  title={A comparison of agent-based models and the parametric g-formula for causal inference},
  author={Murray, Eleanor J and Robins, James M and Seage, George R and Freedberg, Kenneth A and Hern{\'a}n, Miguel A},
  journal={American journal of epidemiology},
  volume={186},
  number={2},
  pages={131--142},
  year={2017},
  publisher={Oxford University Press}
}

@article{li2022generalizing,
  title={Generalizing trial evidence to target populations in non-nested designs: Applications to aids clinical trials},
  author={Li, Fan and Buchanan, Ashley L and Cole, Stephen R},
  journal={Journal of the Royal Statistical Society. Series C, Applied statistics},
  volume={71},
  number={3},
  pages={669},
  year={2022},
  publisher={NIH Public Access}
}

@article{li2021note,
  title={A note on semiparametric efficient generalization of causal effects from randomized trials to target populations},
  author={Li, Fan and Hong, Hwanhee and Stuart, Elizabeth A},
  journal={Communications in Statistics-Theory and Methods},
  pages={1--32},
  year={2021},
  publisher={Taylor \& Francis}
}

@article{pearl_2105,
author = {Pearl, Judea},
year = {2015},
month = {01},
pages = {},
title = {Generalizing Experimental Findings},
volume = {3},
journal = {Journal of Causal Inference},
doi = {10.1515/jci-2015-0025}
}

@article{article,
author = {Mumford, Sunni and Schisterman, Enrique},
year = {2019},
month = {06},
pages = {1-2},
title = {New methods for generalizability and transportability: the new norm},
volume = {34},
journal = {European Journal of Epidemiology},
doi = {10.1007/s10654-019-00532-3}
}

@article{westreich2017transportability,
  title={Transportability of trial results using inverse odds of sampling weights},
  author={Westreich, Daniel and Edwards, Jessie K and Lesko, Catherine R and Stuart, Elizabeth and Cole, Stephen R},
  journal={American journal of epidemiology},
  volume={186},
  number={8},
  pages={1010--1014},
  year={2017},
  publisher={Oxford University Press}
}

@article{forastiere_envhp2020,
title = "Assessing short-term impact of PM10 on mortality using a semiparametric generalized propensity score approach",
author = "Laura Forastiere and Michele Carugno and Michela Baccini",
year = "2020",
month = may,
day = "1",
doi = "10.1186/s12940-020-00599-6",
language = "English",
volume = "19",
journal = "Environmental Health: A Global Access Science Source",
issn = "1476-069X",
publisher = "BioMed Central",
number = "1",
}

@book{imbens2015causal,
  title={Causal inference in statistics, social, and biomedical sciences},
  author={Imbens, Guido W and Rubin, Donald B},
  year={2015},
  publisher={Cambridge University Press}
}

@article{stuart,
	author={Elizabeth A. Stuart and Stephen R. Cole and Catherine P. Bradshaw and Philip J. Leaf},
	year={2011},
	title={The use of propensity scores to assess the generalizability of results from randomized trials},
	journal={Journal of the {R}oyal {S}tatistical {S}ociety: {S}eries A ({S}tatistics in {S}ociety)},
	volume={174},
	number={2},
	pages={369-386}
}

@article{sonnenberg1993markov,
  title={Markov models in medical decision making: a practical guide},
  author={Sonnenberg, Frank A and Beck, J Robert},
  journal={Medical decision making},
  volume={13},
  number={4},
  pages={322--338},
  year={1993},
  publisher={Sage Publications Sage CA: Thousand Oaks, CA}
}

@article{tipton,
	author={Elizabeth Tipton},
	year={2013},
	title={Improving generalizations from experiments using propensity score subclassification: Assumptions, properties, and contexts},
	journal={Journal of {E}ducational and {B}ehavioral {S}tatistics},
	volume={38},
	number={3},
	pages={239-266}
}

@article{tipton14,
	author={Elizabeth Tipton and Larry Hedges and Michael Vaden-Kiernan and Geoffrey Borman and Kate Sullivan and Sarah Caverly},
	year={2014},
	title={Sample selection in randomized experiments: A new method using propensity score stratified sampling},
	journal={Journal of {R}esearch on {E}ducational {E}ffectiveness},
	volume={7},
	number={1},
	pages={114-135}
}

@article{dahabreh2019extending,
  title={Extending inferences from a randomized trial to a target population},
  author={Dahabreh, Issa J and Hern{\'a}n, Miguel A},
  journal={European Journal of Epidemiology},
  volume={34},
  number={8},
  pages={719--722},
  year={2019},
  publisher={Springer}
}

@article{dahabreh2020extending,
  title={Extending inferences from a randomized trial to a new target population},
  author={Dahabreh, Issa J and Robertson, Sarah E and Steingrimsson, Jon A and Stuart, Elizabeth A and Hernan, Miguel A},
  journal={Statistics in medicine},
  volume={39},
  number={14},
  pages={1999--2014},
  year={2020},
  publisher={Wiley Online Library}
}

@article{rudolph2017robust,
  title={Robust estimation of encouragement-design intervention effects transported across sites},
  author={Rudolph, Kara E and van der Laan, Mark J},
  journal={Journal of the Royal Statistical Society. Series B, Statistical methodology},
  volume={79},
  number={5},
  pages={1509--1525},
  year={2017},
  publisher={NIH Public Access}
}

@article{muirch,
	author={Colm O'Muircheartaigh and Larry V. Hedges},
	year={2013},
	title={Generalizing from unrepresentative experiments: {A} stratified propensity score approach},
	journal={Journal of the {R}oyal {S}tatistical {S}ociety: {S}eries {C} ({A}pplied {S}tatistics)},
	volume={63},
	number={2},
	pages={195--210}
}

@article{hartman2015sample,
  title={From sample average treatment effect to population average treatment effect on the treated: combining experimental with observational studies to estimate population treatment effects},
  author={Hartman, Erin and Grieve, Richard and Ramsahai, Roland and Sekhon, Jasjeet S},
  journal={Journal of the {R}oyal {S}tatistical {S}ociety: {S}eries {A} ({S}tatistics in {S}ociety)},
  volume={178},
  number={3},
  pages={757--778},
  year={2015},
  publisher={Wiley Online Library}
}

@article{naimi2017introduction,
  title={An introduction to g methods},
  author={Naimi, Ashley I and Cole, Stephen R and Kennedy, Edward H},
  journal={International journal of epidemiology},
  volume={46},
  number={2},
  pages={756--762},
  year={2017},
  publisher={Oxford University Press}
}

@article{robins1999association,
  title={Association, causation, and marginal structural models},
  author={Robins, James M},
  journal={Synthese},
  pages={151--179},
  year={1999},
  publisher={JSTOR}
}

@article{heckman2007econometric,
  title={Econometric evaluation of social programs, part I: Causal models, structural models and econometric policy evaluation},
  author={Heckman, James J and Vytlacil, Edward J},
  journal={Handbook of econometrics},
  volume={6},
  pages={4779--4874},
  year={2007},
  publisher={Elsevier}
}

@book{whatif,
    author = {Hern\'an, MA and Robins, JM},
    title = {Causal Inference: What If},
    publisher = {Boca Raton: Chapman \& Hall/CRC},
    year = {2020}
}

@book{Rothman2008,
author = {Rothman, K.J. and Greenland, Sander and Lash, T.L.},
year = {2011},
month = {11},
pages = {1-758},
title = {Modern epidemiology: Third edition},
publisher = {Lippincott Williams \& Wilkins}
}

@book{mckenzie2022planning,
  title={Planning, implementing and evaluating health promotion programs},
  author={McKenzie, James F and Neiger, Brad L and Thackeray, Rosemary},
  year={2022},
  publisher={Jones \& Bartlett Learning}
}

@article{evidencebased,
	Author = {Brownson, Ross C. and Chriqui, Jamie F. and Stamatakis, Katherine A.},
	Journal = {American Journal of Public Health},
	Number = {9},
	Pages = {1576-1583},
	Title = {Understanding Evidence-Based Public Health Policy},
	Volume = {99},
	Year = {2009}}

@article{Covid_predictions_Jewell,
    author = {Jewell, Nicholas P. and Lewnard, Joseph A. and Jewell, Britta L.},
    title = "{Predictive Mathematical Models of the COVID-19 Pandemic: Underlying Principles and Value of Projections}",
    journal = {JAMA},
    volume = {323},
    number = {19},
    pages = {1893-1894},
    year = {2020},
    month = {05},
    issn = {0098-7484},
    doi = {10.1001/jama.2020.6585},
    url = {https://doi.org/10.1001/jama.2020.6585},
}

@article{frontiers2021,
	Author = {Yadav, Subhash Kumar and Akhter, Yusuf},
	Journal = {Frontiers in Public Health},
	Title = {Statistical Modeling for the Prediction of Infectious Disease Dissemination With Special Reference to COVID-19 Spread},
	Volume = {9},
	Year = {2021}}

@article{forecasting_book,
author = {Lauer, Stephen and Brown, Alexandria and Reich, Nicholas},
title = {Infectious Disease Forecasting for Public Health},
journal={Population biology of vector-borne diseases (May 2020)},
  volume={45},
  year={2020}
}

@article{GASPARRINI2017,
	Author = {Antonio Gasparrini and Yuming Guo and Francesco Sera and Ana Maria Vicedo-Cabrera and Veronika Huber and Shilu Tong and Micheline {de Sousa Zanotti Stagliorio Coelho} and Paulo Hilario {Nascimento Saldiva} and Eric Lavigne and Patricia {Matus Correa} and Nicolas {Valdes Ortega} and Haidong Kan and Samuel Osorio and Jan Kysel{\'y} and Ale{\v s} Urban and Jouni J K Jaakkola and Niilo R I Ryti and Mathilde Pascal and Patrick G Goodman and Ariana Zeka and Paola Michelozzi and Matteo Scortichini and Masahiro Hashizume and Yasushi Honda and Magali Hurtado-Diaz and Julio {Cesar Cruz} and Xerxes Seposo and Ho Kim and Aurelio Tobias and Carmen I{\~n}iguez and Bertil Forsberg and Daniel Oudin {\AA}str{\"o}m and Martina S Ragettli and Yue Leon Guo and Chang-fu Wu and Antonella Zanobetti and Joel Schwartz and Michelle L Bell and Tran Ngoc Dang and Dung Do Van and Clare Heaviside and Sotiris Vardoulakis and Shakoor Hajat and Andy Haines and Ben Armstrong},
	Journal = {The Lancet Planetary Health},
	Number = {9},
	Pages = {e360-e367},
	Title = {Projections of temperature-related excess mortality under climate change scenarios},
	Volume = {1},
	Year = {2017}}

@article{Baccini64,
	Author = {Baccini, M and Kosatsky, T and Analitis, A and Anderson, H R and D{\textquoteright}Ovidio, M and Menne, B and Michelozzi, P and Biggeri, A and others},
	Journal = {Journal of Epidemiology \& Community Health},
	Number = {1},
	Pages = {64--70},
	Title = {Impact of heat on mortality in 15 European cities: attributable deaths under different weather scenarios},
	Volume = {65},
	Year = {2011}}

@article{VicedoCabrera2019HandsonTO,
  title={Hands-on Tutorial on a Modeling Framework for Projections of Climate Change Impacts on Health},
  author={Ana Maria Vicedo-Cabrera and Francesco Sera and Antonio Gasparrini},
  journal={Epidemiology (Cambridge, Mass.)},
  year={2019},
  volume={30},
  pages={321 - 329}
}

@article{guidelines,
	Author = {Jeremy J. Hess and Nikhil Ranadive and Chris Boyer and Lukasz Aleksandrowicz and Susan C. Anenberg and Kristin Aunan and Kristine Belesova and Michelle L. Bell and Sam Bickersteth and Kathryn Bowen and Marci Burden and Diarmid Campbell-Lendrum and Elizabeth Carlton and Gu{\'e}ladio Ciss{\'e} and Francois Cohen and Hancheng Dai and Alan David Dangour and Purnamita Dasgupta and Howard Frumkin and Peng Gong and Robert J. Gould and Andy Haines and Simon Hales and Ian Hamilton and Tomoko Hasegawa and Masahiro Hashizume and Yasushi Honda and Daniel E. Horton and Alexandra Karambelas and Ho Kim and Satbyul Estella Kim and Patrick L. Kinney and Inza Kone and Kim Knowlton and Jos Lelieveld and Vijay S. Limaye and Qiyong Liu and Lina Madaniyazi and Micaela Elvira Martinez and Denise L. Mauzerall and James Milner and Tara Neville and Mark Nieuwenhuijsen and Shonali Pachauri and Frederica Perera and Helen Pineo and Justin V. Remais and Rebecca K. Saari and Jon Sampedro and Pauline Scheelbeek and Joel Schwartz and Drew Shindell and Priya Shyamsundar and Timothy J. Taylor and Cathryn Tonne and Detlef Van Vuuren and Can Wang and Nicholas Watts and J. Jason West and Paul Wilkinson and Stephen A. Wood and James Woodcock and Alistair Woodward and Yang Xie and Ying Zhang and Kristie L. Ebi},
	Journal = {Environmental Health Perspectives},
	Number = {11},
	Pages = {115001},
	Title = {Guidelines for Modeling and Reporting Health Effects of Climate Change Mitigation Actions},
	Volume = {128},
	Year = {2020}}

@article{adaptation,
	Author = {Alana Hansen and Peng Bi},
	Journal = {The Lancet Planetary Health},
	Number = {9},
	Pages = {e353-e354},
	Title = {Climate change adaptation: no one size fits all},
	Volume = {1},
	Year = {2017}}

@article{adaptation2,
	Author = {Masna Rai and Susanne Breitner and Kathrin Wolf and Annette Peters and Alexandra Schneider and Kai Chen},
	Journal = {The Lancet Planetary Health},
	Number = {10},
	Pages = {e784-e792},
	Title = {Future temperature-related mortality considering physiological and socioeconomic adaptation: a modelling framework},
	Volume = {6},
	Year = {2022}}

@article{ferguson2020impact,
  title={Impact of non-pharmaceutical interventions (NPIs) to reduce COVID-19 mortality and healthcare demand},
  author={Ferguson, Neil M and Laydon, Daniel and Nedjati-Gilani, Gemma and Imai, Natsuko and Ainslie, Kylie and Baguelin, Marc and Bhatia, Sangeeta and Boonyasiri, Adhiratha and Cucunub{\'a}, Zulma and Cuomo-Dannenburg, Gina and others},
  year={2020},
  publisher={Imperial College COVID-19 Response Team London},
  journal={MRC Centre for Global Infectious Disease Analysis-COVID-19 reports [Internet]}
}

@article{Flaxman2020,
	Author = {Flaxman, Seth and Mishra, Swapnil and Gandy, Axel and Unwin, H. Juliette T. and Mellan, Thomas A. and Coupland, Helen and Whittaker, Charles and Zhu, Harrison and Berah, Tresnia and Eaton, Jeffrey W. and Monod, M{\'e}lodie and Perez-Guzman, Pablo N. and Schmit, Nora and Cilloni, Lucia and Ainslie, Kylie E. C. and Baguelin, Marc and Boonyasiri, Adhiratha and Boyd, Olivia and Cattarino, Lorenzo and Cooper, Laura V. and Cucunub{\'a}, Zulma and Cuomo-Dannenburg, Gina and Dighe, Amy and Djaafara, Bimandra and Dorigatti, Ilaria and van Elsland, Sabine L. and FitzJohn, Richard G. and Gaythorpe, Katy A. M. and Geidelberg, Lily and Grassly, Nicholas C. and Green, William D. and Hallett, Timothy and Hamlet, Arran and Hinsley, Wes and Jeffrey, Ben and Knock, Edward and Laydon, Daniel J. and Nedjati-Gilani, Gemma and Nouvellet, Pierre and Parag, Kris V. and Siveroni, Igor and Thompson, Hayley A. and Verity, Robert and Volz, Erik and Walters, Caroline E. and Wang, Haowei and Wang, Yuanrong and Watson, Oliver J. and Winskill, Peter and Xi, Xiaoyue and Walker, Patrick G. T. and Ghani, Azra C. and Donnelly, Christl A. and Riley, Steven and Vollmer, Michaela A. C. and Ferguson, Neil M. and Okell, Lucy C. and Bhatt, Samir and Imperial College COVID-19 Response Team},
	Journal = {Nature},
	Number = {7820},
	Pages = {257--261},
	Title = {Estimating the effects of non-pharmaceutical interventions on COVID-19 in Europe},
	Volume = {584},
	Year = {2020}}

@article{davies2020effects,
  title={Effects of non-pharmaceutical interventions on COVID-19 cases, deaths, and demand for hospital services in the UK: a modelling study},
  author={Davies, Nicholas G and Kucharski, Adam J and Eggo, Rosalind M and Gimma, Amy and Edmunds, W John and Jombart, Thibaut and O'Reilly, Kathleen and Endo, Akira and Hellewell, Joel and Nightingale, Emily S and others},
  journal={The Lancet Public Health},
  volume={5},
  number={7},
  pages={e375--e385},
  year={2020},
  publisher={Elsevier}
}

@article{prem2020effect,
  title={The effect of control strategies to reduce social mixing on outcomes of the COVID-19 epidemic in Wuhan, China: a modelling study},
  author={Prem, Kiesha and Liu, Yang and Russell, Timothy W and Kucharski, Adam J and Eggo, Rosalind M and Davies, Nicholas and Flasche, Stefan and Clifford, Samuel and Pearson, Carl AB and Munday, James D and others},
  journal={The Lancet Public Health},
  volume={5},
  number={5},
  pages={e261--e270},
  year={2020},
  publisher={Elsevier}
}

@article{di2020expected,
  title={Expected impact of school closure and telework to mitigate COVID-19 epidemic in France},
  author={Di Domenico, Laura and Pullano, Giulia and Pullano, G and Hens, Niel and Colizza, Vittoria},
  journal={EPIcx Lab},
  volume={15},
  year={2020}
}

@article{anderson2020will,
  title={How will country-based mitigation measures influence the course of the COVID-19 epidemic?},
  author={Anderson, Roy M and Heesterbeek, Hans and Klinkenberg, Don and Hollingsworth, T D{\'e}irdre},
  journal={The lancet},
  volume={395},
  number={10228},
  pages={931--934},
  year={2020},
  publisher={Elsevier}
}

@article{tuite2020mathematical,
  title={Mathematical modelling of COVID-19 transmission and mitigation strategies in the population of Ontario, Canada},
  author={Tuite, Ashleigh R and Fisman, David N and Greer, Amy L},
  journal={Cmaj},
  volume={192},
  number={19},
  pages={E497--E505},
  year={2020},
  publisher={Can Med Assoc}
}

@article{naturemedicine2021,
	Author = {Reiner, Robert C. and Barber, Ryan M. and Collins, James K. and Zheng, Peng and Adolph, Christopher and Albright, James and Antony, Catherine M. and Aravkin, Aleksandr Y. and Bachmeier, Steven D. and Bang-Jensen, Bree and Bannick, Marlena S. and Bloom, Sabina and Carter, Austin and Castro, Emma and Causey, Kate and Chakrabarti, Suman and Charlson, Fiona J. and Cogen, Rebecca M. and Combs, Emily and Dai, Xiaochen and Dangel, William James and Earl, Lucas and Ewald, Samuel B. and Ezalarab, Maha and Ferrari, Alize J. and Flaxman, Abraham and Frostad, Joseph Jon and Fullman, Nancy and Gakidou, Emmanuela and Gallagher, John and Glenn, Scott D. and Goosmann, Erik A. and He, Jiawei and Henry, Nathaniel J. and Hulland, Erin N. and Hurst, Benjamin and Johanns, Casey and Kendrick, Parkes J. and Khemani, Apurva and Larson, Samantha Leigh and Lazzar-Atwood, Alice and LeGrand, Kate E. and Lescinsky, Haley and Lindstrom, Akiaja and Linebarger, Emily and Lozano, Rafael and Ma, Rui and M{\aa}nsson, Johan and Magistro, Beatrice and Herrera, Ana M. Mantilla and Marczak, Laurie B. and Miller-Petrie, Molly K. and Mokdad, Ali H. and Morgan, Julia Deryn and Naik, Paulami and Odell, Christopher M. and O'Halloran, James K. and Osgood-Zimmerman, Aaron E. and Ostroff, Samuel M. and Pasovic, Maja and Penberthy, Louise and Phipps, Geoffrey and Pigott, David M. and Pollock, Ian and Ramshaw, Rebecca E. and Redford, Sofia Boston and Reinke, Grace and Rolfe, Sam and Santomauro, Damian Francesco and Shackleton, John R. and Shaw, David H. and Sheena, Brittney S. and Sholokhov, Aleksei and Sorensen, Reed J. D. and Sparks, Gianna and Spurlock, Emma Elizabeth and Subart, Michelle L. and Syailendrawati, Ruri and Torre, Anna E. and Troeger, Christopher E. and Vos, Theo and Watson, Alexandrea and Watson, Stefanie and Wiens, Kirsten E. and Woyczynski, Lauren and Xu, Liming and Zhang, Jize and Hay, Simon I. and Lim, Stephen S. and Murray, Christopher J. L. and IHME COVID-19 Forecasting Team},
	Journal = {Nature Medicine},
	Number = {1},
	Pages = {94--105},
	Title = {Modeling COVID-19 scenarios for the United States},
	Volume = {27},
	Year = {2021}}

@article{adam2020,
author = {Adam, David},
year = {2020},
month = {04},
pages = {},
title = {Special report: The simulations driving the world’s response to COVID-19},
volume = {580},
journal = {Nature},
doi = {10.1038/d41586-020-01003-6}
}

@article{BrooksPollock2021MappingSD,
  title={Mapping social distancing measures to the reproduction number for COVID-19},
  author={Ellen Brooks-Pollock and Jonathan M. Read and Angela R. McLean and Matt. J. Keeling and Leon Danon},
  journal={Philosophical Transactions of the Royal Society B: Biological Sciences},
  year={2021},
  volume={376}
}

@article{adam2020guide,
  title={A guide to R--the pandemic's misunderstood metric},
  author={Adam, David},
  journal={Nature},
  volume={583},
  number={7816},
  pages={346--349},
  year={2020},
  publisher={Nature Publishing Group}
}

@article{Inglesby2020PublicHM,
  title={Public Health Measures and the Reproduction Number of SARS-CoV-2.},
  author={Thomas V. Inglesby},
  journal={JAMA},
  year={2020}
}

@article{Rubin2020AssociationOS,
  title={Association of Social Distancing, Population Density, and Temperature With the Instantaneous Reproduction Number of SARS-CoV-2 in Counties Across the United States},
  author={David Rubin and Jing Huang and Brian T Fisher and Antonio Gasparrini and Vicky W Tam and Lihai Song and Xi Wang and Jason Kaufman and Kate Fitzpatrick and Arushi Jain and Heather M Griffis and Koby Crammer and Jeffrey Morris and Gregory E. Tasian},
  journal={JAMA Network Open},
  year={2020},
  volume={3}
}

@article{Chinazzi2020TheEO,
  title={The effect of travel restrictions on the spread of the 2019 novel coronavirus (COVID-19) outbreak},
  author={Matteo Chinazzi and Jessica T Davis and Marco Ajelli and Corrado Gioannini and Maria Litvinova and Stefano Merler and Ana Pastore y Piontti and Kunpeng Mu and Luca Rossi and Kaiyuan Sun and Cecile G. Viboud and Xinyue Xiong and Hongjie Yu and M. Elizabeth Halloran and Ira M. Longini and Alessandro Vespignani},
  journal={Science (New York, N.y.)},
  year={2020},
  volume={368},
  pages={395 - 400}
}

@article{Palguta2021DoEA,
  title={Do elections accelerate the COVID-19 pandemic?},
  author={J{\'a}n Palguta and Ren{\'e} Lev{\'i}nsk{\'y} and Samuel Skoda},
  journal={Journal of Population Economics},
  year={2021},
  volume={35},
  pages={197 - 240}
}

@article{VELIAS2022114538,
	Author = {Alina Velias and Sotiris Georganas and Sotiris Vandoros},
	Journal = {Social Science \& Medicine},
	Pages = {114538},
	Title = {COVID-19: Early evening curfews and mobility},
	Volume = {292},
	Year = {2022}}

@article{Zhang2022EvaluatingTI,
  title={Evaluating the impact of stay-at-home and quarantine measures on COVID-19 spread},
  author={Renquan Zhang and Yu Wang and Zheng Lv and Sen Pei},
  journal={BMC Infectious Diseases},
  year={2022},
  volume={22}
}

@article{Feltham2020,
		author = {Feltham, Eric and Forastiere, Laura and Alexander, Marcus and Christakis, Nicholas A.},
	date = {2023/10/01},
	doi = {10.1038/s41562-023-01654-1},
	id = {Feltham2023},
	isbn = {2397-3374},
	journal = {Nature Human Behaviour},
	number = {10},
	pages = {1708--1728},
	title = {Mass gatherings for political expression had no discernible association with the local course of the COVID-19 pandemic in the USA in 2020 and 2021},
	url = {https://doi.org/10.1038/s41562-023-01654-1},
	volume = {7},
	year = {2023}}

@article{rubin1980discussion,
  title={Discussion of" Randomization analysis of experimental data in the Fisher randomization test" by D. Basu},
  author={Rubin, Donald},
  journal={Journal of the American statistical association},
  volume={75},
  pages={591--593},
  year={1980}
}

@article{Hudgens2008TowardCI,
  title={Toward Causal Inference With Interference},
  author={Michael G. Hudgens and M. Elizabeth Halloran},
  journal={Journal of the American Statistical Association},
  year={2008},
  volume={103},
  pages={832 - 842}
}

@article{Robins2000MarginalSM,
  title={Marginal Structural Models and Causal Inference in Epidemiology},
  author={James M. Robins and Miguel A. Hern{\'a}n and Babette A. Brumback},
  journal={Epidemiology},
  year={2000},
  volume={11},
  pages={550-560}
}

@article{Kim2021MatchingMF,
  title={Matching Methods for Causal Inference with Time‐Series Cross‐Sectional Data},
  author={In Song Kim and Adam Rauh and Erik H. Wang and Kosuke Imai},
  journal={American Journal of Political Science},
  year={2021}
}

@article{perrakis2014controlling,
  title={Controlling for seasonal patterns and time varying confounders in time-series epidemiological models: a simulation study},
  author={Perrakis, Konstantinos and Gryparis, Alexandros and Schwartz, Joel and Tertre, Alain Le and Katsouyanni, Klea and Forastiere, Francesco and Stafoggia, Massimo and Samoli, Evangelia},
  journal={Statistics in Medicine},
  volume={33},
  number={28},
  pages={4904--4918},
  year={2014},
  publisher={Wiley Online Library}
}

@article{italycolors,
	Author = {Bonifazi, Gianluca and Lista, Luca and Menasce, Dario and Mezzetto, Mauro and Pedrini, Daniele and Spighi, Roberto and Zoccoli, Antonio},
	Journal = {The European Physical Journal Plus},
	Number = {12},
	Pages = {1208},
	Title = {Study on the effects of the restrictive measures for containment of the COVID-19 pandemic on the reproduction number {\$}{\$}R{\_}t{\$}{\$}in Italian regions},
	Volume = {136},
	Year = {2021}}

@article{Pelagatti2021AssessingTE,
  title={Assessing the effectiveness of the Italian risk-zones policy during the second wave of COVID-19},
  author={Matteo Pelagatti and Paolo Maranzano},
  journal={Health Policy (Amsterdam, Netherlands)},
  year={2021},
  volume={125},
  pages={1188 - 1199}
}

@article{kim_gu_yu_wang_wang_2021,
	Author = {Myungjin Kim and Zhiling Gu and Shan Yu and Guannan Wang and Li Wang},
	Journal = {Journal of Data Science},
	Number = {2},
	Pages = {219--242},
	Title = {Methods, Challenges, and Practical Issues of COVID-19 Projection: A Data Science Perspective},
	Volume = {19},
	Year = {2021}}

@article{wronguseful2020,
	Author = {Holmdahl, Inga and Buckee, Caroline},
	Journal = {New England Journal of Medicine},
	Number = {4},
	Pages = {303-305},
	Title = {Wrong but Useful --- What Covid-19 Epidemiologic Models Can and Cannot Tell Us},
	Volume = {383},
	Year = {2020}}

@article{Di2017AssociationOS,
  title={Association of Short-term Exposure to Air Pollution With Mortality in Older Adults},
  author={Qian Di and Lingzhen Dai and Yun Wang and Antonella Zanobetti and Christine Choirat and Joel D. Schwartz and Francesca Dominici},
  journal={JAMA},
  year={2017},
  volume={318},
  pages={2446-2456}
}

@article{airpollution,
	Author = {Cynthia A. Garcia and Poh-Sin Yap and Hye-Youn Park and Barbara L. Weller},
	Journal = {International Journal of Environmental Health Research},
	Number = {2},
	Pages = {145-157},
	Title = {Association of long-term PM2.5 exposure with mortality using different air pollution exposure models: impacts in rural and urban California},
	Volume = {26},
	Year = {2016}}

@article{schwartz2017air,
  title={Air pollution and mortality in the Medicare population},
  author={Schwartz, JD},
  journal={New England},
  year={2017}
}

@article{schwartz2017estimating,
  title={Estimating causal effects of local air pollution on daily deaths: effect of low levels},
  author={Schwartz, Joel and Bind, Marie-Abele and Koutrakis, Petros},
  journal={Environmental health perspectives},
  volume={125},
  number={1},
  pages={23--29},
  year={2017},
  publisher={National Institute of Environmental Health Sciences}
}

@article{burnett2018global,
  title={Global estimates of mortality associated with long-term exposure to outdoor fine particulate matter},
  author={Burnett, Richard and Chen, Hong and Szyszkowicz, Mieczys{\l}aw and Fann, Neal and Hubbell, Bryan and Pope Iii, C Arden and Apte, Joshua S and Brauer, Michael and Cohen, Aaron and Weichenthal, Scott and others},
  journal={Proceedings of the National Academy of Sciences},
  volume={115},
  number={38},
  pages={9592--9597},
  year={2018},
  publisher={National Acad Sciences}
}

@article{wei2020causal,
  title={Causal effects of air pollution on mortality rate in Massachusetts},
  author={Wei, Yaguang and Wang, Yan and Wu, Xiao and Di, Qian and Shi, Liuhua and Koutrakis, Petros and Zanobetti, Antonella and Dominici, Francesca and Schwartz, Joel D},
  journal={American journal of epidemiology},
  volume={189},
  number={11},
  pages={1316--1323},
  year={2020},
  publisher={Oxford University Press}
}

@article{ERF_AOAS,
 ISSN = {19326157},
 URL = {http://www.jstor.org/stable/30245134},
 author = {Gavin Shaddick and Duncan Lee and James V. Zidek and Ruth Salway},
 journal = {The Annals of Applied Statistics},
 number = {4},
 pages = {1249--1270},
 publisher = {Institute of Mathematical Statistics},
 title = {Estimating Exposure Response Functions Using Ambient Pollution Concentrations},
 urldate = {2022-11-20},
 volume = {2},
 year = {2008}
}

@article{R0estimate,
	Author = {Alimohamadi Yousef,Taghdir Maryam,Sepandi Mojtaba},
	Journal = {J Prev Med Public Health},
	Number = {3},
	Pages = {151-157},
	Title = {Estimate of the Basic Reproduction Number for COVID-19: A Systematic Review and Meta-analysis},
	Volume = {53},
	Year = {2020}}

@article{FatalityRate,
	Author = {Gideon Meyerowitz-Katz and Lea Merone},
	Journal = {International Journal of Infectious Diseases},
	Pages = {138-148},
	Title = {A systematic review and meta-analysis of published research data on COVID-19 infection fatality rates},
	Volume = {101},
	Year = {2020}}

@article{kendrovski,
	Author = {Vladimir Kendrovski and Michela Baccini and Gerardo Sanchez Martinez and Tanja Wolf and Elizabet Paunovic and Bettina Menne},
	Journal = {Int J Environ Res Public Health},
	Pages = {729},
	Title = {Quantifying Projected Heat Mortality Impacts under 21st-Century Warming Conditions for Selected European Countries},
	Volume = {14},
	Year = {2017}}

@article{sanchez,
	Author = {Gerardo Sanchez Martinez and Michela Baccini and Koen De Ridder and Hans Hooyberghs and Wouter Lefebvre and Vladimir Kendrovski and Kristen Scott and Margarita Spasenovska},
	Journal = {BMC Public Health},
	Pages = {407},
	Title = {Projected heat-related mortality under climate change in the metropolitan area of Skopje},
	Volume = {16},
	Year = {2016}}

@article{odriscol21,
	Author = {M O'Driscol and C Harry and C Donnelli and A Cori and I Dorigatti},
 Journal = {Clin Infect Dis},
 Pages = {e215–e223},
 Title= {A comparative analysis of statistical methods to estimate the reproduction number in emerging epidemics with implications for the current COVID-19 pandemic},
	Volume = {37},
	Year = {2021}}

@article{Li2022BayesianCI,
  title={Bayesian causal inference: a critical review},
  author={Fan Li and Peng Ding and Fabrizia Mealli},
  journal={Philosophical Transactions of the Royal Society A},
  year={2022},
  volume={381},
}

@article{Bartlett,
author = {Bartlett, Jonathan and Granger, Emily and Keogh, Ruth and Zwet, Erik and Daniel, Rhian},
year = {2025},
month = {03},
pages = {9622802251316971},
title = {G-formula with multiple imputation for causal inference with incomplete data},
volume = {34},
journal = {Statistical methods in medical research},
doi = {10.1177/09622802251316971}
}

@article{Zigler2020BipartiteIA,
  title={Bipartite interference and air pollution transport: estimating health effects of power plant interventions.},
  author={Corwin M. Zigler and Vera Liu and Fabrizia Mealli and Laura Forastiere},
  journal={Biostatistics},
  year={2020},
  volume={26 1},
}

@article{Feltham2020MassGF,
  title={Mass gatherings for political expression had no discernible association with the local course of the COVID-19 pandemic in the USA in 2020 and 2021},
  author={Eric Feltham and Laura Forastiere and Marcus Alexander and Nicholas A. Christakis},
  journal={Nature Human Behaviour},
  year={2020},
  volume={7},
  pages={1708 - 1728},
}

\newpage
\appendix

\section*{Appendix}

\section{Impact of a hypothetical intervention modifying the exposure distribution} 
\label{sec:hyp}

We now extend the framework developed in Section \ref{sec:intervention} to the assessment of a hypothetical intervention that would modify the exposure level, either by setting it to a specific value or by shifting the exposure
distribution below a certain threshold. 

In our COVID-19 example, the exposure variable $S_{it}$ can be seen as the number of potentially contagious interactions between infectious and susceptible people in area $i$ at time $t$. 
Thus, $S_{it}$ can be seen as proportional to the effective reproduction number $R_e$
\footnote{The effective reproduction number $R_e$, 
is the expected number of new infections caused at a specific time by an infectious individual in a population \cite{odriscol21}.} 
and the number of circulating infected individuals in the area. For the sake of simplicity, we can consider the exposure variable $S_{it}$ as being equal to the effective reproduction number $R_e$.

We can think of a hypothetical public health policy that would control the spread of COVID-19 by keeping the $R_e$ below 1 for one month.
We do not specify the specific measures that will be put in place to achieve this goal. It is instead a hypothetical intervention on the distribution of the $R_e$.
An intervention of this kind is usually evaluated by first estimating the exposure-response function (mortality rate-$R_e$). Then, we can evaluate the effect of the hypothetical intervention by predicting the expected mortality rate under a shift in the distribution of the exposure.

\subsection{Potential Outcomes}
Under the potential outcome framework, we denote as $Y_{it}(\vs)$ the potential outcome of unit $it$ under an exposure matrix $\vS$ equal to $\vs$. As before, we make the no-interference assumption
between areas and we assume that the outcome at time $t$ can at most depend on the exposures over a period from $t-B-K$ to $t-B$, where $0\leq B<T$ and $ 0 \leq K<t-B$. 
Letting $\overline{\vS}_{it}^{M,L}=\{S_{i,(t-M-
L)},S_{i,(t-M-
L+1)},\dots,S_{i,(t-M)}\}$ be the exposure history of $i$ in the time window $[t-M-L, t-M]$, 
$\overline{\vS}_{it}^{B,K}=[S_{i(t-B)}, S_{i(t-B-1)}, \ldots, S_{i(t-B-K)}]$ represents the lagged exposure vector in the carry-over window for the outcome at time $t$.\footnote{
 We could also introduce a function $g(\cdot)$, which summarizes the exposure vector $\overline{\vS}_{it}^{B,K}$ and maps it into a continuous (possibly multivariate) exposure variable $\mathbf{G}_{it}=g(\overline{\vS}_{it}^{B,K}) \in \mathcal{R}^P$, and assume a SUTVA-like assumption that allows indexing the potential outcomes only by this variable, i.e., $Y_{it}(G_i)$. Note that $g(\cdot)$ could also be the identity function. However, for clarity of exposition, in this section we prefer to index the potential outcomes by the whole exposure vector, without introducing a summarizing function.
 }
 Formally, we make the following SUTVA-type assumption:
 

 \begin{assumption}[SUTVA on the exposure]
\label{ass:SUTVA2}
$ \forall \, \vS,\vS'\in \mathcal{R}^{I\times T}$ such that $\overline{\vS}_{it}^{B,K}=\overline{\vS}_{it}^{B,K'}$, the following equality holds:
\[ \qquad Y_{it}(\vS) = Y_{it}(\vS')\]
\end{assumption}

\noindent Under this assumption we can index the potential outcome only by the lagged exposure vector $\overline{\vS}_{it}^{B,K}$, i.e., $Y_{it}(\overline{\vS}_{it}^{B,K})$. Throughout, we will refer to $Y_{it}(\vs)$,  as the potential outcome that we would observe at unit $it$ if the lagged exposure vector $\overline{\vS}_{it}^{B,K}$ were set equal to $\vs$.
In the literature, the function $Y_{it}(\vs)$ is referred to as exposure-response function (ERF) \citep[e.g.,][]{forastiere_envhp2020}.

\subsection{Actual impact on the observed sample}
Oftentimes the exposure-response function $Y_{it}(\vs)$ is the actual object of interest, rather than the effect of a specific intervention. In fact, 
ERFs are essential to estimate health effects of exposures: they describe quantitatively the extent to which a specific health outcome changes when exposure to the specified agent varies by a given amount.   
For example, there is an extensive literature on the harmful effects of air pollution, and several parametric and semi-parametric approaches have been used to estimate the exposure-response curve between long-term or short-term exposure to ambient air pollution concentrations
and health outcomes \citep{ERF_AOAS, airpollution, schwartz2017air, schwartz2017estimating, Di2017AssociationOS, burnett2018global, wei2020causal}.
The exposure-response relationship has important regulatory implications. In fact, quantifying the relationship between exposure and the resulting effects on health makes it possible to assess the effectiveness of hypothetical measures that reduce the exposure \citep{forastiere_envhp2020}.

In the case of COVID-19, 
the ERF is usually represented by epidemiological models. The different measures that have been implemented to control the epidemic or that can be hypothesized act on the transmission rate in order to reduce the spread from infected to susceptible individuals. 

Let $f_{\overline{\vS}_{it}^{B,K}}(\vs)$ be the observed distribution of the exposure $\overline{\vS}_{it}^{B,K}$ for each day $t$ at location $i$.
Let us consider a hypothetical intervention that sets the distribution of the exposure for each day $t$ at location $i$ to $p^{\star}_{\overline{\vS}_{it}^{B,K}}(\vs)$. For the sake of notational simplicity, we denote the observed distribution by $f_{it}(\vs)$ and the hypothetical distribution by $p^{\star}_{it}(\vs)$.
We define the average impact of the hypothetical intervention $p^{\star}_{it}(\vs)$, denoted by AEE (Average Exposure Effect), as the following quantity:
\begin{equation}
\label{eq:AEE}
\text{AEE}=
\frac{1}{N_1^{obs}}\sum_{it\in{U_1^{obs}}} \Big(\int\E\Big[Y_{it}(\vs)|  \vX_{i0}\Big]f_{it}(\vs)d\vs- \int\E\Big[Y_{it}(\vs)|  \vX_{i0}\Big]p^{\star}_{it}(\vs)d\vs \Big),
\end{equation}
where $U_1^{obs}$ is a sub-sample of $U^{obs}$ and $N_1^{obs}$ is the number of units in $U_1^{obs}$.
For example, $U_1^{obs}$ could be the subset of units $it$ where the exposure $\vs$ falls below a certain threshold $s^{\star}$. 
Then, the AEE (Equation \ref{eq:AEE}) can be seen as the average outcome events that could have been prevented in the sample $U_1^{obs}$ had we implemented the hypothetical intervention. 
With regard to the hypothetical intervention, we could think of a counterfactual scenario where we set all the exposure values during the cross-over period equal to the critical threshold i.e.,
$p^{\star}_{it}(\vs)=\delta(\vs^{\star})$. A more realistic intervention is the one where we shift the distribution to values below the threshold, i.e., $p^{\star}_{it}(\vs)=f_{it}(\vs|\vs<\vs^{\star})$.
For example, in the case of COVID-19 we could think of public policies that would be able to reduce the effective reproductive number $R_e$ to values below 1.
Note that the notation in Equation \eqref{eq:AEE}
implies that we are assuming that the
hypothetical intervention has the only effect of shifting the distribution to the targeted one ($p^{\star}_{it}(\vs)$), without resulting in any other effect. 
For now, let us assume the the hypothetical distribution $p^{\star}_{it}(\vs)$ does not depend on any unit's characteristics or previous exposures.


\subsubsection{Identification under Sequential Ignorability}


Here, we extend the sequential ignorability assumption \ref{ass:Ign}
to the case where potential outcomes are indexed by the lagged exposure treatment vector $\vs$:


\begin{assumption}[Sequential Randomization - no unmeasured confounders]
\label{ass:Ign_exp}
\begin{align}
\hspace{1cm}
Y_{it}(\vs) \ind S_{i\ell} | \overline{\vS}_{i\ell}^{1, \ell-1-H}=\overline{\vs}^{\ell-H+1}, 
\overline{\mathbf{V}}_{i\ell}, \overline{\mathbf{R}}_{it},
U\subseteq U^{obs} \quad \forall \, d, \forall \, \ell\in [t-B-K, t-B]
\end{align}
\end{assumption}
\noindent where 
$\overline{\vS}_{i\ell}^{1, \ell-1-H}$ is the exposure history from $t-B-K$ until $\ell-1$, 
$\overline{\vs}^{\ell-H+1}$ is the sub-vector of the exposure realization $\vs$,
and $U$ is a subset of units contained in $U^{obs}$ where Assumption \ref{ass:Ign_exp} holds.
This sequential ignorability assumption states that for each unit in $U\subseteq U^{obs}$ the distribution of the potential outcomes at time $t$ is independent of the exposure $S_{i\ell}$  in the cross-over time window $[t-B-K, t-B]$, conditional on the exposure history until time $[t-B-K]$, the covariate and the outcome history before and during the cross-over period.

When evaluating exposure-response functions, oftentimes the exposure has a cross-over effect and the outcome depends on the exposure over multiple days, i.e., $K>0$. In this case, the estimation of the exposure-response function could be affected by time-dependent confounders.
When they are not present -- i.e., 
when previous outcomes do not affect future exposures, and when covariates in $\vX_{it}$ are fixed or time-varying covariates are not affected by previous exposures --
under Assumption \ref{ass:Ign_exp}, 
the conditional exposure-response function is identified by:
\begin{equation}
\begin{aligned}
\label{eq:adj_exp}
    \E\Big[Y_{it}(\vs)| \vX_{i0}\Big]=
     \sum_{\overline{\mathbf{c}}}\E\Big[Y^{obs}_{it}|\overline{S}_{it}^{B,K}=\vs, \overline{\mathbf{C}}_{it}=\overline{\mathbf{c}}, \vX_{i0}\Big]    f_{\overline{\mathbf{C}}_{it}}(\overline{\mathbf{c}}|  \vX_{i0}) 
    \end{aligned}
\end{equation}
with $\overline{\mathbf{C}}_{it}=[\overline{\mathbf{R}}_{it}, \overline{\mathbf{V}}_{it}]$. That is, we 
adjust for all type of confounders, including baseline covariates as well as the evolution of time-varying covariates pre-treatment 
and during the cross-over window.
Thus, the average exposure effect on the exposed (AEE) is identified from the observed data as follows:
\begin{equation}
\label{eq:aee_id}
\text{AEE}=\frac{1}{N_1^{obs}}\sum_{it\in{U_1^{obs}}}\Big(\E\Big[Y_{it}^{obs}|\vX_{i0}\Big]-\int \E\Big[Y_{it}(\vs)|  \vX_{i0}\Big]p^{\star}_{it}(\vs)d\vs\Big),
\end{equation}
with $\E[Y_{it}(\vs)|  \vX_{i0}]$ identified by Equation \eqref{eq:adj_exp}. When the exposure is the number of potentially contagious interactions between infectious and susceptible people and the outcome is the mortality rate, time-independent confounders can include population and environmental characteristics. For example, an area with a high average age might show limited interactions between individuals 
and yet a high mortality rate. 

On the other hand, it is possible to conceive time-dependent confounders affected by previous exposures and affecting future ones. For instance, 
the adoption of protective behaviors (e.g., mask wearing, test and treatment seeking behaviors), that are likely to reduce exposure to the virus as well as the risk of death, might be prompted by an increase in COVID-19 cases. Similarly, the worsening of the epidemic could compromise the contact tracing capacity, and its effectiveness in identifying cases, letting them circulate and infect more people. As a consequence, the effective reproductive number $R_e$ could get higher (although the downward bias in the estimated $R_e$ might be more prominent), while the capacity of monitoring the onset of severe symptoms in a timely matter will also be compromised, leading to an increased mortality rate. In addition, a surge in cases can lead the governments to impose stricter restrictions as well as to increase the provision of medical equipment, affecting both the future $R_e$ and mortality rate.

When time-dependent confounders are present, the identification of the conditional exposure-response function  must account for the feedback loops between previous exposure levels, intermediate outcomes, and time-varying covariates. Under the sequential ignorability {Assumption \ref{ass:Ign_exp}}, the expected potential outcome $\E[Y_{it}(\vs) \mid \vX_{i0}]$ is identified by the following $g$-formula:

\begin{equation}
\begin{aligned}
\label{eq:gformula_exp_corrected}
    \E\Big[Y_{it}(\vs) \mid \vX_{i0}\Big] = & \sum_{\overline{\mathbf{r}}} \Bigg[ \sum_{\overline{\vx}, \overline{\vy}} \E\Big[Y^{obs}_{it} \mid \overline{\vS}_{it}^{B,K} = \vs, \overline{\mathbf{V}}_{it} = [\overline{\vx}, \overline{\vy}], \overline{\mathbf{R}}_{it} = \overline{\mathbf{r}}, \vX_{i0}\Big] \\
    & \times \prod_{\ell=H}^{t-B} p(y_{i\ell} \mid \overline{\vS}_{i\ell}^{B, \ell-B-H} = \overline{\vs}^{\ell-B}, \overline{\vY}_{i\ell}^{1, \ell-1-H} = \overline{\vy}^{\ell-1}, \overline{\vX}_{i\ell}^{0, \ell-1-H} = \overline{\vx}^{\ell}, \overline{\mathbf{R}}_{it} = \overline{\mathbf{r}}, \vX_{i0}) \\
    & \times f(\vx_{i\ell} \mid \overline{\vS}_{i\ell}^{1, \ell-1-H} = \overline{\vs}^{\ell-1}, \overline{\vY}_{i\ell}^{1, \ell-1-H} = \overline{\vy}^{\ell-1}, \overline{\vX}_{i\ell}^{1, \ell-1-H} = \overline{\vx}^{\ell-1}, \overline{\mathbf{R}}_{it} = \overline{\mathbf{r}}, \vX_{i0}) \Bigg] \\
    & \times f_{\overline{\mathbf{R}}_{it}}(\overline{\mathbf{r}} \mid \vX_{i0})
\end{aligned}
\end{equation}
where $H = t-B-K$ is the start of the carry-over window, and $\overline{\vs}^k = \{s_H, \dots, s_k\}$ represents the sub-vector of the hypothetical exposure realization $\vs$. 
Given Equation \eqref{eq:gformula_exp_corrected}, we identify the conditional exposure-response function $\E[Y_{it}(\vs) \mid \vX_{i0}]$ by performing a two-stage marginalization:
i) we first average the observed outcome of those with the lagged exposure vector equal to $\vs$ units over the joint distribution of the evolution of outcomes ($\overline{\vy}$) and covariates ($\overline{\vx}$) during the carry-over period under the lagged exposures given in $\vs$ -- calculated through the product of conditional densities --; ii) we then average this result over the distribution of the pre-treatment history $\overline{\mathbf{R}}_{it}$ observed in the population.
Finally, the AEE is then identified by substituting this result into Equation \eqref{eq:aee_id}.

\subsection{Predicting the impact on future observations}
\label{sec:pred_exp}
\subsubsection{Causal Estimands}

Now, let assume that we wish to forecast the impact of the hypothetical intervention on the observed population $\mathcal{I}$ but in a future period.
Instead of asking the question of what would have happened had we been able to shift the distribution of the exposure to $p^{\star}_{it}(\vs)$, we wish to pose the question of what would happen in a future period if we could implement an intervention able to yield shift the exposure with an hypothetical distribution  $p^{\star}_{it}(\vs)$.
Policy-makers usually rely on the observed evidence of the harmful effect of a biological, chemical or physical agent to propose a hypothetical intervention that could reduce the population exposure. In this case, they are typically interested in assessing its future impact, rather than wonder what would have happened in the past had they been able to implement an intervention that was not even conceived.
For example, after the first wave of COVID-19, we had a good estimate of the basic reproductive number and case fatality rate for SARS-CoV-2 \citep{R0estimate, FatalityRate}. This evidence led us to thinking that if we could keep the effective reproductive number to a value lower than 1, we could predict the number of prevented deaths. 

As in Section \ref{sec:pred}, let 
$\mathcal{T}_Z^F=[T+F,\ldots,T+F+T_Z^F]$,  with $F \geq 1, T_Z^F\geq 0$, be the future time window where we wish to apply the hypothetical intervention, and let $\mathcal{T}^F=[T+F+L, \ldots, T+F+T^F]$, with $B \leq L\leq B+K$ and $L \leq T^F\leq T_Z^F+K$
be the future time window
where we wish to forecast the causal effect. 
Let us assume that we still want to predict the effect in the same population $\mathcal{I}$ where we observed a certain variation in the exposure and the outcome variable during the past time window $\mathcal{T}$. 

We can define the average causal effect of the hypothetical intervention on the exposure in the future as the following quantity:
\begin{equation}
\label{eq: AEEF}
 \text{AEE}_F=
  \frac{1}{|\mathcal{I}_F|\times|\mathcal{T}^F|}\sum_{i\in \mathcal{I}_F}\sum_{t\in\mathcal{T}^F} 
\Big(\int\E\Big[Y_{it}(\vs)|  \vX_{i0}\Big]f_{it}(\vs\mid \mathbf{X}_{i0})d\vs- \int\E\Big[Y_{it}(\vs)|  \vX_{i0}\Big]p^{\star}_{it}(\vs)d\vs \Big),
\end{equation}
where $\mathcal{I}^F\subseteq\mathcal{I}$,
$f(\vs\mid \mathbf{X}_{i0})$ is the realized conditional exposure distribution on $U_1^F$,
while $p^{\star}_{it}(\vs)$ is the hypothetical distribution. 
Thus, our future sample of interest is given by $U_1^{F}=\{it: i\in \mathcal{I}_F, t\in \mathcal{T}^F\}$. 
The $AEE_F$ compares the two counterfactual scenarios where the lagged exposure vector  
is what would be observed under its natural evolution 
$f_{it}(\vs\mid \mathbf{X}_{i0})$ or is drawn from the hypothetical distribution $p^{\star}_{it}(\vs)$ on all units $it\in U_1^F$.
Thus, the $AEE_F$ represents the effect of implementing the hypothetical intervention as opposed to letting the 
population be subject to the natural course of the exposure, without such intervention. 
For instance, we may wish to assess the gain (in terms of prevented deaths) of implementing restrictions able to alter the course of the epidemic by reducing the $R_e$ to values below 1 or strictly equal to 1 for all the days in $U_1^F$ where $R_e$ would have been above 1. 

\subsubsection{Identification
under Temporal Transportability}
In the average causal effect of the hypothetical intervention in Equation \eqref{eq: AEEF},
the entire exposure-dose response function $Y_{it}(\vs)$ is missing , and the lagged exposure $\overline{\vS}_{it}^{B,K}$ that would be observed under its natural evolution 
$f_{it}(\vs)$ without any intervention is not observed, and the distribution $f_{it}(\vs\mid \mathbf{X}_{i0})$ is not known.
The hypothetical distribution $p^{\star}_{it}(\vs)$ might be pre-specified or might be a shifted version of the distribution $p(\vs)$.

The simplest solution is to assume that  the exposure-response function, estimated on an observed sample, remains constant over time on the same population. That is, the distribution of the potential outcomes under a lagged exposure vector $\vs$, for any $\vs$, does not change with time.
For example, in the case of COVID-19, we could assume a constant case fatality rate.
If the exposure in the future time window $\mathcal{T}_Z^F$ could be assumed to be the same as in the past time window, either constant over time or following the same trend,
$AEE_F$ could be simply evaluated in the past sample $U_1^{obs}$.
Nevertheless, the exposure usually evolves overtime, and 
exposure-response functions are usually heterogeneous, depending on 
unit's characteristics that may vary over time.

Let us first focus on the problem of heterogeneous exposure-response functions.
If the modifiers were fixed and only depended on the location, then the distribution of the exposure-response function in the future sample would be the same as in the past sample. On the contrary, in the presence of time-varying modifiers, a different distribution of these modifiers between the past and future samples would result in a different distribution of the exposure-response function. However, one may assume that if all the time-varying effect modifiers were included in X and Y, the conditional distribution is time-invariant.
This is formalized in the following temporal transportability assumption similar to Assumption \ref{ass:transp}. 

\begin{assumption}[Temporal Transportability of the Exposure-Response Function] 
\label{ass:transp_exp}
\textit{Let $A_i$ be a random selection variable for unit $i$ taking values in a temporal domain $\mathcal{D}$. Define the path-dependent history $\mathcal{H}_{it}(\vs) = \{ \overline{\mathbf{S}}_{it}^{B, K} = \vs, \overline{\mathbf{X}}_{it}^{B, L_x}, \overline{\mathbf{Y}}_{it}^{B, L_y}, \mathbf{X}_{i0} \}$. Define $Y_i(\vs)$ as the potential outcome and $\mathcal{H}_i(\vs)$ as the history evaluated at the selected time $A_i$:}
\begin{equation*}
    Y_i(\vs) = \sum_{t \in \mathcal{D}} Y_{it}(\vs) \cdot \mathbb{I}(A_i = t); \qquad \mathcal{H}_i(\vs) = \sum_{t \in \mathcal{D}} \mathcal{H}_{it}(\vs) \cdot \mathbb{I}(A_i = t).
\end{equation*}
\textit{For all units $i \in \mathcal{I}_F$, the selection variable $A_i$ is independent of the potential outcomes given the path-dependent history:}
\begin{equation}
    Y_i(\vs) \ind A_i \mid \mathcal{H}_i(\vs)
\end{equation}
\end{assumption}

\noindent In words, Assumption \ref{ass:transp_exp} states that the distribution of the exposure-response function $Y_{it}(\vs)$ is the same across the temporal domain $\mathcal{D}$, provided units share the same baseline covariates $\vX_{i0}$ and the same evolution of outcomes $\overline{\mathbf{Y}}_{it}^{B, L_y}$ and covariates $\overline{\vX}_{it}^{B,L_x}$ both before and during the exposure window under the realization $\vs$. 
By conditioning on $\mathcal{H}_i(\vs)$, we ensure that all time-varying modifiers—including those in the pre-treatment window and those affected by the exposure during the carry-over period—are accounted for. This rules out the presence of unobserved time-varying modifiers that differ between the observed past and the future period. Consequently, if $\mathcal{D}$ spans across future and past time points, the exposure-response function $Y_{it}(\vs)$ for a future time point can be identified by the conditional distribution estimated from past time points with equivalent history.

Another essential component in identifying the $AEE_F$ (Equation \eqref{eq: AEEF}) is the distribution of the exposure in the future time window, $f_{\overline{\mathbf{S}}_{it}^{B,K}}(\vs)$ for $it \in U_1^F$. An approach to identify this distribution is to assume that the evolution of the exposure follows the same structural dynamics in the future as it did in the past. For example, in our COVID-19 application, this corresponds to the assumption that, absent a new intervention, the epidemic follows its "natural course" as defined by the underlying transmission laws. We formalize this as follows:

\begin{assumption}[Temporal Transportability of the Exposure Distribution] 
\label{ass:transp_exp_dis}
\textit{Let $A_i$ be a random selection variable for unit $i$ taking values in a temporal domain $\mathcal{D}$. Define the exposure history $\mathcal{H}_{i\ell}^S = \{ \overline{\mathbf{S}}_{i\ell}^{1,L_{ss}-1}, \overline{\mathbf{X}}_{i\ell}^{0,L_{xs}}, \overline{\mathbf{Y}}_{i\ell}^{1, L_{ys}-1} \}$. Let $S_i$ and $\mathcal{H}_i^S$ be the exposure and history evaluated at the selected time $A_i$:}
\begin{equation*}
    S_i = \sum_{\ell \in \mathcal{D}} S_{i\ell} \cdot \mathbb{I}(A_i = \ell); \qquad \mathcal{H}_i^S = \sum_{\ell \in \mathcal{D}} \mathcal{H}_{i\ell}^S \cdot \mathbb{I}(A_i = \ell).
\end{equation*}
\textit{For every $i \in \mathcal{I}_F$, the selection variable $A_i$ is independent of the exposure given the history:}
\begin{equation}
    S_i \ind A_i \mid \mathcal{H}_i^S
\end{equation}
\end{assumption}

\noindent Assumption \ref{ass:transp_exp_dis} states that the conditional distribution of the exposure $S_{i\ell}$ is invariant across the temporal domain $\mathcal{D}$. This implies that the stochastic process for the exposure is stationary once we condition on previous exposures, covariates, and outcomes. 

By applying this assumption over the interval $\mathcal{T}_{cond}(t) = [t-B-K, t-B]$, we can identify the natural lagged exposure distribution for future units in $U_1^F$. This allows us to predict $\overline{\mathbf{S}}_{it}^{B,K}$ and calculate the first term of the $AEE_F$. This stability of the exposure mechanism is the theoretical cornerstone for using epidemiological models to forecast the "no-intervention" course of an epidemic based on historical transmission dynamics \citep[e.g.,]{Flaxman2020, ferguson2020impact, davies2020effects}.

As discussed in Section \ref{sec:pred}, in general, the time-varying modifiers $\overline{\vY}_{it}^{B, L_y}$ and $\overline{\vX}_{it}^{B, L_x}$ are not observed for future time points. Nevertheless, under a temporal transportability condition one could predict their evolution from the past. In particular, we must assume the following conditions.

\begin{assumption}[Temporal Transportability of Time-Varying Modifiers] 
\label{ass:conf_transp_exp}
\textit{Let $A_i$ be a random selection variable for unit i taking values in a temporal domain $\mathcal{D}$. For a fixed exposure path $\vs$, define the time-varying transition histories as:
 $\mathcal{H}_{i\ell}^X(\vs) = \{ \overline{\mathbf{S}}_{i\ell}^{1, L_s} = \overline{\vs}_{\ell}^{1, L_s}, \overline{\mathbf{X}}_{i\ell}^{1, L_{xx}}, \overline{\mathbf{Y}}_{i\ell}^{1, L_{xy}}, \mathbf{X}_{i0} \}$
    and $\mathcal{H}_{i\ell}^Y(\vs) = \{ \overline{\mathbf{S}}_{i\ell}^{B, K} = \overline{\vs}_{\ell}^{B, K}, \overline{\mathbf{X}}_{i\ell}^{B, L_x}, \overline{\mathbf{Y}}_{i\ell}^{B, L_y}, \mathbf{X}_{i0} \}$, where $\overline{\vs}_{\ell}^{1, L_s}$ and $\overline{\vs}_{\ell}^{B, K}$ denote the specific sub-vectors of the realization path $\vs$ corresponding to the specified lags. Define the selected variables and histories as:}
\begin{equation*}
    X_i = \sum_{\ell \in \mathcal{D}} X_{i\ell} \cdot \mathbb{I}(A_i = \ell); \quad Y_i = \sum_{\ell \in \mathcal{D}} Y_{i\ell} \cdot \mathbb{I}(A_i = \ell);
\end{equation*}
\begin{equation*}
    \mathcal{H}_i^X(\vs) = \sum_{\ell \in \mathcal{D}} \mathcal{H}_{i\ell}^X(\vs) \cdot \mathbb{I}(A_i = \ell); \quad \mathcal{H}_i^Y(\vs) = \sum_{\ell \in \mathcal{D}} \mathcal{H}_{i\ell}^Y(\vs) \cdot \mathbb{I}(A_i = \ell).
\end{equation*}
\textit{For every $i \in \mathcal{I}_F$, the following transition stability conditions hold:}
\begin{enumerate}[label={\textbf{Part~\Alph*}}, ref={\theassumption.\Alph*}, leftmargin=4\parindent]
\item \label{ass_exp-A} {(Confounders):} $\qquad\qquad X_i \ind A_i \mid \mathcal{H}_i^X(\vs)$
\item \label{ass_exp-B} {(Lagged Outcomes):} $\qquad Y_i \ind A_i \mid \mathcal{H}_i^Y(\vs)$
\end{enumerate}
\end{assumption}

\noindent Assumption \ref{ass:conf_transp_exp} captures the temporal stability of the effect modifiers  across the domain $\mathcal{D}$. 
Part \ref{ass_exp-A} states that the conditional distribution of time-varying covariates at any time point $\ell$ is domain-invariant, provided we condition on the past history and the specified exposure path $\vs$. This holds both within the carry-over window (where covariates may be affected by the exposure) and before the carry-over window (where they evolve naturally). 
Similarly, Part \ref{ass_exp-B} ensures that the outcome transition mechanisms are stable over time. By conditioning on the path-dependent histories $\mathcal{H}_i^X(\vs)$ and $\mathcal{H}_i^Y(\vs)$, we account for all time-varying modifiers that characterize the population's response to the exposure. This stability allows us to "impute" the future evolution of the effect modifiers by recursively unrolling densities estimated from the observed historical data. 


In practice, to simulate to predict these time-varying modifiers, the exposure history is missing and must also be imputed.
As for the treatment in Section \ref{sec:pred}, for $k\in \mathcal{T}$ we keep $S_{ik}$ equal to the observed value. 
On the contrary, the exposure history before the carry-over period 
when not observed, must be imputed relying on the Assumption \ref{ass:transp_exp_dis}. 

Following the structural transportability assumptions established above, we now provide the formal identification results for the forecasting estimands. We first present the identification of the conditional exposure-response function, which serves as the fundamental building block for predicting future outcomes under counterfactual regimes.

\begin{proposition}[Identification of Future Exposure-Response Functions]
\label{prop:transp_exp}
\textit{Assume the following conditions hold for a future time point $t \in \mathcal{T}^F$ and a hypothetical exposure path $\mathbf{s}$:}
\begin{enumerate}
    \item {Assumption \ref{ass:SUTVA2}} (SUTVA on Exposure);
    \item {Consistency:} $Y^{obs}_{it} = Y_{it}(\mathbf{s})$ if $\overline{\mathbf{S}}_{it}^{B, K} = \mathbf{s}$ for $t \in \mathcal{T}$;
    \item {Assumption \ref{ass:transp_exp}} (Outcome Transportability) holds for $\mathcal{D} = \mathcal{T} \cup \{t\}$;
    \item {Assumption \ref{ass:transp_exp_dis}} (Exposure Transportability) holds for $\mathcal{D} = \mathcal{T} \cup \{T+1, \dots, t-B-K\}$;
    \item {Assumption \ref{ass:conf_transp_exp}} (Transition Transportability) holds for $\mathcal{D} = \mathcal{T} \cup \{T+1, \dots, t\}$;
    \item {Recursive Positivity:} 
    $\text{supp}(f_{\overline{\mathbf{R}}_{it}, \overline{\mathbf{G}}_{it}}(\cdot \mid t \in \mathcal{T}^F, \mathbf{X}_{i0})) \subseteq \text{supp}(f_{\overline{\mathbf{R}}_{it}, \overline{\mathbf{G}}_{it}}(\cdot \mid t \in \mathcal{T}, \mathbf{X}_{i0}))$, where $\overline{\mathbf{G}}_{it} = \{\mathbf{X}_{ik}, \mathbf{Y}_{ik}, \mathbf{S}_{ik}\}_{k=t-B-K-L_{max}}^{t-B-K-1}$, with $L_{max} = \max(L_x, L_y, L_{xx}, L_{xy})$, is an  intermediate vector of covariates, outcomes, and exposures.
\end{enumerate}
\textit{Then, the expected potential outcome $\mathbb{E}[Y_{it}(\mathbf{s}) \mid t \in \mathcal{T}^F, \mathbf{X}_{i0}]$ is identified as:}
\begin{equation}
\label{eq:ident_future_ERF_collapsed}
    \mathbb{E}\Big[Y_{it}(\mathbf{s}) \mid t \in \mathcal{T}^F, \mathbf{X}_{i0}\Big] = \sum_{\overline{\mathbf{r}}} \Psi(\mathbf{s}, \overline{\mathbf{r}}, \overline{\mathbf{g}}, \mathbf{X}_{i0}) \times f_{\overline{\mathbf{R}}_{it}, \overline{\mathbf{G}}_{it}}(\overline{\mathbf{r}} , \overline{\mathbf{g}}\mid t \in \mathcal{T}^F, \mathbf{X}_{i0})
\end{equation}
\textit{where $\Psi(\mathbf{s}, \overline{\mathbf{r}},\overline{\mathbf{g}},  \mathbf{X}_{i0})$ is the historical identification functional:}
\begin{equation}
\label{eq:psi_functional}
    \Psi(\mathbf{s}, \overline{\mathbf{r}}, \overline{\mathbf{g}}, \mathbf{X}_{i0}) = \mathbb{E}\Big[Y^{obs}_{it} \mid t \in \mathcal{T}, \overline{\mathbf{S}}_{it}^{B,K} = \mathbf{s}, \overline{\mathbf{R}}_{it} = \overline{\mathbf{r}}, \overline{\mathbf{G}}_{it} = \overline{\mathbf{g}}, \mathbf{X}_{i0}\Big]
\end{equation}
\textit{and the future density of the pre-exposure modifiers, $f_{\overline{\mathbf{R}}_{it}, \overline{\mathbf{G}}_{it}}(\overline{\mathbf{r}} , \overline{\mathbf{g}}\mid t \in \mathcal{T}^F, \mathbf{X}_{i0})$, is identified (via Assumptions \ref{ass:transp_exp_dis} and \ref{ass:conf_transp_exp}) by the following historical recursive product for $X$, $Y$, and $S$:
}
\begin{equation}
\label{eq:unrolling_specific}
\begin{aligned}
    f(\mathbf{g}, \mathbf{r} \mid \mathcal{H}_{iT}, t \in \mathcal{T}^F) = \sum_{\mathcal{H}_{iT}, \mathcal{H}_{gap} \setminus \{\mathbf{r}, \mathbf{g}\}} \Bigg[ &\prod_{k=T+1}^{t-B-K-1} f(x_{ik} \mid \mathcal{H}_{ik}^X, t \in \mathcal{T}) f(y_{ik} \mid \mathcal{H}_{ik}^Y, t \in \mathcal{T}) f(s_{ik} \mid \mathcal{H}_{ik}^S, t \in \mathcal{T}) \\
&\times f(\bar{\mathbf{x}}_{iT}, \bar{\mathbf{y}}_{iT}, \bar{\mathbf{s}}_{iT} \mid \mathbf{X}_{i0}, t \in \mathcal{T}) \Bigg]
\end{aligned}
\end{equation}
\textit{where $\mathcal{H}_{iT} = \{\mathbf{x}_{ik}, \mathbf{y}_{ik}, \mathbf{s}_{ik}\}_{k=1}^{T}$ and $\mathcal{H}_{gap} = \{\mathbf{x}_{ik}, \mathbf{y}_{ik}, \mathbf{s}_{ik}\}_{k=T+1}^{t-B-K-1}$ are historical and intermediate gap trajectories, and
 the transition kernels $f(\cdot \mid \dots, t \in \mathcal{T})$ are taken from the observed historical sample $\mathcal{T}$.}

 \noindent Proof is reported in the Appendix.

\end{proposition}

Proposition \ref{prop:transp_exp} establishes the theoretical basis for transporting historical evidence into future predictions. Its structure provides several critical insights.
First, it establishes that the expected value of potential outcomes in the future can be identified directly from the relationship estimated on historical data ($\Psi$), provided the units share the same pre-exposure history $\overline{\mathbf{R}}_{it}$ and $\overline{\mathbf{G}}_{it}$, where the latter is a vector of intermediate covariates, outcomes, and exposures required for Assumption \ref{ass:conf_transp_exp} to identify the evolution of the post-exposure modifiers  $\overline{\mathbf{V}}_{it}=[\overline{\mathbf{X}}_{it}^{B,K-1},\overline{\mathbf{Y}}_{it}^{B,K}]$ and marginalize over it. 
Second, the distribution of the future pre-exposure history  $\overline{\mathbf{R}}_{it}$ and and $\overline{\mathbf{G}}_{it}$ is identified by "unrolling" the historical evolutions of covariates, outcomes, and exposures from the end of history ($T$) to the start of the future intervention window ($t-B-K$), and marginalizing over the distribution of historical data. 
%
Finally, as in Section \ref{sec:pred}, identifying the future exposure-response function requires only the prediction of time-varying modifiers in the pre-exposure period before time $t-B-K$. The evolution of the post-exposure modifiers  $\overline{\mathbf{V}}_{it}=[\overline{\mathbf{X}}_{it}^{B,K-1},\overline{\mathbf{Y}}_{it}^{B,K}]$ does not need to be explicitly predicted for the future, as the transportability of the mechanism allows for averaging over their historical distribution observed under the same exposure realization $\mathbf{s}$. The historical functional $\Psi$ already incorporates how these modifiers respond to the exposure path $\mathbf{s}$, effectively marginalizing over their counterfactual distribution.
Importantly, as opposed to Proposition \ref{prop:transp}, where SUTVA on a binary treatment $D_{it}$ allowed us to identify the potential outcome without explicitly modeling the intervening exposure process, here we strictly require the transportability of the exposure distribution (Assumption \ref{ass:transp_exp_dis}). This assumption is necessary not only to bridge the temporal gap between $T$ and the future window but also to ensure that the marginalization over the post-exposure modifiers $\overline{\mathbf{V}}_{it}$ in the carry-over window is valid. Without the stationarity of the exposure process, the historical functional $\Psi$ would not be sufficient to capture the counterfactual response of the modifiers in the future sample.

We can now characterize the total impact of a hypothetical policy by identifying the Average Exposure Effect in the future ($AEE_F$).

\begin{proposition}[Identification of the Future Average Exposure Effect]
\label{prop:AEEF_identification}
\textit{Under the assumptions of Proposition \ref{prop:transp_exp}, the Average Exposure Effect in the future is identified as:}
\begin{equation*}
\label{eq:AEEF_id_final}
\begin{aligned}
AEE_F =  \frac{1}{|\mathcal{I}_F|\times|\mathcal{T}^F|}\sum_{i\in \mathcal{I}_F}\sum_{t\in\mathcal{T}^F} 
\Big(\int\E\Big[Y_{it}(\vs)|  \vX_{i0}\Big]f_{it}(\vs\mid \mathbf{X}_{i0})d\vs- \int\E\Big[Y_{it}(\vs)|  \vX_{i0}\Big]p^{\star}_{it}(\vs)d\vs \Big)
\end{aligned}
\end{equation*}
\textit{where the expected potential outcome $\mathbb{E}[Y_{it}(\mathbf{s}) \mid t \in \mathcal{T}^F, \mathbf{X}_{i0}]$ is identified by Eq. \eqref{eq:ident_future_ERF_collapsed}, and the natural future exposure distribution $f_{it}(\mathbf{s}\mid \mathbf{X}_{i0})$ of $\mathbf{S}_{it}^{B,K}$ for $ t \in \mathcal{T}^F$ is identified by historical transition kernels:}
\begin{equation}
\label{eq:natural_S_unrolling}
\begin{aligned}
    f_{it}(\vs \mid \mathbf{X}_{i0}) = \sum_{\mathcal{H}_{i,t-B} \setminus \{\vs\}} \Bigg[ \prod_{k=1}^{t-B} & f(x_{ik} \mid \mathcal{H}_{ik}^X, t \in \mathcal{T}) f(y_{ik} \mid \mathcal{H}_{ik}^Y, t \in \mathcal{T}) f(s_{ik} \mid \mathcal{H}_{ik}^S, t \in \mathcal{T}) \Bigg]
\end{aligned}
\end{equation}
\textit{where the transition distributions for $k \in \{1, \dots, T\}$ generate the observed history $\mathcal{H}_{iT}$, and for $k \in \{T+1, \dots, t-B\}$ unroll the future gap and exposure window. The summation $\sum_{\mathcal{H}_{i,t-B} \setminus \{\vs\}}$ denotes marginalization over all covariates and outcomes, as well as exposures outside the window $\overline{\mathbf{S}}_{it}^{B,K}$.}
\end{proposition}

Evaluating the $AEE_F$ requires identifying both the future exposure-response function, which follows Proposition \ref{prop:transp_exp}, and the natural course of the exposure $f_{\overline{\mathbf{S}}_{it}^{B,K}}$. Equation \eqref{eq:natural_S_unrolling} demonstrates that by assuming the stationarity of the exposure process, we can predict the likely path of the exposure in the future carry-over window using the evolution process estimated from the observed past. 

In practice, for each unit in $U_1^F$, the pre-exposure modifiers $\overline{\mathbf{R}}_{it}$ and the baseline exposure vector before $t-B-K$ 
are first predicted by recursively unrolling trends estimated from historical data. Samples of the exposure vector during the treatment period, are then drawn from both the hypothetical distribution $p^{\star}_{it}(\mathbf{s})$ and the predicted natural distribution $f_{\overline{\mathbf{S}}_{it}^{B,K}}$. For both draws, we then estimate the average potential outcomes under the corresponding exposure vector $\vs$, using units in the observed sample exposed to the vector $\overline{S}_{it}^{B,K}=\vs$ and with the same history $\overline{\mathbf{R}}_{it}$ as the predicted one.



 Thus far we have assumed that
 the hypothetical intervention $p^{\star}_{it}(\vs)$ setting the exposure does not depend on time-varying characteristics or outcomes or on previous exposures.
  However, we could think of dynamic conditional hypothetical interventions that could potentially depend on previous outcomes or exposures, i.e.,
  $p^{\star}_{it}(\vs| \overline{S}^{1,L^{\star}_{ss}-1}_{it}, \overline{\vX}_{it}^{0,L^{\star}_{xs}}, \overline{\mathbf{Y}}_{it}^{1, L^{\star}_{ys}-1},  \mathbf{X}_{i0})$.
 An example of these type of interventions is the Italian color code classification system, imposing different restrictions depending on the current effective reproductive number and hospitalization \citep{italycolors}. \cite{davies2020effects} and \cite{naturemedicine2021} conducted model-based projections under similar hypothetical interventions triggered by different conditions. 
 In such cases, the time-varying modifiers $\overline{\mathbf{V}}_{it}$ and exposures within the carry-over period must be sequentially imputed, as the intervention is triggered by the evolving state of the system.
 

For example, as already mentioned, the exposure to SARS-CoV-2 can be represented by the number of infectious cases circulating and the effective reproductive number. The relationship between this exposure and the mortality rate -- the exposure-response function -- depends on the case fatality rate. In fact, given the knowledge of the case fatality rate for SARS-CoV-2, and provided the number of cases in a population, we could compute the number of deaths we could have prevented under a hypothetical reduction of the effective reproductive number. If we were interested in conducting this exercise for a future period, we cannot just rely on the estimated exposure-response function, or in this setting, the estimated case fatality rate. The reason is that the mortality risk of COVID-19 depends on the population characteristics, but also on virus mutations which could affect the deadliness of the virus, the amount of medical supplies, hospital capacity, the development and availability of treatments. All these modifiers are likely to change over time. Thus, the case fatality rate estimated on the past sample might not be representative of the mortality risk for a future sample. However, if we were able to estimate the heterogeneity of the case fatality rate with respect to these modifiers, by relying on the temporal transportality assumption (Proposition \ref{prop:transp_exp}) we would be able to predict on a future sample the case fatality rate and, in turn, the mortality rate under a hypothetical intervention on the $R_e$. 
Nevertheless, we must be careful to consider and include in the analysis all the time-varying modifiers, some of which could be related to unmeasured characteristics. 
Moreover, 
given the limited testing and contact tracing capacity and because a large number of infections with SARS-CoV-2 are pauci-symptomatic or even asymptomatic, the uncertainty around the number of undetected infections result in limited knowledge on the actual infection fatality rate and its heterogeneity.


\begin{remark}[Alternative estimand: conditioning on observed history]
An alternative estimand of practical interest is the history-conditioned future average exposure effect, which conditions the expectations on the full observed history $\mathcal{H}_{iT} = \{\mathbf{X}_{ik}, \mathbf{Y}_{ik}, \mathbf{S}_{ik}\}_{k=1}^{T}$, in addition to baseline $\mathbf{X}_{i0}$, i.e., $\mathbb{E}[Y_{it}(\mathbf{s}) \mid \mathcal{H}_{iT}, \mathbf{X}_{i0}, t \in \mathcal{T}^F]$.

This conditioning simplifies the identification by treating the historical path as a fixed initial condition rather than a random variable to be marginalized. 
In particular, identification is simplified by initiating the recursive unrolling in Propositions \ref{prop:transp_exp} and \ref{prop:AEEF_identification} strictly from $T+1$,
eliminating the need for the marginalization over the observed past, $f(\mathcal{H}_{iT} \mid \mathbf{X}_{i0})$.
\end{remark}




\section{Proofs}

\subsection{Proof of Proposition \ref{prop:transp}}

To identify the expected potential outcome in the future, $\E[Y_{it}(d) \mid t \in \mathcal{T}^F, \vX_{i0}]$, we must bridge the counterfactual future distribution to the observed historical data. The identification relies on the fact that while the values of the modifiers and outcomes change, the structural laws governing their evolution are invariant across domains, as under the Assumptions \ref{ass:transp} and \ref{ass:conf_transp}.  We decompose the identification into a recursive simulation of state variables followed by a transported outcome mapping.
%


\noindent Recall $\mathcal{H}_{it}(d) = \{ \overline{\mathbf{Z}}_{it}^{B, K}: \mathbf{D}_{it} = d, \overline{\mathbf{X}}_{it}^{B, L_x}, \overline{\mathbf{Y}}_{it}^{B, L_y}, \mathbf{X}_{i0} \}$. We utilize the decomposition of the path-dependent history $\mathcal{H}_{it}(d) = \{ \overline{\mathbf{Z}}_{it}^{B, K}: \mathbf{D}_{it} = d, \overline{\mathbf{R}}_{it}, \overline{\mathbf{V}}_{it}]$, where $\overline{\mathbf{Z}}_{it}^{B, K}$ is the treatment path such that $D_{it}=d$, $\overline{\mathbf{R}}_{it}=[\overline{\vX}_{it}^{B+K,L_x-K}, \overline{\vY}_{it}^{B+K+1,L_y-K}]$ is the set of pre-treatment modifiers and $\overline{\mathbf{V}}_{it}=[\overline{\vX}_{it}^{B,K-1},\overline{\vY}_{it}^{B,K}]$ is the set of time-varying modifiers during the treatment window $t-B-K, t-B$. 
Additionally, let $\mathbf{G}_{it} = \{X_{ik}, Y_{ik}, Z_{ik}\}_{k=t-B-K-L_{max}}^{t-B-K-1}$ be the vector of variables in  the intermediate period required to bridge the historical data to the carry-over window. 

Here, we also make explicit that we consider a treatment vector $\overline{\mathbf{Z}}_{1:t-B-K-1} = \tilde{\mathbf{z}}$ from baseline to the pre-treatment period. Therefore, the treatment vector in $\mathbf{G}_{it}$ is set to be equal to this pre-specified treatment history, i.e., $\mathbf{G}_{it} = \{X_{ik}, Y_{ik}, \tilde{z}_{ik}\}_{k=t-B-K-L_{max}}^{t-B-K-1}$, and the distributions of the effect modifiers will be conditioned on it. 

We first prove the following equality:

\begin{equation*}
\begin{aligned}
\label{eq:p3_step1}
   \E\Big[Y_{it}(d) \mid t \in \mathcal{T}^F, \vX_{i0}\Big]
    =& \sum_{\overline{\mathbf{r}}} \sum_{\overline{\mathbf{g}}}
  \E\Big[Y^{obs}_{it} \mid t \in \mathcal{T}, D_{it}=d, \overline{\mathbf{G}}_{it}=\overline{\mathbf{g}}, \vX_{i0}\Big]\\  
 &\times f_{\overline{\mathbf{R}}_{it},\overline{\mathbf{G}}_{it}}(\overline{\mathbf{r}},\overline{\mathbf{g}} \mid \tilde{\mathbf{z}}, t \in \mathcal{T}^F, \vX_{i0}) 
 \end{aligned}
\end{equation*}
The proof proceeds with the following steps.

\begin{enumerate}
    \item[Step 1]
{Marginalization over the full counterfactual history.} 
By the Law of Total Iterated Expectations, we marginalize over the joint distribution of the treatment path and modifiers for the target population ($t \in \mathcal{T}^F$):
\begin{equation}
\begin{aligned}
\label{eq:p3_step1_Z}
    \E\Big[Y_{it}(d) \mid t \in \mathcal{T}^F, \vX_{i0}\Big] = \sum_{\overline{\vz}: \mathbf{D}_{it} = d}\sum_{\overline{\mathbf{r}}}\sum_{\overline{\mathbf{g}}}\sum_{\overline{\mathbf{v}}} &\E\Big[Y_{it}(d) \mid t \in \mathcal{T}^F, \overline{\mathbf{Z}}_{it}^{B, K}=\overline{\vz}, \overline{\mathbf{V}}_{it}=\overline{\mathbf{v}}, \overline{\mathbf{R}}_{it}=\overline{\mathbf{r}}, \overline{\mathbf{G}}_{it}=\overline{\mathbf{g}},\vX_{i0}\Big] \\&\times f(\overline{\vz}, \overline{\mathbf{v}}, \overline{\mathbf{r}},\overline{\mathbf{g}},  \mid \tilde{\mathbf{z}}, t \in \mathcal{T}^F, \vX_{i0})
    \end{aligned}
\end{equation}

\item[Step 2]{ Transporting potential outcomes.} 
Under {Assumption \ref{ass:transp}}, the distribution of potential outcomes $Y_{it}(d)$ time $t \in \mathcal{T}^F$ is equal to that in the observed time frame $A \in \mathcal{T}$. 
\begin{equation*}
 \begin{aligned}
    \E\Big[Y_{it}(d) \mid t \in \mathcal{T}^F, \overline{\mathbf{Z}}_{it}^{B, K}=\overline{\vz}, &\overline{\mathbf{V}}_{it}=\overline{\mathbf{v}}, \overline{\mathbf{R}}_{it}=\overline{\mathbf{r}}, \overline{\mathbf{G}}_{it}=\overline{\mathbf{g}},\vX_{i0}\Big]=\\& \E\Big[Y_{it}(d) \mid t \in \mathcal{T}, \overline{\mathbf{Z}}_{it}^{B, K}=\overline{\vz}, \overline{\mathbf{V}}_{it}=\overline{\mathbf{v}}, \overline{\mathbf{R}}_{it}=\overline{\mathbf{r}}, \overline{\mathbf{G}}_{it}=\overline{\mathbf{g}},\vX_{i0}\Big]
     \end{aligned}
\end{equation*}

\item[Step 3]{Collapsing the joint distribution.} 
We then apply {Assumption \ref{ass:conf_transp}} to the modifiers $\overline{\mathbf{V}}_{it}$ in the treatment window: the future conditional distribution of $\overline{\mathbf{V}}_{it}$ matches that observed in the observed period:
\begin{equation*}
    f(\overline{\mathbf{v}} \mid  t \in \mathcal{T}^F, \overline{\vZ}_{it}^{B,K}: \mathbf{D}_{it} = d, \overline{\mathbf{R}_{it}}=\overline{\mathbf{r}}, \overline{\mathbf{G}}_{it}=\overline{\mathbf{g}},\vX_{i0}) = f(\overline{\mathbf{v}} \mid  t \in \mathcal{T}, \overline{\vZ}_{it}^{B,K}: \mathbf{D}_{it} = d, \overline{\mathbf{R}_{it}}=\overline{\mathbf{r}}, \overline{\mathbf{G}}_{it}=\overline{\mathbf{g}},\vX_{i0})
\end{equation*}
In particular, for each $j \in [t-B-K, t-B]$, let the histories be $\mathcal{H}_{ij}^Y = \{ \overline{\tilde{\vz}}_{ij}^{B, K}, \overline{\mathbf{x}}_{ij}^{B, L_y}, \overline{\mathbf{y}}_{ij}^{B, L_y}, \mathbf{X}_{i0} \}$ and $\mathcal{H}_{ij}^X = \{ \overline{\tilde{\vz}}_{ij}^{1, L_s}, \overline{\mathbf{x}}_{ij}^{1, L_{xx}}, \overline{\mathbf{y}}_{ij}^{1, L_{xy}}, \mathbf{X}_{i0} \}$. These histories are constructed from the  realizations $\mathbf{v}, \mathbf{r}, \mathbf{g}$, and $\tilde{\vz}$. Under {Assumption} \ref{ass:conf_transp}, we have:
\begin{equation*}
f(\overline{\mathbf{v}} \mid  t \in \mathcal{T}^F, \overline{\vZ}_{it}^{B,K}: \mathbf{D}_{it} = d, \overline{\mathbf{R}_{it}}=\overline{\mathbf{r}}, \overline{\mathbf{G}}_{it}=\overline{\mathbf{g}},\vX_{i0}) = \prod_{j=t-B-K}^{t-B} f(x_{ij} \mid \mathcal{H}_{ij}^X, t \in \mathcal{T}) f(y_{ij} \mid \mathcal{H}_{ij}^Y, t \in \mathcal{T})
\end{equation*}
Substituting these into Eq. \ref{eq:p3_step1_Z}, the inner summation over $[\overline{\vz}, \overline{\mathbf{v}}]$ recovers the historical conditional mean:
\begin{equation*}
\begin{aligned}
    \sum_{\overline{\vz}:D_{it}=d} \sum_{\overline{\mathbf{v}}}\sum_{\overline{\mathbf{g}}} &\E\Big[Y_{it}(d) \mid t \in \mathcal{T}, \overline{\mathbf{Z}}_{it}^{B, K}=\overline{\vz}, \overline{\mathbf{V}}_{it}=\overline{\mathbf{v}}, \overline{\mathbf{R}}_{it}=\overline{\mathbf{r}},\overline{\mathbf{G}}_{it}=\overline{\mathbf{g}}, \mathbf{X}_{i0}\Big] \\&\times f(\overline{\mathbf{v}} \mid  t \in \mathcal{T}, \overline{\vZ}_{it}^{B,K}: \mathbf{D}_{it} = d, \overline{\mathbf{R}_{it}}=\overline{\mathbf{r}}, \overline{\mathbf{G}}_{it}=\overline{\mathbf{g}}, \vX_{i0})\\&\times f(\overline{\vz}\mid t \in \mathcal{T}^F, \overline{\mathbf{r}}) =
    \\&\sum_{\overline{\vz}:D_{it}=d}\E\Big[Y_{it}(d) \mid t \in \mathcal{T}, \overline{\mathbf{Z}}_{it}^{B, K}=\overline{\vz},  \overline{\mathbf{R}}_{it}=\overline{\mathbf{r}},\overline{\mathbf{G}}_{it}=\overline{\mathbf{g}}, \vX_{i0}\Big]f(\overline{\vz}\mid t \in \mathcal{T}^F, \overline{\mathbf{r}})= 
    \\&\E\Big[Y_{it}(d) \mid t \in \mathcal{T}, D_{it}= d, \overline{\mathbf{R}}_{it}=\overline{\mathbf{r}}, \overline{\mathbf{G}}_{it}=\overline{\mathbf{g}}, \vX_{i0}\Big]
\end{aligned}
\end{equation*}
where the last equality holds thanks to Assumption \ref{ass:SUTVA_E} (SUTVA).
\item[Step 4] Consistency. Under {Consistency}, the potential outcome $Y_{it}(d)$ is equal to $Y^{obs}_{it}$ when the observed treatment path $\overline{\vZ}_{it}$  is consistent with $d$. Therefore:
\begin{equation*}
\begin{aligned}
\E\Big[Y_{it}(d) \mid t \in \mathcal{T}, D_{it}= d, \overline{\mathbf{R}}_{it}=\overline{\mathbf{r}}, \overline{\mathbf{G}}_{it}=\overline{\mathbf{g}},\vX_{i0}\Big]=
\E\Big[Y^{obs}_{it} \mid t \in \mathcal{T}, D_{it}= d, \overline{\mathbf{R}}_{it}=\overline{\mathbf{r}}, \overline{\mathbf{G}}_{it}=\overline{\mathbf{g}},\vX_{i0}\Big]
\end{aligned}
\end{equation*}

\item[Step 4] Recursive Identification of the Pre-treatment Modifiers. 
Finally, the density $f(\overline{\mathbf{r}}, \overline{\mathbf{g}} \mid \mathbf{\tilde{z}}, t \in \mathcal{T}^F, \mathbf{X}_{i0})$ is identified by recursively applying {Assumption \ref{ass:conf_transp}} for all time points from baseline $\vX_{i0}$ to the start of the treatment window. 

Because the treatment path $\tilde{\mathbf{z}}$ is fixed for this period, we only unroll the domain-invariant transition kernels for $X$ and $Y$ conditioned on the specific values in $\tilde{\mathbf{z}}$:

\begin{equation}
\label{eq:p3_unrolling_fixed_background_final}
\begin{aligned}
f(\overline{\mathbf{r}}, \overline{\mathbf{g}} \mid \tilde{\mathbf{z}}, t \in \mathcal{T}^F, \mathbf{X}_{i0}) = \sum_{\mathcal{H}_{iT}, \mathcal{H}_{gap} \setminus \{\overline{\mathbf{r}}, \overline{\mathbf{g}}\}} \Bigg[ &\prod_{k=T+1}^{t-B-K-1} f(x_{ik} \mid \mathcal{H}_{ik}^X(\tilde{\mathbf{z}}), \tilde{z}_{ik}, t \in \mathcal{T}) f(y_{ik} \mid \mathcal{H}_{ik}^Y(\tilde{\mathbf{z}}), t \in \mathcal{T}) \\ 
&\times f(\mathcal{H}_{iT} \mid \mathbf{x}_{i0}, t \in \mathcal{T}) \Bigg]
\end{aligned}
\end{equation}

To clarify the components of this identification functional, we define the following trajectories:
\begin{itemize}
    \item $\mathcal{H}_{iT} = \{\mathbf{x}_{ik}, \mathbf{y}_{ik}, \tilde{\mathbf{z}}_{ik}\}_{k=1}^{T}$ represents the {observed historical trajectory}, which serves as the empirical initialization for the unrolling process.
    \item $\mathcal{H}_{gap} = \{\mathbf{x}_{ik}, \mathbf{y}_{ik}, \tilde{\mathbf{z}}_{ik}\}_{k=T+1}^{t-B-K-1}$ represents the {intermediate gap trajectory}, where modifiers $\{X, Y\}$ evolve naturally in response to the fixed pre-treatment values $\tilde{z}_{ik}$.
    \item $\mathcal{H}_{ik}^X(\tilde{\mathbf{z}})$ and $\mathcal{H}_{ik}^Y(\tilde{\mathbf{z}})$ represent the {sliding-window conditioning histories} required by the transition kernels. These are constructed using the evolving realizations of modifiers and the fixed values from the vector $\tilde{\mathbf{z}}$, ensuring the transition lags $L_{max}$ are satisfied for each step $k$.
\end{itemize}
The summation $\sum_{\mathcal{H}_{iT}, \mathcal{H}_{gap} \setminus \{\overline{\mathbf{r}}, \overline{\mathbf{g}}\}}$ denotes the {marginalization} over all possible joint trajectories of the past and gap data, excluding the specific sub-vectors $\overline{\mathbf{r}}$ and $\overline{\mathbf{g}}$ that constitute the pre-treatment state and the buffer required for the transportability of the carry-over modifiers.

\end{enumerate}



The {Recursive Positivity} condition ensures that the counterfactual history $H_{it}(d)$ remains within the support of the historical data throughout the unrolling process. Thus, the expected potential outcome is fully identified as a functional of the historical data distribution, completing the proof. \hfill $\square$

\subsection{Proof of Proposition \ref{prop:transp_exp}}

To identify the expected potential outcome in the future, $\mathbb{E}[Y_{it}(\mathbf{s}) \mid t \in \mathcal{T}^F, \mathbf{X}_{i0}]$, we mainly rely on  Assumptions \ref{ass:transp_exp} \ref{ass:transp_exp_dis}, and \ref{ass:conf_transp_exp}. We decompose the identification into a recursive simulation of state variables followed by a transported outcome mapping.

Recall $\mathcal{H}_{it}(\mathbf{s}) = \{ \overline{\mathbf{S}}_{it}^{B, K} = \mathbf{s}, \overline{\mathbf{X}}_{it}^{B, L_x}, \overline{\mathbf{Y}}_{it}^{B, L_y}, \mathbf{X}_{i0} \}$. We utilize the decomposition of the path-dependent history $\mathcal{H}_{it}(\mathbf{s}) = \{ \overline{\mathbf{S}}_{it}^{B, K} = \mathbf{s}, \overline{\mathbf{R}}_{it}, \overline{\mathbf{V}}_{it}\}$, where $\mathbf{s}$ is the hypothetical exposure path, $\overline{\mathbf{R}}_{it}= \{\overline{\mathbf{X}}_{it}^{B+K, L_x-K}, \overline{\mathbf{Y}}_{it}^{B+K, L_y-K}\}$ is the set of pre-exposure modifiers and $\overline{\mathbf{V}}_{it}=[\overline{\mathbf{X}}_{it}^{B,K-1},\overline{\mathbf{Y}}_{it}^{B,K}]$ is the set of time-varying modifiers during the carry-over window $t-B-K, t-B$.
In addition, let $\mathbf{G}_{it} = \{X_{ik}, Y_{ik}, S_{ik}\}_{k=t-B-K-L_{max}}^{t-B-K-1}$ be the intermediate trajectory in the gap required to satisfy the transportability of the effect modifiers, where $L_{max} = \max(L_x, L_y, L_s, L{xx}, L_{xy})$. 

The proof proceeds with the following steps.

\begin{enumerate}
    \item[Step 1] Marginalization over the full counterfactual history.
By the Law of Total Iterated Expectations, we marginalize over the joint distribution of the exposure path and modifiers for the target population ($t \in \mathcal{T}^F$):
\begin{equation}
\begin{aligned}
\label{eq:p6_step1_S}
    \mathbb{E}\Big[Y_{it}(\mathbf{s}) \mid t \in \mathcal{T}^F, \mathbf{X}_{i0}\Big] = \sum_{\overline{\mathbf{r}}}\sum_{\overline{\mathbf{g}}} \sum_{\overline{\mathbf{v}}}&\mathbb{E}\Big[Y_{it}(\mathbf{s}) \mid t \in \mathcal{T}^F, \overline{\mathbf{S}}_{it}^{B, K}=\mathbf{s}, \overline{\mathbf{V}}_{it}=\overline{\mathbf{v}}, \overline{\mathbf{R}}_{it}=\overline{\mathbf{r}}, \overline{\mathbf{G}}_{it}=\overline{\mathbf{g}},\mathbf{X}_{i0}\Big] \\
    &\times f(\overline{\mathbf{v}} \mid t \in \mathcal{T}^F, \overline{\mathbf{S}}_{it}^{B, K}=\mathbf{s}, \overline{\mathbf{R}}_{it}=\overline{\mathbf{r}}, \overline{\mathbf{G}}_{it}=\overline{\mathbf{g}}, \mathbf{X}_{i0}) \\
    &\times  f(\overline{\mathbf{r}}, \overline{\mathbf{g}}\mid t \in \mathcal{T}^F, \mathbf{X}_{i0})
\end{aligned}
\end{equation}

\item[Step 2] Transporting potential outcomes.
Under Assumption \ref{ass:transp_exp}, the distribution of potential outcomes $Y_{it}(\mathbf{s})$ at time $t \in \mathcal{T}^F$ is equal to that in the observed time frame $t \in \mathcal{T}$:
\begin{equation*}
\begin{aligned}
    \mathbb{E}\Big[Y_{it}(\mathbf{s}) \mid t \in \mathcal{T}^F, \overline{\mathbf{S}}_{it}^{B, K}=&\mathbf{s}, \overline{\mathbf{V}}_{it}=\overline{\mathbf{v}}, \overline{\mathbf{R}}_{it}=\overline{\mathbf{r}}, \overline{\mathbf{G}}_{it}=\overline{\mathbf{g}},\mathbf{X}_{i0}\Big] = \\&\mathbb{E}\Big[Y_{it}(\mathbf{s}) \mid t \in \mathcal{T}, \overline{\mathbf{S}}_{it}^{B, K}=\mathbf{s}, \overline{\mathbf{V}}_{it}=\overline{\mathbf{v}}, \overline{\mathbf{R}}_{it}=\overline{\mathbf{r}}, \overline{\mathbf{G}}_{it}=\overline{\mathbf{g}},\mathbf{X}_{i0}\Big]
    \end{aligned}
\end{equation*}

\item [Step 3] Identification of Carry-over Transitions and Collapsing the joint distribution.
We then apply Assumption \ref{ass:conf_transp_exp} 
to the modifiers $\overline{\mathbf{V}}_{it}$ in the carry-over window: the future conditional distribution of $\overline{\mathbf{V}}_{it}$ matches that observed in the past period:
\begin{equation*}
    f(\overline{\mathbf{v}} \mid t \in \mathcal{T}^F, \overline{\mathbf{S}}_{it}^{B,K} = \mathbf{s}, \overline{\mathbf{R}}_{it}=\overline{\mathbf{r}}, \overline{\mathbf{G}}_{it}=\overline{\mathbf{g}}, \mathbf{X}_{i0}) = f(\overline{\mathbf{v}} \mid t \in \mathcal{T}, \overline{\mathbf{S}}_{it}^{B,K} = \mathbf{s}, \overline{\mathbf{R}}_{it}=\overline{\mathbf{r}}, \overline{\mathbf{G}}_{it}=\overline{\mathbf{g}},\mathbf{X}_{i0})
\end{equation*}
In particular, for each $j \in [t-B-K, t-B]$, let the histories be $\mathcal{H}_{ij}^Y = \{ \overline{\vs}_{ij}^{B, K}, \overline{\mathbf{x}}_{ij}^{B, L_y}, \overline{\mathbf{y}}_{ij}^{B, L_y}, \mathbf{X}_{i0} \}$ and $\mathcal{H}_{ij}^X = \{ \overline{\vs}_{ij}^{1, L_s}, \overline{\mathbf{x}}_{ij}^{1, L_{xx}}, \overline{\mathbf{y}}_{ij}^{1, L_{xy}}, \mathbf{X}_{i0} \}$. These histories are constructed from the fixed path $\vs$ and realizations $\mathbf{v}, \mathbf{r}, \mathbf{g}$. Under {Assumption} \ref{ass:conf_transp_exp}, we have:
\begin{equation*}
f(\overline{\mathbf{v}} \mid t \in \mathcal{T}^F, \overline{\mathbf{S}}_{it}^{B,K} = \mathbf{s}, \overline{\mathbf{R}}_{it}=\overline{\mathbf{r}}, \overline{\mathbf{G}}_{it}=\overline{\mathbf{g}}, \mathbf{X}_{i0}) = \prod_{j=t-B-K}^{t-B} f(x_{ij} \mid \mathcal{H}_{ij}^X, t \in \mathcal{T}) f(y_{ij} \mid \mathcal{H}_{ij}^Y, t \in \mathcal{T})
\end{equation*}

Substituting these into Eq. \eqref{eq:p6_step1_S}, the inner summation over $\overline{\mathbf{v}}$ recovers the historical conditional mean:
\begin{equation*}
\begin{aligned}
    \sum_{\overline{\mathbf{v}}} &\mathbb{E}\Big[Y_{it}(\mathbf{s}) \mid t \in \mathcal{T}, \overline{\mathbf{S}}_{it}^{B, K}=\mathbf{s}, \overline{\mathbf{V}}_{it}=\overline{\mathbf{v}}, \overline{\mathbf{R}}_{it}=\overline{\mathbf{r}}, \overline{\mathbf{G}}_{it}=\overline{\mathbf{g}},\mathbf{X}_{i0}\Big] f(\overline{\mathbf{v}} \mid t \in \mathcal{T}, \overline{\mathbf{S}}_{it}^{B, K} = \mathbf{s}, \overline{\mathbf{R}}_{it}=\overline{\mathbf{r}}, \overline{\mathbf{G}}_{it}=\overline{\mathbf{g}},\mathbf{X}_{i0}) \\
    &= \mathbb{E}\Big[Y_{it}(\mathbf{s}) \mid t \in \mathcal{T}, \overline{\mathbf{S}}_{it}^{B, K}= \mathbf{s}, \overline{\mathbf{R}}_{it}=\overline{\mathbf{r}},\overline{\mathbf{G}}_{it}=\overline{\mathbf{g}}, \mathbf{X}_{i0}\Big]
\end{aligned}
\end{equation*}

\item[Step 4] Consistency. Under Consistency, the potential outcome $Y_{it}(\mathbf{s})$ is equal to $Y^{obs}_{it}$ when the observed exposure vector is $\mathbf{s}$. Therefore:
\begin{equation*}
\begin{aligned}
\mathbb{E}\Big[Y_{it}(\mathbf{s}) \mid t \in \mathcal{T}, \overline{\mathbf{S}}_{it}^{B, K}= \mathbf{s}, \overline{\mathbf{R}}_{it}=\overline{\mathbf{r}}, \overline{\mathbf{G}}_{it}=\overline{\mathbf{g}},\mathbf{X}_{i0}\Big]=
\mathbb{E}\Big[Y^{obs}_{it} \mid t \in \mathcal{T}, \overline{\mathbf{S}}_{it}^{B, K} = \mathbf{s}, \overline{\mathbf{R}}_{it}=\overline{\mathbf{r}}, \overline{\mathbf{G}}_{it}=\overline{\mathbf{g}},\mathbf{X}_{i0}\Big]
\end{aligned}
\end{equation*}
This recovered term is the functional $\Psi(\mathbf{s}, \overline{\mathbf{r}}, \overline{\mathbf{g}},\mathbf{X}_{i0})$ as defined in the proposition.

\item[Step 5] Final Unrolling and Marginalization of the Gap. 
The density $f(\mathbf{r}, \mathbf{g} \mid t \in \mathcal{T}^F, \mathbf{X}_{i0})$ is identified by recursively applying Assumptions \ref{ass:conf_transp_exp} and \ref{ass:transp_exp_dis} for all time points from baseline to the start of the carry-over window. 
Let $\mathcal{H}_{gap} = \{X_k, Y_k, S_k\}_{k=T+1}^{t-B-K-1}$ and $\mathcal{H}_{iT} = \{X_k, Y_k, S_k\}_{k=1}^{T}$ be the intermediate and historical trajectories before the treatment window. Then
\begin{equation}
\label{eq:final_id_rigorous}
\begin{aligned}
f(\mathbf{g}, \mathbf{r} \mid \mathcal{H}_{iT}, t \in \mathcal{T}^F) = \sum_{\mathcal{H}_{iT}, \mathcal{H}_{gap} \setminus \{\mathbf{r}, \mathbf{g}\}} \Bigg[ &\prod_{k=T+1}^{t-B-K-1} f(x_{ik} \mid \mathcal{H}_{ik}^X, t \in \mathcal{T}) f(y_{ik} \mid \mathcal{H}_{ik}^Y, t \in \mathcal{T}) f(s_{ik} \mid \mathcal{H}_{ik}^S, t \in \mathcal{T}) \\
&\times f(\bar{\mathbf{x}}_{iT}, \bar{\mathbf{y}}_{iT}, \bar{\mathbf{s}}_{iT} \mid \mathbf{X}_{i0}, t \in \mathcal{T}) \Bigg]
\end{aligned}
\end{equation}
where $f(\bar{\mathbf{x}}_{iT}, \bar{\mathbf{y}}_{iT}, \bar{\mathbf{s}}_{iT} \mid \mathbf{X}_{i0}, t \in \mathcal{T})$ is the density of the observed historical data given baseline, and the summation denotes marginalization over all variables in the observed history $\mathcal{H}_{iT}$ and gap $\mathcal{H}_{gap}$ except for the target sub-vectors $\{\mathbf{r}, \mathbf{g}\}$.

By unrolling through these domain-invariant kernels, the distribution of the covariates and exposures in the temporal gap is identified from the historical data. 
\end{enumerate}

\noindent The Recursive Positivity condition ensures that the counterfactual histories remain within the support of historical data. This completes the proof for the potential outcome. \hfill $\square$

\subsection{Proof of Proposition \ref{prop:AEEF_identification}}
\begin{enumerate}
    \item[Step 1]
 Identification of Counterfactual Outcomes. By Proposition \ref{prop:transp_exp}, the term $\mathbb{E}[Y_{it}(\mathbf{s}) \mid t \in \mathcal{T}^F, \mathbf{X}_{i0}]$ is identified for any $\mathbf{s}$ using historical data. Integrating this over the hypothetical distribution $p^{\star}_{it}(\mathbf{s})$ identifies the intervention component of the $AEE_F$.

\item[Step 2] Identification of the Natural Exposure. Under Assumptions \ref{ass:transp_exp_dis} and \ref{ass:conf_transp_exp}, the transition kernels for the joint process $\{X, Y, S\}$ are time-invariant. We identify the natural future exposure distribution $f_{it}(\mathbf{s} \mid \mathbf{X}_{i0})$ by unrolling the system from $k=1$ to $t-B$. Specifically, we marginalize the joint density of the full trajectory $\mathcal{H}_{i,t-B}$ over all variables except the window of interest $\overline{\mathbf{S}}_{it}^{B,K} = \mathbf{s}$:
\begin{equation*}
f_{it}(\mathbf{s} \mid \mathbf{x}_{i0}) = \sum_{\mathcal{H}_{i,t-B} \setminus \{\mathbf{s}\}} \prod_{k=1}^{t-B} f(x_{ik} \mid \mathcal{H}_{ik}^X, t \in \mathcal{T}) f(y_{ik} \mid \mathcal{H}_{ik}^Y, t \in \mathcal{T}) f(s_{ik} \mid \mathcal{H}_{ik}^S, t \in \mathcal{T})
\end{equation*}
where the product includes the observed history for $k \leq T$ and the simulated gap and window for $k > T$. This accounts for the feedback loops where future modifiers $\overline{\mathbf{V}}_{it}$ influence the natural exposure path.

\item[Step 3] Averaging over the Future Domain. The final $AEE_F$ 
 is obtained by taking the difference between the expected potential outcomes under the identified natural distribution $f_{it}$ and the policy $p^{\star}$ and then by taking the sample average across the future unit-time domain $\mathcal{I}_F \times \mathcal{T}^F$.
\end{enumerate}
\hfill $\square$

\end{document}